\documentclass[12pt]{article}
\pdfoutput=1
\usepackage{jheppub, amsmath,amssymb,amsfonts,amsxtra, mathrsfs, makeidx,graphics,graphicx,amsthm,epsfig, ytableau,bm,longtable,float, color,tikz,mathtools,bbm}
\usetikzlibrary{positioning}

\newcommand\longdownarrow{\rotatebox[origin=c]{-90}{\scalebox{1.5}{$\longrightarrow$}}}

\newcommand{\eg}{\textit{e.g.}}

\newcommand{\ie}{\textit{i.e.}}

\numberwithin{equation}{section}

\newcommand{\nn}{\nonumber}

\newcommand{\be}{\begin{equation}} \newcommand{\ee}{\end{equation}}
\newcommand{\bea}{\begin{equation} \begin{aligned}} \newcommand{\eea}{\end{aligned} \end{equation}}

\def\tilde{\widetilde}

\def\bar{\overline}

\def\rt2{\sqrt{2}}

\def\det{\mathop{\rm det}}

\def\Tr{\mathop{\rm Tr}}

\def\CG{{\cal G}}
\def\CH{{\cal H}}

\def\CN{{\cal N}}

\def\1{{\ds 1}}

\newcommand{\cN}{\mathcal{N}}
\newcommand{\cO}{\mathcal{O}}

\newcommand{\bC}{\mathbb{C}}

\newcommand{\bZ}{\mathbb{Z}}

\def\SU{\mathrm{SU}}

\def\repa{\raise4pt\hbox{$\square$}\mkern-14mu\raise-4pt\hbox{$\square$}}
\def\repab{\overline{\raise4pt\hbox{$\square$}\mkern-14mu\raise-4pt\hbox{$\square$}\mkern-1mu}}

\def\smileface{\ensuremath{\hbox{\large$\bigcirc$}\mkern-15mu\raise-1pt\hbox{\scriptsize$\smallsmile$}%
\mkern-10mu\raise4pt\hbox{..}\mkern4mu}}
\def\frownface{\ensuremath{\hbox{\large$\bigcirc$}\mkern-15mu\raise-1pt\hbox{\scriptsize$\smallfrown$}%
\mkern-10mu\raise4pt\hbox{..}\mkern4mu}}

\newcommand{\ba}{\begin{array}}
\newcommand{\ea}{\end{array}}
\newcommand{\bi}{\begin{itemize}}
\newcommand{\ei}{\end{itemize}}
\def\vec#1{\bm{#1}}
\def\bea#1\eea{\allowdisplaybreaks \begin{align}#1\end{align}}
 \newcommand{\ben}{\begin{enumerate}}
\newcommand{\een}{\end{enumerate}}
\newcommand{\bean}{\begin{eqnarray*}}
\newcommand{\eean}{\end{eqnarray*}}
\newcommand{\eref}[1]{(\ref{#1})}

\newcommand{\fref}[1]{Figure~\ref{#1}}
\newcommand{\PE}{\mathop{\rm PE}}

\newcommand{\res}{\mathop{\rm Res}}

\newcommand{\BC}{\mathbb{C}}

\newcommand{\BZ}{\mathbb{Z}}

\newcommand{\comment}[1]{}

\newcommand{\adj}{\mathbf{Adj}}

\newcommand{\Adj}{\mathbf{\adj}}

\newcommand{\fflat}{\mathcal{F}^\flat}

\title{Coulomb branch Hilbert series \\
and Hall-Littlewood polynomials}
\author[a]{Stefano Cremonesi,}
\author[a]{Amihay Hanany,}
\author[b]{Noppadol Mekareeya,}
\author[c,d]{\\and Alberto Zaffaroni}

\affiliation[a]{Theoretical Physics Group, Imperial College London, \\
Prince Consort Road, London, SW7 2AZ, UK}
\affiliation[b]{Theory Division, Physics Department, CERN, \\CH-1211, Geneva 23, Switzerland}
\affiliation[c]{Dipartimento di Fisica, Universit\`a di Milano-Bicocca,  \\ I-20126 Milano, Italy}
\affiliation[d]{INFN, sezione di Milano-Bicocca, I-20126 Milano, Italy}

\emailAdd{s.cremonesi@imperial.ac.uk}
\emailAdd{a.hanany@imperial.ac.uk}
\emailAdd{noppadol.mekareeya@cern.ch}
\emailAdd{alberto.zaffaroni@mib.infn.it}

\preprint{
{\small
\begin{flushright}
CERN-PH-TH/2013-278\\
IMPERIAL-TP-14-SC-01
\end{flushright}
}
}

\abstract{
There has been a recent progress in understanding the chiral ring of 3d ${\cal N}=4$ superconformal gauge theories by explicitly constructing an exact generating function (Hilbert series) counting BPS operators on the Coulomb branch.  
In this paper we introduce Coulomb branch Hilbert series in the presence of background magnetic charges for flavor symmetries, which are useful for computing the Hilbert series of more general theories through gluing techniques. We find a simple formula of the Hilbert series with background magnetic charges for $T_{\vec \rho}(G)$ theories in terms of Hall-Littlewood polynomials.  Here $G$ is a classical group and $\vec \rho$ is a certain partition related to the dual group of $G$. The Hilbert series for vanishing background magnetic charges show that Coulomb branches of $T_{\vec \rho}(G)$ theories are complete intersections. We also demonstrate that mirror symmetry maps background magnetic charges to baryonic charges.
}

\begin{document}
\maketitle

\section{Introduction}

Identifying the chiral ring and moduli space on the Coulomb branch of an $\CN = 4$ supersymmetric gauge theory in $2+1$ dimensions has been a long standing problem.  On a generic point of the Coulomb branch, the triplet of scalars in the $\CN = 4$ vector multiplets acquires a vacuum expectation value, and the gauge fields that remain massless are abelian and can be dualized to scalar fields.  Semiclassically, the Coulomb branch is parametrised by the vacuum expectation values of these four scalars.  This classical description, however, receives quantum corrections.    
The chiral ring associated with the Coulomb branch, in fact, has a complicated structure involving monopole operators in addition to the classical fields in the Lagrangian.  

In spite of the complicated structure of the chiral ring and quantum corrections on the Coulomb branch, it is still possible to enumerate in an exact way the gauge invariant BPS operators that have a non-zero expectation value along the Coulomb branch \cite{Cremonesi:2013lqa}.  The idea is that the chiral ring of the quantum Coulomb branch can be described in terms of monopole operators dressed with scalar fields from the vector multiplet.
The generating function that enumerates such BPS operators according to their quantum numbers is called the {\it Coulomb branch Hilbert series}. This function can be computed for any  $3d$ $\cN=4$ supersymmetric gauge theory that has a Lagrangian description and that are good or ugly in the sense of \cite{Gaiotto:2008ak}.  We review the method proposed in \cite{Cremonesi:2013lqa}, henceforth called the {\it monopole formula} for Coulomb branch Hilbert series, in section \ref{sec:general}.

Another way to compute the Coulomb branch Hilbert series of a given theory is to use mirror symmetry as a working assumption (see \eg~\cite{Hanany:2011db}).  Mirror symmetry exchanges the Coulomb branch of the theory in question with the Higgs branch of another theory \cite{Intriligator:1996ex}, where the latter does not receive quantum corrections.  The Higgs branch Hilbert series can be computed in a conventional way from the Lagrangian of the mirror theory using Molien integrals.  This method has certain limitations, for example, when the Lagrangian of the mirror theory is not available. Even when the latter is known, if the theory contains gauge groups of large ranks or large number of hypermultiplets, the computation of the Molien integrals can become very cumbersome in practice.

One of the aims of this paper (and its companion \cite{Cremonesi:2014vla}) is to develop a machinery for efficiently computing Coulomb branch Hilbert series for several classes of $\cN=4$ gauge theories. 
We can obtain the Hilbert series of the theories in question by `gluing' together the Hilbert series of building blocks. A similar method has been applied successfully to the computation of Higgs branch Hilbert series. The gluing procedure  consists in gauging a common global flavor symmetry of the building blocks. For the gluing machinery to work, we need to define and compute  Coulomb branch Hilbert series in the presence of background magnetic fluxes associated to monopole operators for the global symmetry. The gluing is performed by summing over the background monopole fluxes with an appropriate weight, as discussed in detail in section \ref{sec:general}.

In this  paper we discuss the general properties of the Hilbert series with background fluxes and we provide computations for a class of simple theories,   the  three-dimensional superconformal field theories  known as $T_{\vec \rho}(G)$ \cite{Gaiotto:2008ak}.
The latter are  linear quiver theories with non-decreasing ranks associated with a partition $\vec \rho$ and a flavor symmetry $G$ and  were defined in terms of boundary conditions for 4d $\cN=4$ SYM with gauge group $G$ \cite{Gaiotto:2008ak}. We are able to give a closed analytic expression for the Coulomb branch Hilbert series of $T_{\vec \rho}(G)$. These expressions serve as  basic building blocks for constructing a large class of  more complicated theories. In a companion paper  \cite{Cremonesi:2014vla} we discuss the particularly interesting case of the mirror of Sicilian theories arising from twisted compactification of the $6d$ $(2,0)$ theory on a circle times a Riemann surface with punctures, which can be obtained by gluing copies of $T_{\vec \rho}(G)$ theories \cite{Benini:2010uu}.

One of the main results of this paper  is an intriguing relation between  the Coulomb branch Hilbert series of the $T_{\vec \rho}(G)$ and a class  of symmetric functions, as discussed in sections \ref{sec:HLSclass} and \ref{sec:TSOSp}. We conjecture  that,  given a classical group $G$ and a corresponding partition $\vec \rho$ of the dual group, the Coulomb branch Hilbert series of $T_{\vec \rho}(G)$ with background monopole fluxes for the flavor symmetry $G$ can be written in terms of Hall-Littlewood polynomials (see, \eg~\cite{macdonald1998symmetric} and Appendix \ref{sec:HLpoly}).  
We give several pieces of evidence in support of our conjecture and others are given in the companion paper \cite{Cremonesi:2014vla}. Our general formula is \eref{mainHL}, and for the special case of $G=SU(N)$ the formula is given by \eref{HLformulaT}. We shall henceforth refer to this form of the Coulomb branch Hilbert series as the {\it Hall-Littlewood formula}.%
\footnote{Modified Hall-Littlewood polynomials have appeared in the context of Hilbert series of affinized flag varieties in \cite{Bouwknegt:1999jr}. It would be interesting to relate that formalism to the one of this paper.} 
Hall-Littlewood polynomials have also appeared in the recent literature in the context of the superconformal index of four dimensional ${\cal N}=2$  Sicilian theories  \cite{Gadde:2011uv}. As we will see in \cite{Cremonesi:2014vla}, this is not a coincidence.  Our conjecture is actually inspired by the results in  \cite{Gadde:2011uv}.

Turning off the background monopole fluxes for the flavor group in \eqref{mainHL}, we obtain a simple expression for the Hilbert series of the Coulomb branch of $T_{\vec\rho}(G)$ for any classical $G$, formula \eref{HTrhoCIgeneral}, which shows that Coulomb branches of $T_{\vec\rho}(G)$ theories are complete intersections.

In the rest of the paper we examine the structure of the Coulomb branch Hilbert series of the theory and the physical meaning of the background monopole fluxes.

In section \ref{sec:bary} we study the action of mirror symmetry on the background monopole charges under the flavor symmetry of a theory, in order to shed light on their physical meaning.   These charges are mapped to baryonic charges on the Higgs branch of the mirror theory. Indeed, in many examples we consider, we compute the generating function of the Coulomb branch Hilbert series and match it with the baryonic generating function \cite{Forcella:2007wk} on the Higgs branch of the mirror theory.  This relation is given by \eref{bar}.

We continue and examine the analytic properties of the Coulomb branch Hilbert series of $T_{\vec \rho} (SU(N))$ in section \ref{sec:analyt}.  Similarly to the observation of \cite{Gaiotto:2012uq} in the context of superconformal indices, we find that the Coulomb branch Hilbert series of $T_{\rho}(SU(N))$ has a pole whose residue corresponds to that of a new theory $T_{\vec \rho'}(SU(N))$, where the Young diagram of $\vec \rho'$ can be obtained from that of $\vec \rho$ by moving one box to a different position. The analytic structure further substantiates our conjecture that the Hall-Littlewood formula computes the Coulomb branch Hilbert series of $T_{\rho}(G)$ theories.

Let us summarize the key results of this paper below.
\bi
\item  The Coulomb branch Hilbert series for theories arising from the `gluing' precedure is given by \eref{gluing}.
\item   The Hall-Littlewood formula for a general $T_{\vec \rho}(G)$ is given by \eref{mainHL}, and by \eref{HLformulaT} in the special case of $G=SU(N)$. \item Turning off background fluxes, these formulae reduce to \eref{HTrhoCIgeneral} and \eref{HTrhoCI}. 
\item The relations between the generating function of Coulomb branch Hilbert series and the baryonic generating function of the mirror theory is given by \eref{bar}.
\ei

In the next section we review the monopole formula for the Coulomb branch Hilbert series \cite{Cremonesi:2013lqa}.

\paragraph{Note added:} 1. After the submission of version 1 of this paper to arXiv, we learnt from the discussions in \href{http://mathoverflow.net/questions/160131/hall-littlewood-functions-and-functions-on-the-nilpotent-cone}{MathOverflow} that there are related works by mathematicians on the Hall-Littlewood formula. We would like to acknowledge the contributors in such discussions. 2. One might ask whether there is any relation between the Coulomb branch Hilbert series and the $3d$ superconformal index \cite{Kim:2009wb, Imamura:2011su, Krattenthaler:2011da}. Indeed, there is a recent work \cite{Razamat:2014pta} showing that, under a particular limit, the superconformal index of a $3d$ $\CN=4$ theory reduces to the Hilbert series.  

\section{Coulomb branch Hilbert series of a 3d $\cN=4$ gauge theory}\label{sec:general}
We are interested in the Coulomb branch of three-dimensional $\cN=4$ superconformal field theories which have a Lagrangian ultraviolet description as gauge theories of vector multiplets and hypermultiplets.
The branch is parameterized by the vacuum expectation value of the triplet of scalars in the $\cN=4$ vector multiplets and by the vacuum expectation value of the dual photons, at a generic point where the gauge group is spontaneously broken to its maximal torus. This  results in a singular HyperK\"ahler cone of quaternionic dimension equal to the rank of the gauge group $G$. The Coulomb branch is not protected against quantum corrections and the associated  chiral ring  has a complicated structure involving monopole operators in addition to the classical fields in the Lagrangian.

\subsection{The monopole formula for the Coulomb branch Hilbert series} 
In \cite{Cremonesi:2013lqa} a general formula for the Hilbert series of the Coulomb branch of an $\cN=4$ theory was proposed, which we now review.
We will work in the $\cN=2$ formulation,  where the $\cN=4$ vector multiplet decomposes into an $\cN=2$ vector multiplet and a chiral multiplet  $\Phi$ transforming in the adjoint representation of the gauge group. 
The Hilbert series is the generating function of the chiral ring, which  enumerates gauge invariant BPS operators which have a non-zero expectation value along the Coulomb branch modulo holomorphic relations.
The $\cN=2$ vector multiplets are replaced in the description of the chiral ring by monopole operators, which are subject to relations that arise at the quantum level.  The magnetic charges of the monopoles $m$ are labeled by the weight lattice of the GNO dual gauge group $G^\vee$ \cite{goddard:1976qe}.  The monopoles can be dressed with the scalar components $\phi$  of the chiral multiplet $\Phi$  that preserve some supersymmetry. As can be seen from the supersymmetry transformations of an $\cN=4$ theory \cite{Cremonesi:2013lqa},  the components $\phi$ that are BPS live in the Lie algebra   of the group $H_m$ which is left unbroken by the monopole flux. The residual gauge symmetry which is left in the monopole background consists of a continuous part $H_m$  and of a discrete part corresponding to the Weyl group $W_{G^\vee}$ of $G^\vee$, which acts on both the monopole flux $m$ and on the $\phi$.  Due to the action of the Weyl group, the
gauge invariant operators can be labeled by a flux belonging to a Weyl Chamber of the weight lattice $\Gamma_{G^\vee}$. They will be dressed by all possible products of $\phi$ invariant under the action of the
residual group $H_m$.

The final formula counts all gauge invariant BPS operators according to their  dimension and reads: 
\be\label{Hilbert_series}
H_G(t)=\sum_{\vec m\,\in\, \Gamma_{G^\vee}/W_{G^\vee}} t^{\Delta(\vec m)} P_G(t;\vec m) \;.
\ee
The sum is over a Weyl Chamber of the weight lattice $\Gamma_{G^\vee}$ of the GNO dual group \cite{goddard:1976qe}. $P_G(t;m)$ is a factor which counts the gauge invariants of the residual gauge group $H_{\vec m}$ made with the adjoint $\phi$, according to their dimension. It is given by 
\be\label{classical_dressing}
P_G(t; \vec m)=\prod_{i=1}^r \frac{1}{1-t^{d_i(\vec m)}} \;,
\ee
where $d_i(\vec m)$, $i=1,\dots,{\rm rank}\; H_{\vec m}$ are the degrees of the independent Casimir invariants of  $H_{\vec m}$, also known as exponents of $H_{\vec m}$.   $t^{\Delta(m)}$ is the quantum dimension of the monopole operator, which is given by \cite{Borokhov:2002cg,Gaiotto:2008ak,Benna:2009xd,Bashkirov:2010kz}
\be\label{dimension_formula}
\Delta(\vec m)=-\sum_{\vec \alpha \in \Delta_+(G)} |\alpha(\vec m)| + \frac{1}{2}\sum_{i=1}^n\sum_{\vec \rho_i \in R_i}|\vec \rho_i(\vec m)|\;,
\ee
where the first sum over positive roots $\vec \alpha\in\Delta_+(G)$ of $G$ is the contribution of $\cN=4$ vector multiplets and the second sum over the weights of the matter field representation $R_i$ under the gauge group is the contribution of the $\cN=4$ hypermultiplets $H_i$, $i=1,\dots,n$. Half-hypermultiplets contribute to $\Delta(m)$ with a factor of $\frac{1}{4}$ instead of $\frac{1}{2}$.

If the gauge group $G$ is not simply connected there is a nontrivial topological symmetry group under which the monopole operators may be charged, the center $Z(G^\vee)$ of $G^\vee$. Let $z$ be a fugacity valued in the topological symmetry group and $J(\vec m)$ the topological charge of a monopole operator of GNO charges $\vec m$. The Hilbert series of the Coulomb branch \eqref{Hilbert_series} can then be refined to 
\be\label{Hilbert_series_refined}
H_G(t,z)=\sum_{\vec m\,\in\, \Gamma_{G^\vee}/W_{G^\vee}} z^{J(\vec m)} t^{\Delta(\vec m)} P_G(t;\vec m) \;.
\ee

The formula can be applied to `good' or `ugly' theories (according to the classification in  \cite{Gaiotto:2008ak}) where the dimension of all monopole operators satisfies the unitarity bound $\Delta\ge 1/2$. This  ensures that the Hilbert series \eqref{Hilbert_series} is a Taylor series of the form $1+\cO(t^{1/2})$ at $t\to 0$.

The formula bypasses previous techniques for determining the Coulomb moduli space, which were based on compactification of 4d $\cN=2$ theories, the computation of the quantum corrections to the metric of the moduli space or the use of mirror symmetry. The latter method  is useful only when the mirror gauge group is sufficiently small. 
We demonstrated the utility of the Coloumb branch formula with many explicit examples  in \cite{Cremonesi:2013lqa}. On the other hand, even the Coloumb branch formula is difficult to evaluate when the gauge group becomes large. The main problems are the number of independent sums, which is equal to the rank of the gauge group, and the presence of absolute values in ~(\ref{Hilbert_series_refined}). So we need to find alternative tools for evaluating the formula. A quite efficient way is to look for an analytic formula for the Coulomb branch Hilbert series in the presence of background magnetic fluxes for the flavor symmetry group. Such Coulomb branch Hilbert series with background magnetic fluxes can then be used in the gluing technique to derive the Coulomb branch Hilbert series of more general theories.

\subsection{The Hilbert series with background magnetic fluxes}

Hilbert series with background monopole fluxes are defined as follows.  If a theory has gauge group $G$ and a global flavor symmetry $G_F$ acting on the matter fields we can define a Hilbert series as in  (\ref{Hilbert_series}) but in the presence of background monopole fluxes for the global symmetry group $G_F$:
 \be\label{Hilbert_seriesBackground}
H_{G,G_F}(t,{\vec m_F})=\sum_{\vec m\,\in\, \Gamma_{G^\vee}/W_{G^\vee}} t^{\Delta(\vec m, \vec {\vec m_F})} P_G(t;\vec m) \;.
\ee
In this formula ${\vec m_F}$ is a weight of the dual group $G^\vee_F$. By using the full global symmetry we can restrict its value to a Weyl chamber of $G^\vee_F$ and take ${\vec m_F}\in \Gamma^*_{G^\vee_F}/W_{G^\vee_F}$.
The sum in (\ref{Hilbert_seriesBackground}) is only over the magnetic fluxes $\vec m$ of the gauge group $G$. The background fluxes ${\vec m_F}$ enter explicitly in the dimension formula (\ref{dimension_formula}) through all the hypermultiplets that are charged under the global symmetry $G_F$, which acts on the Higgs branch of our theory.  Note that background fluxes in the context of $3d$ superconformal indices are studied in \cite{Kapustin:2011jm}.

A general way of constructing complicated theories is to start with a collection of theories and gauge some common global symmetry $G_F$ that they share. The Hilbert series of the final theory is given by summing over the monopoles of $G_F$ and
including the contribution to the dimension formula of the $\cN=4$ dynamical vector multiplets associated with $G_F$:
\be\label{gluing}
H(t)=\sum_{{\vec m_F}\,\in\, \Gamma_{G^\vee_F}/W_{G^\vee_F}} t^{-\sum_{\vec \alpha_F \in \Delta_+(G_F)} \alpha_F({\vec m_F})} P_{G_F}(t;{\vec m_F}) \prod_i  H^{(i)}_{G,G_F}(t,{\vec m_F})\; ,
\ee
where $\alpha_F$ are the positive roots of $G_F$ and the product with the index $i$ runs over the Hilbert series of the $i$-th theory that is taken into the gluing procedure. If we have explicit analytic formulae for the Hilbert series $H^{(i)}_{G,G_F}(t,{\vec m_F})$ with background fluxes of the original theories, that resum the RHS of \eref{Hilbert_seriesBackground}, the evaluation of $H(t)$ requires to perform a sum without any absolute value, since we can always make $\alpha_F({\vec m_F})$ positive by choosing ${\vec m_F}$ in the main Weyl chamber.

The formulae (\ref{Hilbert_seriesBackground}) and (\ref{gluing}) can be immediately generalized to include fugacities for the topological symmetries acting on the Coulomb branch.

The Hilbert series with background magnetic fluxes are interesting objects per se. We may ask what happens to the background fluxes under mirror symmetry.  
As we will see in section \ref{sec:bary}, the magnetic fluxes ${\vec m_F}$ are mapped to baryonic charges in the mirror theory and the Hilbert series with background magnetic charges are mapped to baryonic generating functions \cite{Butti:2006au,Forcella:2007wk,Butti:2007jv} for the mirror theory with certain $U(1)$ gauge factors removed. We will provide many explicit examples in Section  \ref{sec:bary}. 

In the next sections we will provide explicit and general formulae for an interesting class of 3d ${\cal N}=4$  superconformal theories which may serve as building blocks
for the construction of more general theories. In a companion paper \cite{Cremonesi:2014vla} we will apply the result to the mirrors of M5-brane theories compactified on a circle times a Riemann surface with punctures.




\section{Hilbert series and Hall-Littlewood polynomials} \label{sec:HLSclass}

In this and the following section we discuss  the Coulomb branch Hilbert series of a certain class of $\CN=4$ supersymmetric gauge theories in three dimensions,  the theories called $T_{\vec \rho}(G)$ \cite{Gaiotto:2008ak}, with $G$ a classical group and $\rho$ a partition of a certain number discussed in detail below.%
\footnote{We adhere to the standard notation where $G$ is a Lie group. More precisely, the theory is specified by a choice of the Lie algebra of $G$.}
  Such theories can be naturally realized using brane configurations as in \cite{Hanany:1996ie}.  In this section we discuss the Coulomb branch Hilbert series of $T_{\vec \rho}(SU(N))$. Other classical groups $G$ are discussed in section \ref{sec:quivSOSP}. 

We shall discuss how the Coulomb branch Hilbert series of $T_{\vec \rho}(SU(N))$ in the presence of background magnetic fluxes for $SU(N)$ can be expressed in terms of certain symmetric functions known as Hall-Littlewood polynomials (see \eg~ \cite{macdonald1998symmetric}). The main formula of this section is the Coulomb branch Hilbert series \eref{HLformulaT} for the basic building block $T_{\vec \rho}(SU(N))$.    We summarize key information about Hall-Littlewood polynomials in Appendix \ref{sec:HLpoly}.


\subsection{The theory $T_{\vec \rho}(SU(N))$ }

\begin{figure}[H]
\begin{center}
\includegraphics[scale=0.83]{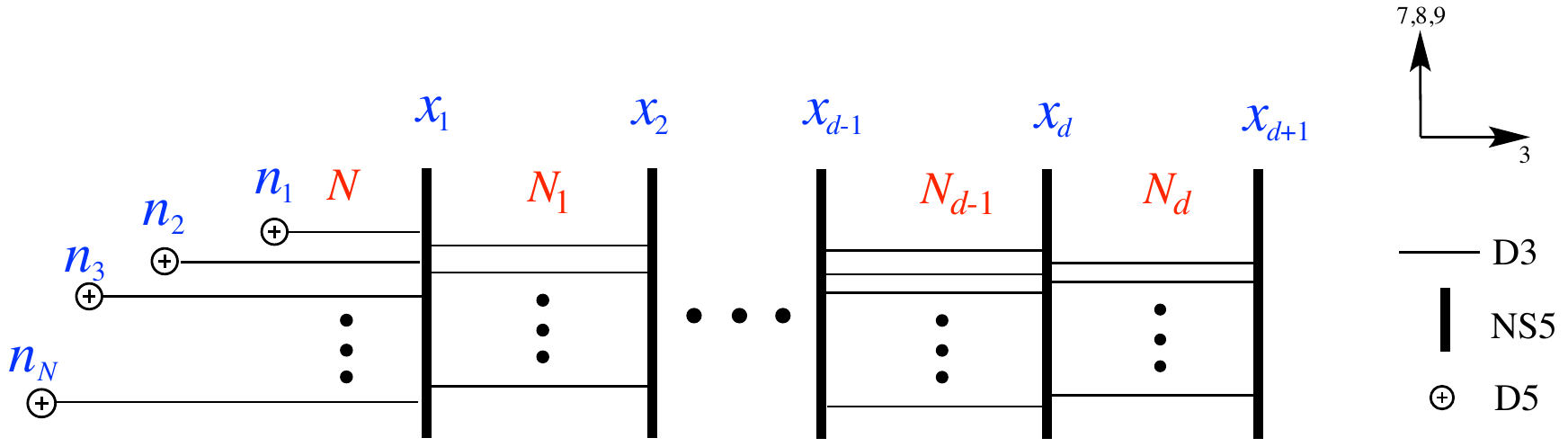}
\caption{The brane configuration of $T_{\vec \rho}(SU(N))$.  The numbers in red indicates the number of D3-branes in each intervals. 
The labels in blue denote the fugacities $x_i$ for NS5-branes and the background fluxes $n_j$ for D5-branes.
}
\label{fig:TrhoG}
\end{center}
\end{figure}

The quiver diagram for $T_{\vec \rho}(SU(N))$ is%
\footnote{We use standard notations where the links denote hypermultiplets transforming in the fundamental representations of the groups they connect.  $(G)$ denotes a gauge group and  $[G]$ a flavor symmetry.}
\bea
[U(N)]-(U(N_1))-(U(N_2))- \cdots -(U(N_d)), \label{quiv:TrhoG}
\eea
where the partition $\vec \rho$ of $N$ is given by
\bea \label{rhopartition}
\vec \rho = (N-N_1, N_{1}-N_{2}, N_{2} - N_{3}, \ldots, N_{d-1}-N_d, N_d)~,
\eea
with the restriction that $\vec \rho$ is a non-increasing sequence:
\bea
N-N_1 \geq N_{1}-N_{2} \geq N_{2} - N_{3} \geq \cdots \geq N_{d-1}-N_d \geq N_d > 0~. \label{resTrhoSUN}
\eea
This condition ensures that the $T_{\vec \rho} (SU(N))$ is a `good' theory \cite{Gaiotto:2008ak}.   
The brane configuration of $T_{\vec \rho}(SU(N))$ is depicted in \fref{fig:TrhoG}. 
$\vec \rho$ corresponds to a collection of the linking numbers of each NS5-brane. The theory associated with the partition $\vec \rho=(1,\cdots ,1)$ 
is usually called $T (SU(N))$ without any further specification.

The $U(1)$ center of the $U(N)$ flavor node in \eqref{quiv:TrhoG} is actually gauged, consequently the flavor symmetry is $U(N)/U(1)$ rather than $U(N)$. 

In addition to this flavor symmetry, the theory has a manifest $U(1)^d$ topological symmetry associated to the center of the dual gauge group, with conserved currents $J_i=\Tr(\ast F_i)$, where $i=1,\dots,d$ and $F_i$ is the field strength of the $i$-th gauge group. 
The topological symmetry is enhanced by quantum corrections to a non-abelian global symmetry which is determined as follows \cite{Gaiotto:2008ak}.
 We refer to $\rho_i$ as the parts of the partition $\vec \rho$. Let $r_k$ be the number of times that part $k$ appears in the partition $\vec \rho$.
The Coulomb branch global symmetry associated with the theory $T_{\vec \rho}(SU(N))$ is
\bea
G_{\vec \rho} = S \left( \prod_{k} U(r_k)  \right)~, \label{flvpunc}
\eea 
where $S$ denotes the removal of the overall $U(1)$.  For example, the flavor symmetry associated with $\vec \rho = (5,5,4,4,4,2,1,1)$ is $S(U(2)\times U(3) \times U(1) \times U(2))$. From \eref{rhopartition}, the number of gauge groups $d$ is related to $r_k$ in \eref{flvpunc} by
\bea
d = \sum_{k} r_k - 1~;
\eea
this is the rank of the global symmetry $G_{\vec \rho}$ acting on the Coulomb branch. The maximal torus $U(1)^d$ of $G_{\vec \rho}$, which is manifest as a topological symmetry in the quiver, enhances to a non-abelian $G_{\vec \rho}$ due to hidden symmetry generators whose associated conserved currents are monopole operators. There is also a manifest $SU(N)$ flavor symmetry acting on the Higgs branch.

In the following we provide two formulae to compute the Coulomb branch Hilbert series of $T_{\vec \rho}(SU(N))$ with background fluxes. One is the first principle formula given in (\ref{Hilbert_seriesBackground}), that we present in \eref{stdformulaT} and refer to as the {\bf monopole formula}; the other is the formula involving the Hall-Littlewood polynomial, that we present in \eref{HLformulaT} and refer to as the {\bf Hall-Littlewood formula}. We conjecture the equivalence of the two formulae, which we have analytically checked for small values of $N$ and tested perturbatively at very large order in $t$ in many different cases. As shown in section \ref{sec:analyt}, the equivalence of  the monopole formula (\ref{Hilbert_seriesBackground}) with the Hall-Littlewood formula \eref{HLformulaT} for the case of the maximal partition $\rho = (1,1,\cdots,1)$ implies the equivalence of the two formulae for a generic partition $\rho$, due to their common analytic structure.

\subsubsection{Monopole formula for the Coulomb branch Hilbert series}

As discussed in section \ref{sec:general}, we  can define a Coulomb branch Hilbert series of the theory depending on magnetic background fluxes $n_i$ for the $SU(N)$ flavor symmetry and refined by fugacities $z_i$  for the $U(1)^d$ topological symmetry. 

The monopole formula (\ref{Hilbert_seriesBackground}) for the Coulomb branch Hilbert series of $T_{\vec \rho} (SU(N))$ with background fluxes reads
\bea \label{stdformulaT}
& H[T_{\vec \rho} (SU(N))](t; z_0, z_1, \ldots, z_{d}; n_1, \ldots, n_N) \nn \\
&= z_0^{\sum_{j=1}^N n_j} \sum_{m_{1,N_1} \geq m_{2,N_1} \geq \ldots \geq m_{N_1,N_1} > -\infty}\cdots \sum_{m_{1,N_d} \geq m_{2,N_d} \geq \ldots \geq m_{N_d,N_d} > -\infty} t^{\Delta\left(\vec n; \{ m_{i, \ell}\}_{\ell=1}^{N_d} \right) }  \times\nn \\
& \quad    \prod_{k=1}^d  z_k^{\sum_{j=1}^{N_k} m_{j, N_k}} P_{U(N_k)} (t; m_{1,N_k}, \cdots, m_{N_k,N_k}) ~,
\eea
where the dimension of bare monopole operators is given by \eref{dimension_formula}
\bea
\Delta\left(\vec n; \{ m_{i,k}\}_{i=1}^k \right) &= \frac{1}{2} \left( \sum_{i'=1}^{N} \sum_{i=1}^{N_{1}} |n_{i'}-m_{i,N_{1}}| +\sum_{j=1}^{d-1} \sum_{i=1}^{N_j} \sum_{i'=1}^{N_{j+1}} |m_{i,N_{j}}-m_{i',N_{j+1}}| \right) \nn \\
& \qquad - \sum_{j=1}^d \sum_{1 \leq i <i' \leq N_j} |m_{i, N_j}-m_{i', N_{j}} | ~.
\eea

In \eref{stdformulaT}, the integers $(m_{1,N_i},\cdots, m_{N_i,N_i})$ are the GNO magnetic fluxes for the group $U(N_i)$ and the sum is restricted to the fundamental Weyl chamber by restricting to ordered $N_i$-tuples $m_{1,N_i}\ge \cdots, \ge m_{N_i,N_i}$. The integers $(n_{1},\cdots, n_{N})$ are instead the background  magnetic fluxes for the flavor symmetry group $SU(N)$. The three sets of  sums in the dimension $\Delta$   take into account the contribution  of the fundamental hypermultiplets, the  bi-fundamental hypermultiplets and the vector multiplets respectively.  Notice that  the background fluxes $n_i$ enter explicitly in the dimension $\Delta$ through the contribution of the $N$ fundamental hypermultiplets. Finally,  the classical factors  $ P_{U(N_k)} $ are defined in \eref{classical_dressing} and more details are given  in Appendix A of  \cite{Cremonesi:2013lqa}. In the same paper the reader can find many simple examples of the use of the monopole formula. 

In \eref{stdformulaT}, we also use 
\bi
\item the fugacities $z_k$ to keep track of the topological charges $\sum_{j=1}^{N_k} m_{j, N_k}$ of $U(1)\subset U(N_{k})$ gauge group, with $k=1, \ldots, d$. 
\item an extra fugacity $z_0$ which keeps track of the background charge $n_1+\ldots+ n_N$ of $U(1)=Z(U(N))$.
\ei 
The flavor symmetry $G_F$ is in fact $U(N)/U(1)$ rather than $U(N)$, therefore there is no associated topological symmetry even when $G_F$ is (weakly) gauged: $z_0$ is not a physically independent  fugacity. We remove the extra topological $U(1)$ by imposing the constraint
\bea \label{constrz}
z_{0}^N \prod_{k=1}^{d} z_k^{N_k} = 1 ~.
\eea
With this convenient choice the Hilbert series \eref{stdformulaT} is invariant under a common shift of the fluxes $n_i$, corresponding to the $U(1)$ which is not part of the flavor symmetry. 
The formula is also invariant under permutations of the $n_i$, the Weyl group of $SU(N)$. Combining the two invariances, we can always restrict the values of the fluxes to $n_1 \geq n_2 \geq \cdots \geq n_N \geq 0$, which will allow to compare with the Hall-Littlewood formula \eref{HLformulaT}. Using the shift symmetry we could further set $n_N=0$.

\subsubsection{Hall-Littlewood formula for the Coulomb branch Hilbert series}

We claim that the Coulomb branch Hilbert series of this theory \eref{stdformulaT} can be written in terms of HL polynomials as%
\footnote{The variables $x_i$ are usually taken to be independent in the literature on Hall-Littlewood polynomials. The monopole formula \eqref{stdformulaT} without the constraint \eqref{constrz} reproduces the HL formula \eqref{HLformulaT} without the constraint \eqref{constrx}, under the fugacity map \eqref{mapzxa}.}
\be \label{HLformulaT}
\begin{split}
&H[{T_\rho (SU(N))}] (t; x_1, \ldots, x_{d+1} ; n_1, \ldots, n_N) \\
&= t^{\frac{1}{2} \delta_{U(N)}(\vec n)}  (1-t)^N K^{U(N)}_{\vec \rho} (\vec x;t)\Psi_{U(N)}^{\vec n}(\vec x t^{\frac{1}{2}\vec w_{\vec \rho}}; t) ~,
\end{split}
\ee
where we explain the notations below:
\ben
\item $n_1, \ldots, n_N$ are the background GNO charges for $U(N)$ group, with 
\bea n_1 \geq n_2 \geq \cdots \geq n_N \geq 0~.\eea
\item The Hall-Littlewood polynomial associated with the group $U(N)$ is given by
\bea
\Psi^{\vec n}_{U(N)} (x_1,\dots,x_N;t)=\sum_{\sigma \in S_N}
x_{\sigma(1)}^{n_1} \dots x_{\sigma(N)}^{n_N}
\prod_{1 \leq i<j \leq N}   \frac{  1-t x_{\sigma(i)}^{-1} x_{\sigma(j)} } {1-x_{\sigma(i)}^{-1} x_{\sigma(j)}}~.
\eea
\item The notation $\delta_{U(N)}$ denotes the sum over positive roots of the group $U(N)$ acting on the {\it background} charges $n_i$:
\bea
\delta_{U(N)}(\vec n) = \sum_{1\leq i < j \leq N} (n_i - n_j) = \sum_{j=1}^{N} (N+1-2j) n_j~.
\eea
Note that this is minus the contribution of the background vector multiplet in the monopole dimension formula.
\item  The fugacities $x_1, \ldots, x_{d+1}$ are naturally associated to the NS5-branes in the brane construction as depicted in \fref{fig:TrhoG}. They are related to the fugacities $z_0, \ldots, z_{d}$ for the manifest topological symmetry group $U(1)^{d+1}/U(1)$ by the fugacity map 
\bea \label{mapzxa}
z_{0} &=x_1~, \qquad z_k = x_{k+1} / x_{k}~, \qquad k=1,\ldots, d~.
\eea
The symmetry $U(1)^{d+1}/U(1)$ enhances to the non-abelian $G_{\vec \rho}$ due to monopole operators. $x_1, \ldots, x_{d+1}$ are subject to the constraint which fixes the overall $U(1)$. Using the map \eqref{mapzxa}, the constraint \eref{constrz} is rewritten as
\bea \label{constrx}
\prod_{i=1}^{d+1} x_i^{\rho_i} =1~.
\eea

\item $\vec w_{{r}}$ denotes the weights of the $SU(2)$ representation of dimension $r$:
\bea
\vec w_{{r}} = (r-1, r-3, \ldots, 3-r, 1-r)~.
\eea
Hence the notation $t^{\frac{1}{2}\vec w_{r}}$ represents the vector
\bea
t^{\frac{1}{2}\vec w_{r}} = (t^{\frac{1}{2}(r-1)}, t^{\frac{1}{2}(r-3)}, \ldots, t^{-\frac{1}{2}(r-3)},t^{-\frac{1}{2}(r-1)})~.
\eea
In \eref{HLformulaT} and from now on, we abbreviate
\bea
\Psi_{U(N)}^{\vec n}(\vec x t^{\frac{1}{2}\vec w_{\vec \rho}}; t) := \Psi_{U(N)}^{(n_1, \ldots, n_N)}(x_1 t^{\frac{1}{2}\vec w_{\rho_1}}, x_2 t^{\frac{1}{2}\vec w_{\rho_2}} , \ldots, x_{d+1} t^{\frac{1}{2}\vec w_{\rho_{d+1}}};t)~.
\eea
The reader interested in a graphical illustration of the previous definition of $\vec x t^{\frac{1}{2}\vec w_{\rho}}$ may refer to Fig. 2 of \cite{Gadde:2011uv}.
\item The prefactor $K^{U(N)}_{\vec \rho} (\vec x;t)$  is given by
\bea \label{KUN}
K^{U(N)}_{\vec \rho} (\vec x;t) = \prod_{i=1}^{\text{length}({\vec \rho}^T)} \prod_{j,k=1}^{\rho^T_i} \frac{1}{1-a^i_j \bar{a}^i_k}~,
\eea
where $\rho^T$ denotes the transpose of the partition $\rho$ and we associate the factors
\be
\begin{split}
a^i_j &= x_j \;\; t^{\frac{1}{2} (\rho_j-i+1)}~, \qquad  i=1,\dots,\rho_j \\
{\bar a}^i_k &= x_k^{-1} t^{\frac{1}{2} (\rho_k-i+1)}~, \qquad  i=1,\dots,\rho_k \label{defaabar}
\end{split}
\ee
to each box in the Young tableau.%
\footnote{A graphical illustration of $a^i_j$ is given by Fig. 3 of \cite{Gadde:2011uv}, where $j$ labels the column from left to right and $i$ labels the rows from the bottom to the top of the Young tableau.} 
The powers of $t$ inside $a^i_j$ and ${\bar a}^i_k$ are positive by construction.  
For instance:
\ben
\item For the full puncture%
\footnote{We often write `puncture' in analogy to the literature on M5 branes on a Riemann surface, but take it to mean `partition' in this paper. We use the shorthand notation $(r^s)=(\underbrace{
r,\cdots,r}_\text{$s$ times})$ for partitions.} $\vec \rho= (1^N)$, we have $\vec \rho^T = (N)$ and so
\bea
K^{U(N)}_{(1^N)} (\vec x; t)= \prod_{1\leq j, k\leq N} \frac{1}{1- x_j x_k^{-1} t} =  \PE[t \chi^{U(N)}_{\bf Adj} (\vec x) ]~,
\eea
where $\PE$ denotes the \emph{plethystic exponential}.%
\footnote{The \textit{plethystic exponential} of a multivariate function $f(t_1, . . . , t_n)$ that vanishes at the origin is defined as ${\rm PE} \left[ f(t_1, t_2, \ldots, t_n) \right] = \exp \left( \sum_{k=1}^\infty \frac{1}{k} f(t_1^k, \cdots, t_n^k) \right)$. For instance $ \PE[n t^m]=(1-t^m)^{-n}$.}
\item For the simple puncture $\vec \rho=(N-1,1)$, we have $\vec \rho^T = (2,1^{N-2})$ and so 
\bea
K^{U(N)}_{(N-1,1)} (\vec x; t)= \PE \left[t^{N/2} (x_1 x_2^{-1} + x_2 x_1^{-1}) +t+ \sum_{j=1}^{N-1} t^j \right]~.
\eea
\een

\een
The representation theoretic explanations for $\vec x t^{\frac{1}{2}\vec w_{\rho}}$ and the prefactor $K^{U(N)}_{\vec \rho} (\vec x;t)$ are presented in Sec. 4.1 of \cite{Mekareeya:2012tn}.  We summarize this in Sec. \ref{sec:HLgenG} of this paper.

We have explicitly checked in a large number of examples that the HL formula \eref{HLformulaT} coincides with the monopole formula \eref{stdformulaT}. 
For instance, we can consider the case $N=3$. There are two relevant partitions, $(1,1,1)$ and $(2,1)$.
The $T_{(1,1,1)}(SU(3))$ theory, also known simply as $T(SU(3))$, has $SU(3)$ global symmetry acting on the Coulomb branch,  as it follows from (\ref{flvpunc}).
The monopole formula (\ref{stdformulaT}) reads
\be \label{std111SU3}
\begin{split}
& H[T_{(1,1,1)}(SU(3))] (t; x_1, x_2, x_3; n_1, n_2)  \\
& =x_1^{n_1+n_2 }\sum_{m_{1,1} \in \BZ} ~\sum_{m_{2,2}\ge m_{1,2} > -\infty} (x_2 x_1^{-1})^{m_{1,2}+m_{2,2}} (x_3 x_2^{-1})^{m_{1,1}}  \times  \\
& \hspace{2cm} t^{\frac{1}{2}\Delta_{(1,1,1)}(m_{1,1}; m_{1,2},m_{2,2};\vec n)} P_{U(2)}(t;m_{1,2},m_{2,2}) P_{U(1)}(t)~,
\end{split}
\ee
where $m_{i,k}$, with $i=1, \ldots, k$, denote the GNO charges for the $U(k)$ gauge group, $x_1 x_2 x_3=1$, and $\Delta_{(1,1,1)}$ is twice the dimension of monopole operators in the $[3]-(2)-(1)$ quiver:
\be
\Delta_{(1,1,1)} =\sum_{i=1}^2|m_{1,1}-m_{i,2}|+\sum_{i=1}^2\sum_{j=1}^3 |m_{i,2}-n_j| -2|m_{1,2}-m_{2,2}|~, \quad n_3=0~.
\ee
We  used the fugacity map (\ref{mapzxa}) and we imposed the constraint (\ref{constrx}).
One can check that \eref{std111SU3} reproduces the Hall-Littlewood formula
\be
\begin{split}
& H[T_{(1,1,1)}(SU(3))] (t; x_2,x_2,x_3; n_1, n_2) \\
&=t^{ n_1} (1-t)^3  {\rm PE} \left[t \sum_{1 \leq i, j \leq 3} x_i x_j^{-1} \right]  
 \Psi^{(n_1,n_2,0)}_{U(3)}(x_1,x_2,x_3;t) ~.  \label{H111SU3}
\end{split} 
\ee
The $T_{(2,1)}(SU(3))$ theory has Coulomb branch symmetry $U(1)$ and  the monopole formula reads
\be \label{CouT21}
H[T_{(2,1)}(SU(3))] (t; x_1, x_2; n_1, n_2,n_3) = \frac{x_1^{n_1+n_2}}{1-t} \sum_{m\in\bZ} t^{\frac{1}{2}\Delta_{(2,1)}(m; \vec n)}(x_2 x_1^{-1})^m~,
\ee
where $m$ denotes the GNO charge for the $U(1)$ gauge group, $x_1^2 x_2=1$, and $\Delta_{(2,1)}$ is twice the dimension of monopole operators in the $[3]-(1)$ quiver:
\be
\Delta_{(2,1)} = \sum_{i=1}^3 |m-n_i|~, \quad n_3=0~.
\ee
We again used the fugacity map (\ref{mapzxa}) and we imposed the constraint (\ref{constrx}). Again one can check that \eref{CouT21} reproduces the Hall-Littlewood formula
\be
\begin{split}
&H[T_{(2,1)}(SU(3))] (t; x_1, x_2; n_1,n_2) \\
&= t^{n_1} (1-t)^3 {\rm PE} [2t +t^\frac{3}{2} ( x_1 x_2^{-1} +x_1^{-1} x_2) +t^2]   \Psi^{(n_1,n_2,0)}_{U(3)}( x_1 t^{\frac{1}{2}}, x_1 t^{-\frac{1}{2}}, x_2; t)~. \label{H21SU3}
\end{split}
\ee

We will further demonstrate the HL formula \eref{HLformulaT} in a number of examples in a companion paper \cite{Cremonesi:2014vla}, where we will successfully compare our formula for the Hilbert series of the Coulomb branch of mirrors of genus $0$ 3d Sicilian theories with the Hilbert series of the Higgs branch of the Sicilian theories themselves, computed as the Hall-Littlewood limit of the superconformal index of the 4d theories in \cite{Gadde:2011uv,Lemos:2012ph}.

\subsection{The Coulomb branch of $T_{\vec \rho}(SU(N))$ is a complete intersection} \label{sec:CISUN}

Coulomb branches of $T_{\vec \rho}(SU(N))$ theories for various partitions $\vec \rho$ were studied in \cite{Hanany:2011db} using mirror symmetry; it was found that many of these algebraic varieties are complete intersections. In this section, we provide a direct argument that for {\it any} partition $\vec \rho$, the Coulomb branch of $T_{\vec \rho}(SU(N))$ is a complete intersection.

Setting $n_1=n_2= \ldots=n_N=0$ in \eref{HLformulaT} and using the identity
\be
(1-t)^{N}  \Psi_{U(N)}^{(0,\ldots,0)}(\vec x;t) = \prod_{k=1}^{N} (1-t^k)~, \label{HL000}
\ee
we obtain
\be
\begin{split}
H[T_{\vec \rho}(SU(N))](t; \vec x; \vec 0) &= K^{U(N)}_{\vec \rho}(\vec x; t)   \prod_{k=1}^{N} (1-t^k)  \\
&= \PE \left[  \sum_{i=1}^{\text{length}({\vec \rho}^T)} \sum_{j,k=1}^{\rho^T_i} x_j x^{-1}_k t^{\frac{1}{2}(\rho_j+\rho_k)-i+1} -\sum_{k=1}^N t^k\right]  ~, \label{HTrhoCI}
\end{split}
\ee
where we have used \eref{KUN} in the second equality.
It follows from the remark below \eref{defaabar} that the powers of $t$ appearing inside the $\PE$ are strictly positive.  

The form of \eref{HTrhoCI} shows that the Coulomb branch, denoted by ${\rm C}[T_{\vec \rho}(SU(N))]$, is a complete intersection, \emph{i.e.}  it is described by a number $n$ of generators subject to a number $r$ of relations equal to the complex codimension of the variety in the embedding space $\bC^n$. Its complex dimension $n-r$ is given by 
\bea
\dim_{\BC} {\text{\sc C}}[T_{\vec \rho} (SU(N))] = \sum_{i=1}^{\text{length}({\vec \rho}^T)} (\rho_i^T)^2 -N ~,
\eea
where the positive contribution counts the number of positive terms (representing generators) and the negative contribution counts the number of negative terms (representing relations) inside the $\PE$ in \eref{HTrhoCI}. This result is in agreement with (2.3) and (2.4) of \cite{Chacaltana:2012zy}.  In fact, one can cancel a common factor $\PE[t-t]=1$ in \eqref{HTrhoCI}, suggesting that a putative generator is eliminated by a relation: we conclude that there are  $\sum_{i=1}^{\text{length}({\vec \rho}^T)} (\rho_i^T)^2-1$ generators subject to $N-1$ relations, one per Casimir invariant of $SU(N)$.


\section{Coulomb branch of $T_\rho (SO(N))$ and $T_\rho (USp(2N))$} \label{sec:TSOSp}

In this section we generalize the results on $T_{\vec \rho}(G^\vee)$ to other classical groups beyond $G=SU(N)$, namely $SO(N)$ and $USp(2N)$.  One of the key results in this section is the formula \eref{mainHL} for the  
Coulomb branch Hilbert series, involving the Hall-Littlewood polynomial.  We demonstrate in many examples below that this formula is uniform for all classical gauge groups. We conjecture that it can be used also for large classes of `bad' theories, where the monopole formula is not working, as we shall discuss in several examples below and in the Appendix.%
\footnote{Recall that a theory is called \emph{bad} when, using the ultraviolet R-symmetry as in \eqref{dimension_formula}, there are monopole operators with $\Delta<\frac{1}{2}$. When that is the case, one cannot assume that the ultraviolet R-symmetry computes the conformal dimension of BPS operators, because the unitarity bound is violated. }

In the following construction, we will also need the GNO, or Langlands, dual of $G$, denoted as $G^\vee$ \cite{goddard:1976qe}. Recall that the Lie algebras $A$ and $D$ are self-dual, while $B$ and $C$ are exchanged by GNO duality.

As pointed out in \cite{Gaiotto:2008ak} and \cite{Chacaltana:2012zy}, $T_{\vec \rho}(G^\vee)$ is constructed as a boundary theory of $4d$ $\CN=4$ super Yang-Mills  on a half-space, with the half-BPS boundary condition specified by a homomorphism $\vec \rho: \mathrm{Lie}(SU(2)) \rightarrow \mathrm{Lie}(G)$. The homomorphisms $\vec \rho$ can be classified, up to conjugation, by the nilpotent orbits of the Lie algebra of $G$.  This classification puts certain restrictions on the partition $\vec \rho$ which we discuss in section \ref{sec:BCDpartitions}.  The quiver diagrams of the corresponding $T_{\vec \rho}(G^\vee)$ theories are presented in section \ref{sec:quivSOSP}.  We compute their Coulomb branch Hilbert series in section \ref{sec:HLgenG}.

\subsection{$B$, $C$ and $D$ partitions} \label{sec:BCDpartitions}

The partition $\vec \rho = (\rho_1, \rho_2, \ldots)$ defines a homomorphism $\vec \rho: \mathrm{Lie}(SU(2)) \rightarrow \mathrm{Lie}(G)$, such that the fundamental representation of $G$ decomposes into a direct sum of irreducible representations of $SU(2)$ of dimensions $\rho_1,  \rho_2, \rho_3,  \ldots$.  We call $\rho_i$ {\it parts} of the partition $\vec \rho$.

Due to the Jacobson-Morozov theorem, such embedding can be classified up to a conjugacy by the nilpotent orbit of $G$.  As discussed in \cite{collingwood1993nilpotent} and section 2.1 of \cite{Chacaltana:2012zy}, the possible cases are as follows:
\bi
\item For $G=SO(N)$, the partition $\vec \rho$ of $N$ satisfies the condition that any even part in $\vec \rho$ must appear an even number of times.   The partition $\vec \rho$ is called a {\it B-} or a {\it D-partition} if $N$ is odd or even, respectively. For instance, the {\it B-}partitions for $SO(3)$ are $(3)$ and $(1,1,1)$; the {\it D-}partitions for $SO(4)$ are $(4)$, $(3,1)$, $(2,2)$ and $(1,1,1,1)$.
Given a partition $\vec \rho$ satisfying this condition, there is a unique nilpotent orbit associated to it, except for the case when all the parts $\rho_i$ are even and each even integer appears even times.  Such a partition is referred to as a {\it very even} partition, whose distinct nilpotent orbits are exchanged by the outer automorphism of $SO(N)$.  An example of a very even partition is $\vec \rho=(4,4)$ for $G=SO(8)$.
\item For $G=USp(2N)$,  the partition $\vec \rho$ of $2N$ satisfies the condition that any odd part in $\vec \rho$ must appear an even number of times.  Such a partition is called a {\it C-partition}. In this case each partition corresponds to a unique nilpotent orbit. For instance, the {\it C-}partitions $\vec\rho$  for $USp(2)$ are $(2)$ and $(1,1)$; for $USp(4)$, 
$(4)$, $( 2,2)$ and $( 1,1,1,1)$; for $USp(6)$, $(6)$, $(4,2)$, $(3,3)$, $(2,2,2)$,  $(2,2,1,1)$, $(2,1,1,1,1)$ and $(1,1,1,1,1,1)$.
\ei
Below we present quiver diagrams for $T_{\vec \rho} (G^\vee)$ for $G=SO(N)$ and $USp(2N)$. Such quivers already appeared as subquivers in the work of \cite{Feng:2000eq}.

\subsection{Quiver diagrams} \label{sec:quivSOSP}

The quiver diagram for $T_{\vec \rho} (SO(2N))$ and $T_{\vec \rho} (SO(2N+1))$ for a given partition $\vec \rho$ is presented in (6.3) and (6.5) of \cite{Benini:2010uu}.  The quiver for $T_{\vec \rho} (USp(2N))$ can be easily obtained by generalising that of $T_{\vec \rho} (SO(2N))$.  We summarize the necessary information below.

\paragraph{$T_{\vec \rho}(SO(2N))$ theory.} In this case $\vec \rho = (\rho_1, \ldots, \rho_\ell)$ is a $D$-partition of $2N$. The quiver diagram for $T_{\vec \rho} (SO(2N))$ is
\bea
[SO(2N)] - (USp(s_1)) - (O(s_2)) - \cdots-(O(s_{\ell-2})) - (USp(s_{\ell-1}))~,
\eea
where $\ell$ is even and
\bea
s_i = \left[\sum_{j=i+1}^\ell \rho_{j} \right]_{+, -} \qquad \text{$+$ for $O$ and  $-$ for $USp$}~,
\eea
with $[n]_{+(\text{{\it resp.}}-)}$ the smallest ({\it resp.} largest) even integer $\geq n$ ({\it resp.} $\leq n$) and the node $USp(0)$ being removed.

\paragraph{$T_{\vec \rho}(SO(2N+1))$ theory.} In this case we consider the Langlands dual of $B_N = SO(2N+1)$, namely $C_N = USp(2N)$.  The partition $\vec \rho = (\rho_1, \ldots, \rho_\ell)$ is a $C$-partition of $2N$.  The corresponding quiver diagram for $T_{\vec \rho} (SO(2N+1))$ is
\bea
[SO(2N+1)] -(USp(s_1)) - (O(s_2)) - \cdots - (O(s_{[\ell]_-}))~,
\eea
where
\bea
s_i = \left[1+  \sum_{j=i+1}^\ell \rho_{j} \right]_{\tilde{+}, -} \qquad \text{$\tilde{+}$ for $O$ and  $-$ for $USp$}~,
\eea
with $[n]_{{\tilde +}}$ the smallest odd integer $\geq n$ and $[n]_\pm$ defined as above.

\paragraph{$T_{\vec \rho}(USp(2N))$ theory.}  In this case we consider the Langlands dual of $C_N = USp(2N)$, namely $B_N = SO(2N+1)$.  The partition $\vec \rho = (\rho_1, \ldots, \rho_\ell)$ is a $B$-partition of $2N+1$.  The corresponding quiver diagram for $T_{\vec \rho} (USp(2N))$ is
\bea
[USp(2N)] -(O(s_1)) - (USp(s_2)) - \cdots - (USp(s_{\ell-1}))~,
\eea
where $\ell$ is odd and
\bea
s_i = \left[\sum_{j=i+1}^\ell \rho_{j} \right]_{+, -} \qquad \text{$+$ for $O$ and  $-$ for $USp$}~,
\eea
with $[n]_\pm$ defined as above and the node $USp(0)$ being removed.

In the previous theories some of the gauge groups of type $O$  can be replaced by groups of type $SO$. The distinction between $SO(s)$ and $O(s)$ gauge  groups is important. Theories with $SO(s)$ gauge groups have typically more BPS gauge invariant operators compared with the same theory with gauge group $O(s)$ and we have different theories according to the choice of $O/SO$ factors. 

\subsection{Hall-Littlewood formula for $T_{\vec \rho} (G^\vee)$ with a classical group $G$} \label{sec:HLgenG}

In this section we generalize the Hall-Littlewood formula \eref{HLformulaT} to a more general group $G$.  We follow closely a similar discussion in section 4.1 of \cite{Mekareeya:2012tn}; see also \cite{Chacaltana:2012zy} for a comprehensive presentation.  Several explicit examples are presented in subsequent subsections and in Appendix \ref{sec:nonmaximal}.

As discussed earlier, the partition $\vec \rho$ defines a homomorphism $\vec \rho: \mathrm{Lie}(SU(2)) \rightarrow \mathrm{Lie}(G)$ such that
\bea
[1,0,\ldots,0]_G =  \bigoplus_i [\rho_i-1]_{SU(2)}~.
\eea
The global symmetry $G_{\vec \rho}$ associated to the puncture and acting on the Coulomb branch is given by the commutant of ${\vec \rho}(\SU(2))$ in $G$.
Explicitly, for a given group $G=U(N)$, $SO(N)$ or $USp(2N)$ and a puncture $\vec \rho = [\rho_i]$ with $r_k$ the number of times that part $k$ appears in the partition $\vec \rho$, we have 
\bea \label{sympunc}
G_{\vec \rho} = \begin{cases} S \left( \prod_{k} U(r_k)  \right) & \qquad G= U(N)~, \\
 \prod_{k~\text{odd}} SO(r_k) \times \prod_{k~\text{even}} USp(r_k) & \qquad G= SO(2N+1)~\text{or}~SO(2N)~, \\
 \prod_{k~\text{odd}} USp(r_k) \times \prod_{k~\text{even}} SO(r_k) & \qquad G= USp(2N)~.
\end{cases}
\eea

Let $x_1, x_2, \ldots$ be fugacities for the global symmetry $G_{\vec \rho}$, and $r(G)$ the rank of $G$.  We conjecture that the Coulomb branch Hilbert series is given by the HL formula
\be
H[T_{\vec \rho}(G^\vee)](t; \vec x;n_1,\ldots, n_{r(G)}) = t^{\frac{1}{2} \delta_{G^\vee}(\vec n)} (1-t)^{r(G)} K^{G}_{\vec \rho}  (\vec x;t) \Psi^{\vec n}_{G}(\vec a(t, \vec x);t)~, \label{mainHL}
\ee
where we explain the notations below.
\ben
\item The Hall-Littlewood polynomial $\Psi^{\vec n}_{G}(\vec a(t, \vec x);t)$ associated with a classical group $G$ is given in Appendix \ref{sec:HLpoly}.
\item The power $\delta_{G^\vee}(\vec n)$ is the sum over positive roots $\vec \alpha\in\Delta_+(G^\vee)$  of the flavor group $G^\vee$ acting on the {\it background} monopole charges $\vec n$:
\bea
\delta_{G^\vee}(\vec n) = \sum_{\vec \alpha \in \Delta_+(G^\vee)} \vec \alpha(\vec n)~.
\eea
Explicitly, for classical groups $G$, these are given by
\bea \label{powers}
\delta_{G^\vee} (\vec n) &= \begin{cases} 
\sum_{j=1}^N (N+1-2j) n_j & \qquad G^\vee =G=U(N), \\
\sum_{j=1}^{N} (2N+1-2j)n_j  & \qquad G^\vee=B_N,~ G=C_N\\
\sum_{j=1}^{N} (2N+2-2j)n_j  & \qquad G^\vee=C_N,~ G=B_N \\
\sum_{j=1}^{N-1} (2N-2j)n_j  & \qquad G^\vee =G=D_N~.
\end{cases}
\eea
\item The argument $\vec a(t, \vec x)$, which we shall henceforth abbreviate as $\vec a$, of the HL polynomial is determined by the following decomposition of the fundamental representation of $G$ to $G_{\vec \rho} \times {\vec \rho} (SU(2))$:
\bea \label{decompfund}
\chi^G_{{\bf fund}} (\vec a) = \sum_{k}  \chi^{G_{\rho_k}}_{{\bf fund}}( \vec x_k) \chi^{SU(2)}_{[\rho_k -1]} (t^{1/2})~,
\eea
where $G_{\rho_k}$ denotes a subgroup of $G_{\vec \rho}$ corresponding to the part $k$ of the partition $\vec \rho$ that appears $r_k$ times.  Formula \eref{decompfund} determines $\vec a$ as a function of $t$ and $\{ \vec x_k \}$ as required.  Of course, there are many possible choices for $\vec a$; the choices that are related to each other by outer automorphisms of $G$ are equivalent.

We provide some examples, such as two inequivalent choices \eref{embedding44} for $\vec \rho=(4,4)$ of $G=SO(8)$, in the subsequent subsections.
\item  $K^G_{\vec \rho}(\vec x; t)$ is a prefactor independent of ${\vec n}$, determined as follows.  The embedding associated with $\vec \rho$ induces the decomposition of the adjoint representation of $G$
\begin{equation}
\chi^G_{\bf Adj} (\vec a) = \sum_{j =0, \frac{1}{2}, 1, \frac{3}{2}, \ldots}  \chi^{G_{\vec \rho}}_{R_j}(\vec x_j)  \chi^{SU(2)}_{[2j]}(t^{1/2})~, \label{decompadj} 
\end{equation} 
where $\vec a$ on the left hand side is the same $\vec a$ as in \eref{decompfund}.  Note that $\bigoplus_j R_j$ gives the decomposition of the Slodowy slice \cite{Chacaltana:2012zy}. Each component in the slice gives rise to a plethystic exponential, giving 
\begin{equation}
K^G_{\vec \rho}(\vec x; t)=\PE \left[\sum_{j =0, \frac{1}{2}, 1, \frac{3}{2}, \ldots}  t^{j+1} \chi^{G_{\vec \rho}}_{R_j}({\vec x}_j )\right].  \label{K}
\end{equation}
We provide a number of examples below; see for example \eref{decomp3311} for $\vec \rho=(3,3,1,1)$ of $G=SO(8)$ and \eref{decomp3221}  for $\vec \rho=(3,2,2,1)$ of $G=SO(8)$.
\een

\subsubsection{The Coulomb branch of $T_{\vec \rho}(G^\vee)$ is a complete intersection}

In this subsection we generalize the computation in Section \ref{sec:CISUN} for any classical group $G$.  We will show that formula \eref{mainHL} implies that the Coulomb branch of $T_{\vec \rho}(G^\vee)$ is a complete intersection for any  classical group $G$ and any partition $\vec \rho$.  

Turning off the background monopole fluxes $n_1=n_2 =\ldots = n_{r(G)} =0$, the HL polynomials reduce to
\bea
\Psi^{(0,\ldots,0)}_{G} (\vec x ;t) = (1-t)^{-r(G)} \prod_{i=1}^{r(G)} (1-t^{d_i(G)})~,
\eea
where $d_i(G)$, with $i=1, \ldots, r(G)$, are the exponents of $G$:
\bea
d_i(G) = 
\begin{cases} 
1, 2, 3, \ldots, N, &\qquad G= U(N) \\ 
2, 4, 6, \ldots, 2N, & \qquad  G=SO(2N+1), ~USp(2N) \\
N, 2, 4, 6, \ldots, 2N-2, & \qquad G=SO(2N)~.
\end{cases}
\eea
Thus we obtain
\be\label{HTrhoCIgeneral}
\begin{split}
H[T_{\vec \rho}(G^\vee)](t; \vec x;\vec 0)&= K^{G}_{\vec \rho} (\vec x; t)  \prod_{i=1}^{r(G)} (1-t^{d_i(G)}) \\
&= \PE \left[\sum_{j =0, \frac{1}{2}, 1, \frac{3}{2}, \ldots}  t^{j+1} \chi^{G_{\vec \rho}}_{R_j}({\vec x}_j )-  \sum_{i=1}^{r(G)} t^{d_i(G)}\right]~,
\end{split}
\ee
where the representations $R_j$ of the group $G_{\vec \rho}$ are given by \eref{decompadj}.  This shows that the Coulomb branch, denoted by ${\text{\sc C}}(T_{\vec \rho} (G^\vee))$, of $T_{\vec \rho}(G^\vee)$ is indeed a complete intersection: there are $\sum_{j} \dim_{G_{\vec \rho}}(R_j)$ generators subject to $r(G)$ relations, one per independent Casimir invariant of $G$.    
The complex dimension of the Coulomb branch is given by
\bea \label{dimCouT}
\dim_{\BC} {\text{\sc C}}[T_{\vec \rho} (G^\vee)] =\sum_{j} \dim_{G_{\vec \rho}}(R_j) - r(G)~,
\eea
where $\dim_{G_{\vec \rho}}(R_j)$ denotes the dimension of representation $R_j$ of the group $G_{\vec \rho}$.  

According to Theorem 6.1.3 of \cite{collingwood1993nilpotent},  the first term in \eref{dimCouT} can be related to the partition as 
\bea \label{dimGR}
\sum_{j} \dim_{G_{\vec \rho}}(R_j)=\begin{cases} \sum_{i} {(\rho^T_i)^2} & G= U(N) \\ \frac{1}{2}\sum_{i} {(\rho^T_i)^2} -\frac{1}{2}  \sum_{i~\text{odd}} r_i & G=B_N, ~D_N \\  \frac{1}{2}\sum_{i} {(\rho^T_i)^2} +\frac{1}{2}  \sum_{i~\text{odd}} r_i & G=C_N~,  \end{cases} 
\eea
where $r_k$ is the number of times that $k$ appears in the partition $\vec \rho$.  Thus, from \eref{dimCouT} and \eref{dimGR}, we have the dimension formula in accordance with (2.3) of \cite{Chacaltana:2012zy}:
\bea \label{dimCou}
\dim_{\BC} {\text{\sc C}}[T_{\vec \rho} (G^\vee)] =\begin{cases} \sum_{i} {(\rho^T_i)^2}-N \qquad & G= U(N) \\ \frac{1}{2}\sum_{i} {(\rho^T_i)^2} -\frac{1}{2}  \sum_{i~\text{odd}} r_i -N \qquad & G=B_N, \; D_N \\  \frac{1}{2}\sum_{i} {(\rho^T_i)^2} +\frac{1}{2}  \sum_{i~\text{odd}} r_i -N \qquad & G=C_N~.\end{cases} 
\eea

\subsection{$T(SO(N))$ and $T(USp(2N))$}

In this subsection we focus on $\vec \rho =(1,1, \ldots, 1)$. The case of non maximal punctures is discussed  in  Appendix \ref{sec:nonmaximal}. 
From now on, we shall abbreviate $T_{(1,\ldots,1)}(G)$ as $T(G)$.
The quiver diagrams for $T(SO(N))$ and $T(USp(2N))$ are given in Fig. 54 of \cite{Gaiotto:2008ak}; we summarize them below.%
\footnote{With this choice of $SO(N)$ gauge groups for $T(D_N)$ and  $T(C_N)$ the Coulomb and Higgs moduli space are precisely   the nilpotent cone of the corresponding $D_N$ and $C_N$ group \cite{Gaiotto:2008ak}.}
{\small \bea
&T(D_N):  \nn\\
& \qquad   (SO(2))-(USp(2))- \cdots - (SO(2N-2))-(USp(2N-2))-[SO(2N)]  \\
&T(C_N): \nn\\
& \qquad   (SO(2))-(USp(2))- \cdots-(USp(2N-2))-(SO(2N))-[USp(2N)]  \\
& T(B_N):  \nn \\
& \qquad  (O(1))-(USp(2))-(O(3))- \cdots - (O(2N-1))-(USp(2N))-[SO(2N+1)]~. \label{TBN}
\eea}
Note that edges connecting $O$ and $USp$ represent bifundamental \emph{half}-hypermultiplets. Remark also that the quiver in \eref{TBN} is a `bad theory'; for example, the number of flavors under the $USp(2)$ gauge group is $2$  (because there are 4 half-hypers charged under this gauge group), which is smaller than $2(1)+1=3$ (see (5.9) of \cite{Gaiotto:2008ak}). 

Mirror symmetry exchanges $G$ with the Langlands or GNO dual $G^\vee$, therefore 
$T(D_N)$ is self-dual, whereas $T(C_N)$ and $T(B_N)$ form a mirror pair \cite{Gaiotto:2008ak}.

From many examples presented in the following subsections, we deduce that the Hilbert series of Coulomb branch for each of such theories is given by
\bea \label{HLTG}
H[T(G^\vee)](t;x_1, \ldots, x_N;n_1,\ldots, n_N)&= t^{\frac{1}{2} \delta_{G^\vee}(\vec n)} (1-t)^{r(G)} K^{G}  (\vec x;t) \Psi^{\vec n}_{G}(\vec x;t)~.
\eea
where $n_1, \ldots, n_N$ are the background charges, $\delta_{G^\vee}(\vec n)$ is given by \eref{powers}, and
\bea
K^G ( \vec x;t) &= \PE \left[t (\chi^G_{\bf Adj} (\vec x) )\right]~.
\eea

Since the quiver \eref{TBN} for $T(B_N)$ is a bad quiver,  we cannot compute the Coulomb branch Hilbert series from the monopole formula, which diverges. On the other hand one may expect that the quiver flows to an interacting conformal fixed point in the infrared. Since the formula \eref{HLTG} involving the Hall-Littlewood polynomial is well defined, we conjecture that it computes the Hilbert series of the Coulomb  branch of the infrared SCFT.  Below we demonstrate this by computing the Coulomb branch Hilbert series of $T(SO(5))$, which is a bad theory, using the Hall-Littlewood formula and find a matching \eref{matchC2B2} with that of $T(USp(4))$, which is a good theory.

When comparing with the monopole formula, we will face the problem of matching fugacities. Unlike in the case of unitary groups, in the Coulomb branch of the $T(G^\vee)$ theories with orthogonal and symplectic groups the Cartan subalgebra of the global symmetry is not fully manifest. Some Cartan generators do not correspond to topological symmetry of the Coulomb branch, but arise instead as monopole operators \cite{Gaiotto:2008ak}. It is not clear to us how to introduce fugacities for these generators in the monopole formula. 

Let us demonstrate these formulae in the examples below.

\subsubsection{$T(SO(4))$}

The quiver of $T(SO(4))$ is
\bea
(SO(2)) - (USp(2)) - [SO(4)] , \label{quiv:TSO4}
\eea
with a Coulomb branch symmetry $SO(4)$. In addition there is  a flavor symmetry $SO(4)$ acting on the Higgs branch.
The conjectured HL formula for the Coulomb branch Hilbert series of $T(SO(4))$ is
\bea \label{HLTSO4}
H[T(SO(4)](t; \vec x; \vec n)= t^{n_1} \PE \left[t \left(\frac{1}{x_1 x_2}+\frac{x_1}{x_2}+\frac{x_2}{x_1}+x_1 x_2 \right) \right] \Psi^{(n_1,n_2)} _{D_2} (\vec x; t)~,
\eea
where explicit expression for small values of $n_1$ and $n_2$ are given in Appendix \ref{sec:HLpoly}.  

For $n_1=n_2=0$, we have 
\bea
\Psi^{(0,0)} _{D_2} (\vec x; t) = (1+t)^2 = \frac{\left(1-t^2\right)^2}{(1-t)^2}~,
\eea
therefore 
\be
\begin{split}
H[T(SO(4)](t; x_1 x_2, x_1 x_2^{-1}; 0,0) &= (1-t^2)^2\PE \left[t (x_1^2+x_2^2 +x_1^{-2}+x_2^{-2}+2) \right] \\
&=g_{\BC^2/\BZ_2} (t, x_1) g_{\BC^2/\BZ_2} (t, x_2)~,
\end{split}
\ee
in terms of the Hilbert series of $\BC^2/\BZ_2$
\bea
g_{\BC^2/\BZ_2} (t, x) = (1-t^2) \PE \left[t( x^2 + 1 +x^{-2}) \right] ~.
\eea
Hence the Coulomb branch of $T(SO(4))$ is $(\BC^2/\BZ_2)^2$; this is identical to the Higgs branch of the same theory, in agreement with the self-mirror property.

\paragraph{Comparison with monopole formula.} 
We can compare the conjectured formula \eref{HLTSO4} with the monopole formula for Coulomb branch Hilbert series 
\bea \label{HstdTSO4}
H_{\rm mon}[T(SO(4)](t; x_1; \vec n)= \sum_{m=-\infty}^\infty \sum_{k=0}^\infty x_1^{2m} t^{\Delta(m,k,\vec n)} P_{USp(2)}(t;k) P_{SO(2)}(t;m)~,
\eea
where $m$ is the topological charge for $SO(2)$ gauge group with $x_1$ the corresponding fugacity, and $k$ is the monopole charges in $USp(2)$ gauge group.  Here
\be
\begin{split}
\Delta(m,k, \vec n) &= \frac{1}{2} \left(|m-k|+|m+k|+ \sum_{i=1}^2 \left(|n_i +k|+ |n_i -k| \right)\right) - |2k|~,  \\
P_{USp(2)}(t;k) &= P_{SU(2)} (t;k) =\begin{cases} \frac{1}{1-t^2}~, \qquad & k=0 \\ \frac{1}{1-t}~, \qquad& k >0~. \end{cases} \\
P_{SO(2)}(t;m) &= P_{U(1)}(t;m)=\frac{1}{1-t}~.
\end{split}
\ee
It can be checked that
\bea
H[T(SO(4)](t; x_1^2, x_2=1; n_1, n_2) = H_{\rm mon}[T(SO(4)](t; x_1; n_1,n_2)~. \label{mismatch1}
\eea

Note that in the monopole formula \eref{HstdTSO4} we can refine only one fugacity $x_1$ of $SO(4)$, whereas in the conjectured Hall-Littlewood formula \eref{HLTSO4} for the Hilbert series both fugacities $x_1$ and $x_2$ appear. This requires some explanation. Although the Coulomb branch symmetry is $SO(4)$, only a $U(1)$ subgroup is manifest. The reason is that the only manifest symmetries in the Coulomb branch are the topological symmetries and the quiver $(SO(2))-(USp(2))-[SO(4)]$ has a single abelian factor. The remaining $SO(4)$ generators, including the other generator of the Cartan subalgebra, correspond to monopole operators \cite{Gaiotto:2008ak}. In particular the other generator of the Cartan subalgebra is provided by the monopole operator with $m=0$, $k=1$ along with $n_i=0$.
It would be interesting to understand how to include fugacities for the full Cartan subalgebra in the monopole formula for the Coulomb branch Hilbert series. We will encounter this phenomenon many times in the following.

\subsubsection{$T(USp(4))$ and $T(SO(5))$}

The HL formula for the Coulomb branch of $T(SO(5))$ is given by
\be \label{HLTSO5}
H[T(SO(5))](t; \vec x; \vec m) = t^{\frac{1}{2}(3m_1+m_2)} \PE \left[t (\chi_{[2,0]}^{C_2} (x_1, x_2)-2) \right]  \Psi^{(m_1,m_2)} _{C_2} (x_1,x_2; t),
\ee
where the character of the adjoint representation $[2,0]$ of $USp(4)$ is
\bea
\chi_{[2,0]}^{C_2} (x_1, x_2) =2+\frac{1}{x_1^2}+x_1^2+\frac{1}{x_2^2}+\frac{1}{x_1 x_2}+\frac{x_1}{x_2}+\frac{x_2}{x_1}+x_1 x_2+x_2^2~.
\eea
Since the Lie algebra of $USp(4)$ is isomorphic to that of $SO(5)$, we also have
\bea \label{HLTUSp4}
H[T(USp(4))](t; \vec y; \vec n)= t^{\frac{1}{2}(4n_1+2n_2)} \PE \left[t (\chi_{[0,2]}^{B_2} (y_1, y_2)-2) \right] \Psi^{(n_1,n_2)} _{B_2} (y_1, y_2; t)~,
\eea
where the character of adjoint representation $[0,2]$ of $SO(5)$ is
\bea
\chi_{[0,2]}^{B_2} (x_1, x_2)=2+\frac{1}{x_1}+x_1+\frac{1}{x_2}+\frac{1}{x_1 x_2}+\frac{x_1}{x_2}+x_2+\frac{x_2}{x_1}+x_1 x_2~.
\eea
These two expressions can be matched as follows:
\bea  \label{matchC2B2}
H[T(SO(5))](t; x_1, x_2; n_1+n_2, n_1-n_2)=H[T(USp(4))](t; x_1 x_2, x_1 x_2^{-1}; n_1, n_2)~.
\eea
The matching reflects the translation between representations of $USp(4)$ and $SO(5)$:
\be
\begin{split}
\chi^{C_2}_{(n_1+n_2, n_1-n_2)} (x_1,x_2) &= \chi^{B_2}_{(n_1,n_2)} (x_1 x_2, x_1 x_2^{-1})~, 
\end{split}
\ee
where $(n_1, n_2)$ denotes the representation of $B_2=SO(5)$ in the standard $e$-basis,%
\footnote{This representation corresponds to Dynkin labels $[n_1-n_2, n_1+n_2]_{SO(5)}$ or $[n_1+n_2, n_1-n_2]_{USp(4)}$.} where $n_1 \geq n_2 \geq 0$ with $n_1, n_2$ all integers or all half-integers.  

One can explicitly check  that \eref{HLTUSp4} with $y_2=1$ can also be obtained from the monopole formula \eref{Hilbert_series} for the Coulomb branch Hilbert series. Again, only one fugacity corresponding to the topological charge of the $SO(2)$ group is manifest.

\paragraph{Comparison with gluing.}  We can obtain $T(USp(4))$ by gluing $T(SO(4))$ theory with $[SO(4)]-[USp(4)]$ via the $SO(4)$ group.  We have the Coulomb branch Hilbert series:
\be \label{TUSp4glued}
\begin{split}
& H_{\text{glued}}[T(USp(4))] (t; x_1, x_2; n_1, n_2) \\
&=\sum_{\substack{k_1 \geq |k_2|\\ k_2 \in \BZ}} H[T(SO(4))] (t; x_1, x_2 ; k_1, k_2) \times \\
& \qquad \qquad  t^{-\delta_{D_2}(k_1,k_2)+\frac{1}{2}\delta_{D_2-C_2}(k_1,k_2, n_1,n_2;t)} P_{SO(4)} (t;k_1,k_2)~.
\end{split}
\ee
where $P_{SO(4)} (t;k_1,k_2)$ is given by (A.10) of \cite{Cremonesi:2013lqa} and 
\be
\begin{split}
\delta_{D_2}(k_1,k_2) &=\sum_{j=1}^2 \sum_{i=1}^{j-1} |k_i-k_j| +|k_i+k_j|~, \\
\delta_{D_2-C_2}(k_1,k_2, n_1,n_2;t) &= \sum_{s=0}^1 \sum_{i=1}^2 \sum_{j=1}^2 |(-1)^s n_i - k_j|~.
\end{split}
\ee
Then, \eref{TUSp4glued} can be matched with \eref{HLTSO5} as follows:
\be \label{compglueTUSp4}
\begin{split}
&H[T(SO(5)](t; x_1,x_2; n_1+n_2, n_1-n_2) \\
&= H_{\text{glued}}[T(USp(4))] (t; x_1 x_2, x_1 x_2^{-1}; n_1, n_2)~.
\end{split}
\ee
Comparing \eref{compglueTUSp4} with \eref{matchC2B2}, we see that 
\bea
H_{\text{glued}}[T(USp(4))] (t; x_1, x_2; n_1, n_2)= H[T(USp(4))](t; x_1, x_2; n_1, n_2)~.
\eea

\subsubsection{$T(SO(6))$ and comparison with $T(SU(4))$} \label{sec:checkTSO6andTSU4}

The HL formula for the Coulomb branch of $T(SO(6))$ is given by
\bea \label{HLTSO6}
H[T(SO(6))(t; \vec x; \vec n)= t^{\frac{1}{2}(4n_1+2n_2)} \PE \left[t \left(\chi_{[0,1,1]}^{D_3} ({\vec x})-3 \right) \right] \Psi^{(n_1,n_2,n_3)} _{D_3} (\vec x; t)~.
\eea
Since the Lie algebra of $SO(6)$ is isomorphic to that of $SU(4)$, we expect the matching between the Coulomb branch Hilbert series of $T[SU(4)]$ and that of $T[SO(6)]$.  
Indeed, the Coulomb branch Hilbert series \eref{HLformulaT} for $G=SU(4)$ and $\rho=(1^4)$,
\be
\begin{split}
& H[T(SU(4))](t; \vec y; n_1,n_2,n_3,0 )  \\
&= t^{\frac{1}{2}(3n_1+n_2-n_3)}  \PE \left[ t \left(\chi^{SU(4)}_{[1,0,1]} (\vec y) -4 \right) \right]  \Psi^{(n_1,n_2,n_3,0)}_{U(4)} (\vec y;t)~,
\end{split}
\ee
agrees with \eref{HLTSO6} upon a suitable translation. Explicitly, for any $a_1, a_2, a_3 \geq 0$,
\bea
H[T(SO(6)] \left(t;x_1,x_2,x_3; \vec m (\vec a) \right) = H[T(SU(4))](t; y_1,y_2,y_3; \vec n(\vec a) ) ~,
\eea
with
\be
\begin{split}
\vec m (\vec a) &= \left(\frac{1}{2}a_1+ a_2+ \frac{1}{2}a_3, \; \frac{1}{2}a_1+\frac{1}{2}a_3, \; -\frac{1}{2}a_1+\frac{1}{2}a_3 \right)~, \qquad \\
\vec n (\vec a) &= (a_1+a_2+a_3, \; a_2+a_3, \; a_3, \;0)~,
\end{split}
\ee
and the fugacity map
\bea
y_1^2=\frac{x_1}{x_2 x_3}~, \qquad y_2^2=\frac{x_2}{x_1 x_3}~, \qquad y_3^2=\frac{x_3}{x_1 x_2}~,\qquad y_4^2 = x_1 x_2 x_3~.
\eea

For the case of $SO(6)$, \eref{HLTSO6} with $x_2=x_3=1$ can also be reproduced from the monopole formula \eref{Hilbert_series} for the Coulomb branch Hilbert series. Again, only one fugacity corresponding to topological charge can be made manifest in the latter.


\section{Background magnetic charges and baryonic charges} \label{sec:bary}

Having computed the Coulomb branch Hilbert series of many theories with background magnetic charges turned on, it is natural to ask which quantities on the Higgs branch are such background charges mapped to under mirror symmetry. In this section we show that the answer is the {\it baryonic charges} in the mirror of the theory in question. So, in a sense, as expected from mirror symmetry which acts as S-duality on brane configurations in type IIB string theory, the magnetic background fluxes are mapped to background electric fluxes. 

The relation between the Hilbert series with background fluxes of a theory and its mirror can be made  precise as following. If the gauge  group $G^M$ of the mirror theory contains a $U(1)$ factor, it also has a $U(1)$ topological symmetry acting on its Coulomb branch. In the original theory, this $U(1)$ will be acting on the Higgs branch.   Ungauging the $U(1)$ gauge symmetry in the mirror is equivalent to gauging the corresponding global  $U(1)$ symmetry in the original theory.   
We thus have the following equality
\be\label{bar}
\CG[G^M/U(1)] (t,b) = \sum_{m\in \mathbb{Z}} \CG[G^M/U(1),m](t) b^m =  \frac{1}{1-t} \sum_{m\in \mathbb{Z}}  H[G,U(1)](t,m) b^m~,
\ee
where we explain the notation as follows.
\bi
\item 
$\CG[G^M/U(1)](t,b) $ is  the Hilbert series for the {\it Higgs branch} of the mirror theory with gauge group $G^M/U(1)$, graded according to
the dimension and the baryonic symmetry corresponding to the ungauged $U(1)$.  This function is known as the {\bf bayonic generating function} of the mirror theory \cite{Butti:2006au,Forcella:2007wk,Butti:2007jv}.
$\CG[G^M/U(1)](t,b)$ can be decomposed into sectors of definite baryonic charge by writing the Hilbert series as a formal Laurent series in the baryonic fugacity $b$.   We denote such a function in the $m$-th sector by $\CG[G^M/U(1),m](t)$.
\item $H[G,U(1)](t,m)$ is the Hilbert series of the {\it Coulomb branch} of the original theory with a background magnetic flux for the global $U(1)$. The right hand side of the previous formula is the result of gauging this global $U(1)$:  $b$ is a fugacity for the topological symmetry that arises after gauging.
\ei

Even though the correspondence \eref{bar} holds separately for each $U(1)$ group, for simplicity we will mostly present examples where we gauge (resp. ungauge) the maximal torus of the symmetry group acting on the Coulomb branch (resp. Higgs branch of the mirror theory). Given a theory $T$, we use the notations $T_{\vec B}$ for the theory obtained by replacing all $U(N_i)$ gauge groups in $T$ by $SU(N_i)$ gauge groups, and $T_{\vec J}$ for the theory obtained by gauging the maximal torus of the flavor symmetry. $\vec B$ and $\vec J$ label the baryonic and topological symmetries gained in the two processes. If theories $T$ and $T^M$ are mirror theories, so are $T_{\vec B}$ and  $T^M_{\vec J}$. The equivalence of ungauging a $U(1)$ on the Coulomb branch of a theory and gauging a $U(1)$ on the Higgs branch of the mirror theory is straightforward to see at the level of Hilbert series \cite{Cremonesi:2013lqa}. The equivalence of gauging a $U(1)$ on the Coulomb branch of a theory and ungauging a $U(1)$ on the Higgs branch of the mirror theory is much less trivial to see at the level of Hilbert series. We show it in a number of examples in the rest of the section.

\subsection{$T(SU(N))$ theory: $(1)-(2)-\cdots-[N]$ quiver}

Let us first consider $T(SU(N))$ theories, which are self-mirror.  In the following we provide explicit examples of $N=2$ and $N=3$, from which a general formula for any $N$ can be concluded.
 
\subsubsection{$T(SU(2))$ theory}

The baryonic generating function of $(1)-[2]$ quiver is defined as the Hilbert series of the Higgs branch of the same theory with the $U(1)$ node ungauged  \cite{Forcella:2007wk}, \ie~ that of $[1]-[2]$ quiver.  The charge of the hypermultiplets under this ungauged $U(1)$ has an interpretation of the baryonic charge of the $(1)-[2]$ theory.  

Let $b$ be the fugacity associated with the baryonic charges. The baryonic generating function is then given by
\bea
\CG[T(SU(2))_B](t; x; b) = \PE[ (b+b^{-1})(x+x^{-1}) t^{1/2}]~,
\eea
where $x$ is the fugacity for $SU(2)$.  After a rescaling $t \rightarrow t^2$ needed to compare with  4d quantities, the function 
\bea (1-t)\CG[T(SU(2))_B](t; x; b) \nn \eea 
is indeed the $F$-flat Hilbert series \cite{Forcella:2008bb, Forcella:2008eh} of $(1)-[2]$ quiver and the baryonic generating function of $\BC^2/\BZ_2$; see (4.6) of \cite{Forcella:2007wk}.

The baryonic generating function can be related to the Coulomb branch Hilbert series of $T(SU(2))$ with the background flux turned on. Indeed, we find that
\be
\begin{split}
\CG[T(SU(2))_B](t; x; b) 
&=  \PE[ (b+b^{-1})(x+x^{-1}) t^{1/2}]  \\
&= \frac{1}{1-t} \sum_{n=-\infty}^\infty  H[T(SU(2))] (t; x, x^{-1}; |n|, 0) \; b^{n} \\
&= H[T(SU(2))_J](t;x;b) ~, \label{bary12}
\end{split}
\ee
where $H[T(SU(2))]$ is given by \eref{HLformulaT} with $\vec \rho=(1,1)$, which extends to the Weyl chamber $n<0$ by replacing $n\to -n$, or equivalently by the monopole formula \eref{stdformulaT}:
\bea
H[T(SU(2))](t; x_1, x_2; n_1, n_2) = x_1^{n_1+n_2}  \sum_{u=-\infty}^\infty t^{\frac{1}{2}(|u-n_1|+|u-n_2|)} \frac{1}{1-t} (x_2 x_1^{-1})^{u}~.
\eea
The summation in \eref{bary12} has an interpretation of {\it generating function} for the Coulomb branch Hilbert series with background fluxes.  
Equation \eref{bary12} relates this generating function to the baryonic generating function on the Higgs branch of the theory. For reference, we provide the expressions for such a function in the regions $n \geq 0$ and $n<0$:
\bea
\sum_{n\geq 0} H[T(SU(2))] (t; x, x^{-1}; |n|, 0) \; b^{n} &=\frac{\left(1+t\right) -t^{3/2} \left(x^{-1}+x\right) b}{\left(1-tx^{-2}\right) \left(1-t x^2\right) \left(1-t^{1/2} b x^{-1}\right) (1-t^{1/2} b x)} ~,\nn \\
\sum_{n<0} H[T(SU(2))] (t; x, x^{-1}; |n|, 0) \; b^{n} &= \frac{t^{1/2} b^{-2} \left[\left(x^{-1}+x\right) b-t^{1/2} \left(1+t\right)\right]}{\left(1-tx^{-2}\right) \left(1-t x^2\right) \left(1-t^{1/2} x b^{-1}\right) \left(1-t^{1/2} x^{-1} b^{-1}\right)}~.\nn \\
\eea

As anticipated, \eref{bary12} shows that ungauging the $U(1)$ gauge symmetry on the Higgs branch side corresponds to gauging a $U(1)$ global symmetry of the Coulomb branch of the mirror, and the baryonic charges are just the topological charge in the Coulomb branch. In this gauging process we introduce the factor $(1-t)^{-1}$ on the right hand side of \eref{bary12}.
The generating function of the Coulomb branch Hilbert series of the quiver $(1)-[2]$ is nothing but the Hilbert series of the Coulomb branch of the theory obtained by gauging the $U(1)$ Cartan subgroup of the flavor symmetry $U(2)/U(1)$, namely the quiver $(1)-(1)-[1]$, or equivalently $(1)-(1)-(1)$ with the overall $U(1)$ gauge group factored out.

\subsubsection{$T(SU(3))$ theory}

As before, the baryonic generating function of $(1)-(2)-[3]$ quiver is equal to the Higgs branch Hilbert series of the same theory with the two unitary gauge groups replaced by special unitary groups:
\be \label{baryonTSU3}
\begin{split}
& \CG[T(SU(3))_{\vec B}](t; x_1, x_2, x_3; b_1, b_2) \\
&= \int {\rm d} \mu_{SU(2)} ( z) \frac{1}{\PE[  t(z^2+1+z^{-2}) ]}  \PE \Big[(b_1^{-1}b_2 (z+z^{-1}) +b_1 b_2^{-1} (z+z^{-1})  \\
& \qquad + b_2(z+z^{-1})\sum_{i=1}^3 x_i^{-1} +b_2^{-1}(z+z^{-1})\sum_{i=1}^3 x_i ) t^{1/2} \Big]~,
\end{split}
\ee
where 
\bi
\item $b_1$ and $b_2$ are baryonic fugacities; they can also be viewed  as electric fugacities for the $U(1)$ node and for the $U(1)$ center of the $U(2)$ node respectively,
\item  $x_1,x_2,x_3$, with $x_1 x_2 x_3=1$, are fugacities of the $SU(3)$ flavor symmetry,
\item the Haar measure of $SU(2)$ is
\bea
\int {\rm d} \mu_{SU(2)} ( z)  = \frac{1}{2 \pi i} \oint_{|z|=1} \frac{1-z^2}{z}~.
\eea
\ei

\subsubsection*{Coulomb branch Hilbert series and baryonic generating function}

We are interested in relating \eref{baryonTSU3} to the Coulomb branch Hilbert series of $T(SU(3))$. It is convenient to write the latter using the monopole formula \eref{stdformulaT} as follows:
\be \label{HTSU3std}
\begin{split}
& H[T(SU(3))](t; x_1, x_2, x_3; n_1,n_2, n_3 =0 )  \\
&=x_1^{n_1+n_2}  \sum_{m_{1,1}=-\infty}^\infty \sum_{m_{1,2} \geq m_{2,2} > -\infty}^\infty   (x_2 x_1^{-1})^{m_{1,2}+m_{2,2}} (x_3 x_2^{-1})^{m_{1,1}} \times  \\
& \qquad t^{\Delta(m_{1,1}; m_{1,2},m_{2,2}; n_1, n_2,0)}  P_{U(2)}(m_{1,2},m_{2,2};t)P_{U(1)}(t) ~,
\end{split}
\ee
with $x_1 x_2 x_3=1$, $P_{U(2)}(\vec m; t)$ and $P_{U(1)}(t)$ given by \eref{classical_dressing}, and
\be
\begin{split}
&\Delta(m_{1,1}; m_{1,2},m_{2,2}; n_1, n_2,n_3) \\
&= \frac{1}{2}\sum_{i=1}^2  |m_{1,1}-m_{i,2}|+ \frac{1}{2}\sum_{i=1}^2  \sum_{j=1}^3 |m_{i,2}-n_j|-|m_{1,2}-m_{2,2}|~.
\end{split}
\ee
Indeed we find that \eqref{baryonTSU3} and \eqref{HTSU3std} are related as
\be\label{bary123}
\begin{split}
& \CG[T(SU(3))_{\vec B}](t, \vec x, \vec b) \\
& = (1-t)^{-2} \sum_{n_1, n_2 \in \BZ} H[T(SU(3))](t; x_1^{-1}, x_2^{-1},x_3^{-1};n_1,n_2,0) b_1^{n_1} (b_2^2/b_1)^{n_2}\\
& = H[T(SU(3))_{\vec J}](t; x_1^{-1}, x_2^{-1},x_3^{-1};b_1, b_2^2 b_1^{-1})
~.
\end{split}
\ee
This indeed confirms the relation between the generating function of the Coulomb branch Hilbert series and the baryonic generating function, or the Higgs branch Hilbert series of $T(SU(3))_{\vec B}$ and the Coulomb branch Hilbert series of $T(SU(3))_{\vec J}$.  

Note that $b_2$ appears only as $b_2^2$ in \eqref{bary123} because it is the baryonic fugacity of an $SU(2)$ gauge group normalized in such a way that gauge invariants have charges in $2\bZ$.

\subsubsection*{Hall-Littlewood formula}

Alternatively to the monopole formula \eref{HTSU3std}, the Coulomb branch Hilbert series is also given by the HL formula \eref{HLformulaT}:
\bea
H_3(t; x_1,x_2, x_3; n_1,n_2,0) = t^{n_1} (1-t)^3 \PE \left[t \sum_{i,j=1}^3 x_i x_j^{-1} \right] \Psi^{(n_1,n_2,0)}_{U(3)} (\vec x; t)~,
\eea
where we emphasize that this formula is valid only if $n_1 \geq n_2 \geq 0$ and we take $x_1 x_2 x_3=1$.  The relation with \eref{baryonTSU3} can therefore be separated into 6 regions, as in Appendix A.2 of \cite{Butti:2007jv}.  Explicitly, we obtain
\bea \label{G123}
\CG[T(SU(3))_{\vec B}](t; \vec x; \vec b) &= (1-t)^{-2} \times \nn\\
&\Bigg[ \sum_{m=0}^\infty \sum_{n=0}^\infty H_3(t; x_1^{-1},x_2^{-1},x_3^{-1}; m+n,n) b_1^m b_2^{2n} \nn \\
&+  \sum_{m=1}^\infty \sum_{n=1}^\infty  H_3(t; x_1,x_2,x_3; m+n,n) b_1^{-m} b_2^{-2n} \nn\\
&+  \sum_{n=1}^\infty \sum_{m=n+1}^\infty  H_3(t; x_1^{-1},x_2^{-1},x_3^{-1}; m,n) b_1^{m} b_2^{-2n} \\
&+  \sum_{n=0}^\infty \sum_{m=n+1}^\infty H_3(t; x_1,x_2,x_3; m,n) b_1^{-m} b_2^{2n} \nn\\
&+  \sum_{m=0}^\infty \sum_{\substack{n=m \\ n\neq 0}}^\infty  H_3(t; x_1^{-1},x_2^{-1},x_3^{-1}; n,m) b_1^{m} b_2^{-2n} \nn\\
&+  \sum_{m=1}^\infty \sum_{n=m}^\infty H_3(t; x_1,x_2,x_3; n,m) b_1^{-m} b_2^{2n}  \Bigg]~, \nn
\eea
where $x_1x_2x_3=1$. 
Each summand indicates the baryonic generating function, in terms of the Coulomb branch Hilbert series, in one of the $6$ Weyl chambers of $SU(3)$.

\subsubsection{$T(SU(N))$ theory}

\begin{figure}[H]
\begin{center}
\includegraphics[scale=0.6]{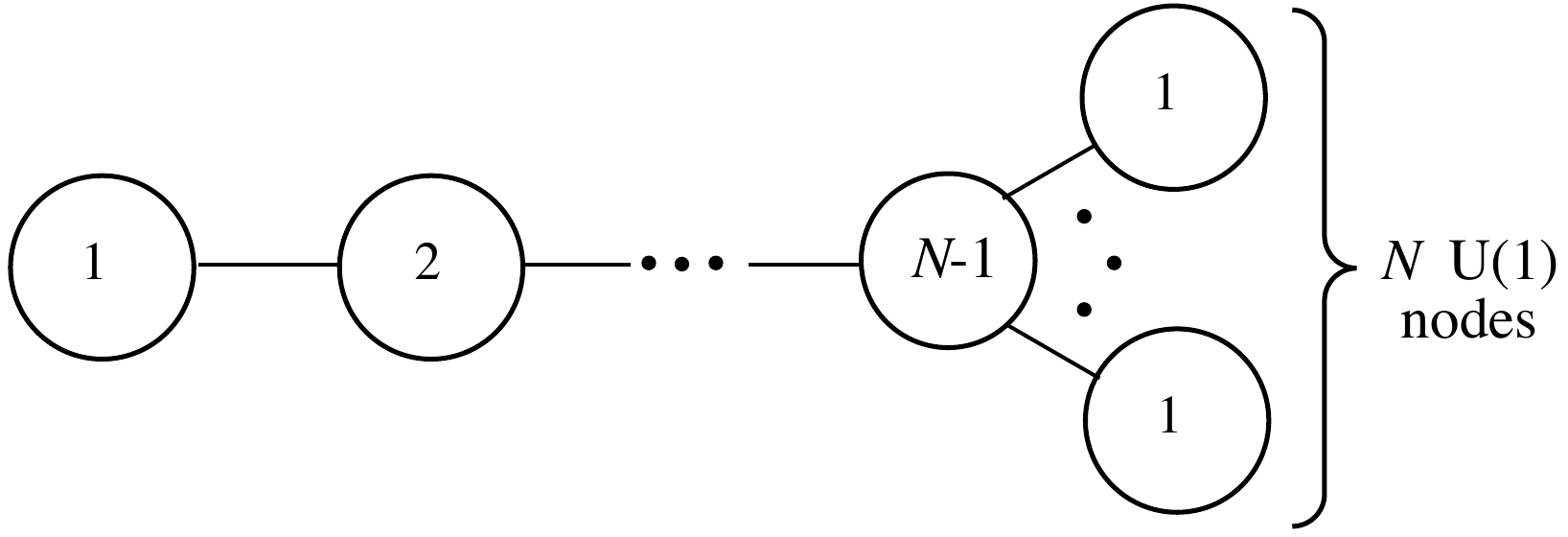}
\caption{The quiver diagram of $T(SU(N))_{\vec J}$.  In this quiver, the flavor group $U(N)$ of $T(SU(N))$ is replaced by $N$ $U(1)$ nodes joined to the $U(N-1)$ node, with the overall $U(1)$ gauge group factored out.}
\label{fig:TSUNgaugeflv}
\end{center}
\end{figure}

It is straightforward to derive a similar relation to \eref{bary12} and \eref{bary123} for a general $T(SU(N))$ theory.  Using the same notation as before, we obtain
\bea\label{bary123N}
& \CG[T(SU(N))_{\vec B}](t, x_1, \ldots, x_N, b_1, \ldots, b_{N-1}) = \frac{1}{(1-t)^{N-1}} \times \nn \\
&  \sum_{n_1, \cdots, n_{N-1} \in \BZ} H[T(SU(N))](t; x_1^{\pm}, x_{2}^{\pm},\cdots,x_{N}^{\pm};n_1,n_2, \cdots, n_{N-1},0)  b_1^{n_1} \prod_{k=2}^{N-1} (b_{k}^{k} b_{k-1}^{1-k})^{n_{k}}~ = \nn \\
 &~~~= H[T(SU(N))_{\vec J}](t; x_1^{\pm}, x_{2}^{\pm},\cdots,x_{N}^{\pm};b_1, b_2^2 b_1^{-1}, \cdots, b_{N-1}^{N-1} b_{N-2}^{-(N-2)})
, 
\eea
where the power $\pm$ is $+$ for $N$ even and $-$ for $N$ odd. \eref{bary123N} shows that the Hilbert series of the Higgs branch of $T(SU(N))_{\vec B}$ and the Coulomb branch of $T(SU(N))_{\vec J}$ coincide. The latter is described by a quiver where the flavor group is replaced by $N$ $U(1)$ nodes joined to the $U(N-1)$ node, with the overall $U(1)$ gauge group factored out. The corresponding quiver diagram is depicted in \fref{fig:TSUNgaugeflv}.

\subsection{$T_{(2,1)}(SU(3))$ theory: $(1)-[3]$ quiver}

The mirror theory of $T_{(2,1)}(SU(3))$ is
\bea  \label{mirT12SU3}
\begin{tikzpicture}[font=\scriptsize]
\begin{scope}[auto,%
  every node/.style={draw, minimum size=1cm}, node distance=1cm];
\node[circle] (U1a) at (0, 0) {$U(1)$};
\node[circle, right=of U1a] (U1b) {$U(1)$};
\node[rectangle, below=of U1a] (U1f1)  {$U(1)$};
\node[rectangle, below=of U1b] (U1f2)  {$U(1)$};
\end{scope}
\draw (U1a) -- (U1b)
(U1a)--(U1f1)
(U1b)--(U1f2);
\end{tikzpicture}
\eea
The baryonic generating function of the mirror of $T_{(2,1)}(SU(3))$ is given by
\be \label{barymirT21}
\begin{split}
& \CG[(\text{mirror $T_{(2,1)}(SU(3))$})_{\vec B}] (t; x_1, x_2; b_1, b_2) \\
& = \PE \left[ ( b_1 b_2^{-1}+b_1^{-1} b_2 + b_1 x_1^{-1} +b_1^{-1} x_1+ b_2 x_2^{-1} +b_2^{-1} x_2) t^{1/2} \right]~.
\end{split}
\ee
 On the other hand, the Hilbert series of the Coulomb branch of $T_{(2,1)}(SU(3)): (1)-[3]$ is given by \eref{CouT21}. Formulae 
\eref{barymirT21} and \eref{CouT21} are related as
\be 
\begin{split}
& \CG[(\text{mirror $T_{(2,1)}(SU(3))$})_{\vec B}] (t; x_1, x_2; b_1, b_2) \\
& = \frac{1}{(1 - t)^{2}}\sum_{n_1,n_2 \in \BZ}  (x_1 x_2)^{-n_2} H[T_{(2,1)}(SU(3))] (t; x_1, x_2; n_1, n_2,0) b_1^{-n_1} b_2^{n_2} \\
& = \frac{1}{(1 - t)^{3}}\sum_{m,n_1,n_2 \in \BZ} \left(\frac{x_2}{x_1}\right)^{m}\left(\frac{x_1}{b_1}\right)^{n_1} \left(\frac{b_2}{x_2}\right)^{n_2} t^{\frac{1}{2}\left(\sum_{i=1}^2 |m-n_i|+|m|\right)} ~.
\end{split}
\ee
This again confirms the relation between the generating function of the Coulomb branch Hilbert series and the baryonic generating function of the mirror theory.

\subsection{$T_{(3,1,1)}(USp(4))$ theory: $(SO(2))-[USp(4)]$ quiver}
\label{sec:baryUsp}

We now consider the $T_{(3,1,1)}(USp(4))$ theory, which corresponds to $SO(2)$ gauge theory with 4 flavors of half-hypermultiplets in the two-dimensional vector representation.
The aim of this section is to present certain subtleties that do not appear in $T_{\vec \rho} (SU(N))$ theory.  The Coulomb branch Hilbert series of this theory is discussed in appendix \ref{sec:T311Usp4}. In the following we shall compute the baryonic generating of the Higgs branch of the mirror theory and establish some relations with such a Coulomb branch Hilbert series.

\subsubsection*{The Coulomb branch Hilbert series of $T_{(3,1,1)}(USp(4))$}

The monopole formula for the Coulomb branch Hilbert series of $T_{(3,1,1)}(USp(4))$ is
\bea  \label{CouT311USp4}
H[T_{(3,1,1)}(USp(4))](t; z; n_1, n_2) = \sum_{m=-\infty}^{\infty} t^{\Delta(n_1,n_2; m)} P_{SO(2)}(t) z^{2m}~,
\eea
where $z^2$ keeps track of the topological charge of the gauge group $SO(2)$, and
\bea
P_{SO(2)}(t) = \frac{1}{1-t}~, \qquad \Delta(n_1,n_2; m) = \frac{1}{2}\sum_{i=1}^2 \sum_{s=0}^1 |(-1)^s n_i +m|~.
\eea
If we set $n_1=n_2=0$, the Coulomb branch Hilbert series simplifies to
\bea \label{CouT311USp400}
H[T_{(3,1,1)}(USp(4))](t; z; 0,0) = \PE \left[ t + t^2( z^2+ z^{-2}) -t^4\right] ~.
\eea
Note that this is the Hilbert series of $\BC^2/\BZ_4$, the Coulomb branch of the $U(1)$ gauge theory with $4$ flavors of hypermultiplets of charge $1$, which is the same as the $SO(2)$ gauge theory with $4$ flavors of half-hypermultiplets in the vector representation of $SO(2)$. Next we compute the baryonic generating function of the mirror of $T_{(3,1,1)}(USp(4))$.

\subsubsection*{Mirror of $T_{(3,1,1)}(USp(4))$ and the baryonic generating function}

The mirror theory of $T_{(3,1,1)}(USp(4))$ theory was discussed in section 6.2 of \cite{Feng:2000eq}.  Its quiver is given by
\be \label{mirT311Usp4}
\begin{split}
& \text{Mirror of $T_{(3,1,1)}(USp(4))$:} \qquad \\
& \qquad [SO(2)]-(USp(2))-(SO(2))-(USp(2))-[SO(2)]~.
\end{split}
\ee
This is a `bad' theory, since the number of flavors of each $USp(2)$ gauge node is $2$, which is less than $2(2)+1=5$.  In fact, this theory flows to the mirror theory of $U(1)$ gauge theory with $4$ flavors, as we shall demonstrate below \eref{HiggsmirT311}. Here we are interested in the Higgs branch, which is protected from quantum corrections, so we can use the `bad' quiver to compute it. 

The computation of the $F$-flat Hilbert series and the baryonic generating function of this mirror theory, obtained by ungauging the $SO(2)$ group, is rather techical, and we relegate it to Appendix \ref{app:HiggsT311Usp4}.  We present only the end result of the baryonic generating function, which is given by \eref{GmirT311USp4}:
\be \label{baryT311USp4}
\begin{split}
\CG(t; x, y; b)&:=\CG[(\text{Mirror of $T_{(3,1,1)}(USp(4))$})_B](t; x, y; b) \\
&= \PE \left[ \{ 2+(b x^{-1} + b^{-1} x) + (b y^{-1} + b^{-1} y) \} t -  2t^2\right] \\
&= \PE \left[ \{ 2+(b_1+b_1^{-1}) + (b_2+b_2^{-1}) \} t -  2t^2 \right]~.
\end{split}
\ee
where $x$ and $y$ are the fugacities of the two $SO(2)$ flavor symmetries, and $b$ is the baryonic fugacity corresponding to the ungauged $SO(2)$ in \eref{mirT311Usp4}.  Two combinations of the baryonic and flavor fugacities appear in this generating function, namely
\bea \label{twobaryT311Usp4}
b_1 := b x^{-1}~, \qquad b_2 := b y^{-1}~.
\eea
Note that \eref{baryT311USp4} is, in fact, the Hilbert series of $(\BC^2/\BZ_2)^2$, where $b_1$ and $b_2$ are fugacities for the $SU(2)$ isometry associated with each copy of $\BC^2/\BZ_2$.

The Higgs branch Hilbert series of quiver \eref{mirT311Usp4} is given by  \eref{HiggsmirT311USp4}:
\be
\begin{split}
\oint_{|b|=1} \frac{{\rm d} b}{2 \pi i b} (1-t) \CG(t; x, y; b) &= \PE \left[ t+t^2( x y^{-1} + y x^{-1}) - t^4\right]  \\
&= H[T_{(3,1,1)}(USp(4))](t; x^{1/2}y^{-1/2}; 0, 0)~; \label{HiggsmirT311}
\end{split}
\ee
this is equal to the Coulomb branch Hilbert series \eref{CouT311USp4} with vanishing background fluxes $n_1=n_2=0$, as predicted by mirror symmetry.  Indeed, this indicates that the `bad' theory \eref{baryT311USp4} flows to the mirror theory of $U(1)$ gauge theory with $4$ flavors in the infra-red.

\subsubsection*{Relation between \eref{CouT311USp4} and \eref{baryT311USp4}}

We find that
\bea
\CG_{\BZ_2}(t; x=z, y=z^{-1}; b)= (1-t)^{-1} \sum_{n=-\infty}^\infty H[T_{(3,1,1)}(USp(4))](t; z; n, n) b^{2n}~, \label{relGH}
\eea
where $\CG_{\BZ_2}(t; z, z^{-1}; b)$ is a $\BZ_2$ projection of $\CG(t; z, z^{-1}; b)$ defined as follows:
\bea
\CG_{\BZ_2}(t; z, z^{-1}; b) := \frac{1}{2} \left[ \CG(t; z, z^{-1}; b)+\CG(t; z, z^{-1}; -b) \right] ~.
\eea
Note that this $\BZ_2$ projection is {\it not} to be identified with the parity of the orthogonal gauge group $O(2)$, cf. \eref{parityO2}, which does not commute with the Cartan elements of $SO(2)$. It is rather the $\BZ_2$ subgroup of $SO(2)$ that consists of gauge rotations by multiples of $\pi$. 
As such, it projects out the odd powers of $b$ in $\CG(t; z, z^{-1}; b)$.

This example displays several subtleties that are not present for $T_{\vec \rho}(SU(N))$.  Let us comment and list certain open questions as follows:
\ben
\item  
There are two background fluxes $n_1$ and $n_2$ for the flavor symmetry $USp(4)$ of the theory $T_{(3,1,1)} (USp(4))$ but only one baryonic charge $b$ (and correspondingly only one manifest  topological $SO(2)$   symmetry) in the mirror theory. This mismatch  is very much like the situation discussed in \eref{mismatch1}.  This is related to the absence of Cartan generators of the  symmetry $USp(4)$ on the Coulomb branch of the mirror of $T_{(3,1,1)} (USp(4))$, where only $SO(2)$ is manifest.  The remaining Cartan generators of $USp(4)$ correspond to monopole operators.

\item  The coefficients of odd powers of $b$ in \eref{baryT311USp4} (\ie~ those with odd baryonic charges) {\it cannot} be matched with the Coulomb branch Hilbert series \eref{CouT311USp4} for any background fluxes.  Only those of even powers of $b$ can be matched with the Coulomb branch Hilbert series \eref{CouT311USp4}; this happens when $n_1$ is set to be equal to $n_2$.  The $\BZ_2$ projection is there to get rid of the odd powers of $b$ in \eref{baryT311USp4} and hence the matching can be done as in \eref{relGH}.
We also observe that by gauging the $U(1)$'s associated to $n_1$ and $n_2$ and ungauging the $U(1)$ associated to $m$ on the Coulomb branch side, one can reproduce the full $\CG$ in \eref{baryT311USp4}.
\een

\section{Analytic structure of the Coulomb branch Hilbert series} \label{sec:analyt}

In this section we examine the analytic structure of the Coulomb branch Hilbert series of $T_{\vec \rho}(SU(N))$, from the complementary perspectives of the Hall-Littlewood formula and of the monopole formula. 
The study of the analytic structure further substantiates our conjecture that the Hall-Littlewood formula computes the Coulomb branch Hilbert series of $T_{\vec \rho}(SU(N))$ theories.

The analysis of the HL formula is very similar to that of HL index of 4d Sicilian theories in \cite{Gaiotto:2012uq}. The prefactor $K^{U(N)}_{\vec \rho}(\vec x; t)$ in the HL formula \eref{HLformulaT} has a pole corresponding to a particular box corresponding to the rightmost part of the partition $\vec \rho$. Taking the residue of the Hilbert series at this pole isolates the Coulomb branch Hilbert series of a new theory $T_{\vec \rho'}(SU(N))$, where $\vec \rho'$ is obtained from $\vec \rho$ by moving such a box to a previous column. 

We can identify the same simple pole in the monopole formula for the Coulomb branch Hilbert series with background fluxes of $T_{\vec \rho}(SU(N))$. Computing the residue yields the monopole formula for $T_{\vec \rho'}(SU(N))$, in complete analogy with the analysis of the HL formula. It thus follows that 
the equivalence of  the monopole formula (\ref{Hilbert_seriesBackground}) with the Hall-Littlewood formula \eref{HLformulaT} in the case the maximal partition $\rho = (1,1,\cdots,1)$ implies the equivalence of the two formulae for a generic partition $\rho$.
 
Let us demonstrate the idea in the following example. The general case is discussed in Section \ref{sec:boxmove} and Appendix \ref{sec:analytgeneral}. Another example of the application of the same technique is discussed in Appendix \ref{sec:reduceflav}.

\subsection{Obtaining $T_{(2,1)}(SU(3))$ from $T_{(1,1,1)}(SU(3))$} \label{sec:analyticSU3}

Let us compare the Hilbert series with background fluxes for the theories $T_{(1,1,1)}(SU(3)): [3]-(2)-(1)$ and $T_{(2,1)}(SU(3)): [3]-(1)$.  The latter theory can be obtained from the former by moving one box as follows:
\bea
 {\begin{ytableau} ~& ~&  *(blue!20)  \end{ytableau}} & \qquad \longrightarrow \qquad  {\begin{ytableau} ~& ~\\ *(blue!20)  \end{ytableau}} \nn \\
 \qquad \qquad \vec \rho=(1,1,1)  & \qquad \qquad \qquad \vec \rho^\prime =(2,1)
\eea
The relation between 4d superconformal indices of Sicilian theories including 
the former or the latter puncture was studied in \cite{Gaiotto:2012uq}.  In this section, we perform a similar computation to find the relation between the Coulomb branch Hilbert series for the two $T_{\vec\rho}(SU(3))$ theories.

From the HL formula \eref{HLformulaT}, we obtain the expressions \eref{H111SU3} and \eref{H21SU3} for the Coulomb branch Hilbert series of $T_{(1,1,1)}(SU(3))$ and $T_{(2,1)}(SU(3))$ respectively.

Formula \eref{H111SU3} has simple poles at $x_i x_j^{-1}\to t$ for $i\neq j$, due to the plethystic exponential factor in the second line (the HL polynomials are regular). In view of the permutation symmetry, we can focus on the pole at $x_1 x_3^{-1}\to t$. We set 
\be\label{pole1}
x_1 = y_1 t^{1/2} z~, \qquad x_2=y_2~, \qquad x_3 = y_1 (t^{1/2} z)^{-1}~
\ee 
and consider the limit where $z\to 1$. Similarly to the discussion in section 2.2 of \cite{Gaiotto:2012uq} for the HL limit of superconformal indices, the residue of \eref{H111SU3} as $z\to 1$ in \eref{pole1} reproduces the Hilbert series \eref{H21SU3} with a certain prefactor that is easily determined:
\be
\begin{split}
& \underset{z\rightarrow 1}{\rm Res} ~ H[T_{(1,1,1)}(SU(3))] (t; y_1 t^{1/2} z, y_2, y_1 t^{-1/2} z^{-1}; n_1, n_2) \\
&=\frac{1}{2} \PE \left[t^{1/2} (y_1 y_2^{-1}+y_2 y_1^{-1})+t \right]  H[T_{(2,1)}(SU(3))] (t; y_1, y_2; n_1,n_2)~.
\label{final}
\end{split}
\ee

\subsubsection*{Analytic structure of the monopole formula}

The aim of this section is to show that the residue formula \eref{final} can also be derived by means of the monopole formula of the Hilbert series \eref{std111SU3}. 
The idea is that when evaluating this expression at the values of $x_i$ corresponding to a pole, one introduces extra powers of $t$ depending on the monopole fluxes which make the series diverge along certain directions in the GNO lattice. 
Taking the residue, some gauge groups effectively disappear, reproducing the quiver corresponding to a different partition.

\paragraph{Example: pole at $x_1 x_3^{-1}\to t$ in \eref{std111SU3}.} The monopole formula \eref{std111SU3} has a simple pole when the fugacities $\vec x$ are as in \eqref{pole1} and $z\to 1$, because there is a region of the summation in \eref{std111SU3} where the series diverges. To see this, we 
consider the region where 
\be\label{dangerous_direction}
m_{1,1}, m_{2,2} \gg m_{1,2},n_j~.
\ee  
Changing summation variable from $m_{1,1}$ to
\bea
q = m_{1,1} - m_{2,2}~,
\eea
the power of $t^{1/2}$ in \eref{std111SU3} becomes
\be
\begin{split}
&\sum_{j=1}^3 n_j-(m_{1,1}+m_{1,2}+m_{2,2})+ \Delta_{(1,1,1)}(m_{1,1} = m_{2,2}+q; m_{1,2},m_{2,2}; \vec n)  \\
&= \sum_{j=1}^3 |m_{1,2}-n_j| + |q| = \Delta_{(2,1)}(m_{1,2}; \vec n) + |q| ~, \end{split}
\ee
where 
\bea 
\Delta_{(2,1)}(m_{1,2}; \vec n) = \sum_{j=1}^3 |m_{1,2}-n_j| ~, \quad n_3=0~,
\eea 
is twice the dimension of monopole operators in the quiver $[3]-(1)$.
In this region, the summations in \eref{std111SU3} diverge when $z\to 1$ because $m_{2,2}$ disappears from the summands. The singularity depends on the summation over $m_{2,2}$, which depends only on $z$:
\be \label{Resz}
 \underset{z \rightarrow 1}{\rm Res} \sum_{m_{2,2}=L}^\infty z^{n_1+n_2-q-m_{1,2}-2m_{2,2}} = \underset{z \rightarrow 1}{\rm Res}~\frac{z^{2-2L+n_1+n_2-q-m_{1,2}}}{z^2-1} = \frac{1}{2}~,
\ee
where $L$ is a lower cutoff larger than $m_{1,2}$ and $n_j$.

Because $P_{U(2)}(t; m_{1,2}, m_{2,2})$ becomes $1/(1-t)^2$ for $m_{2,2} \gg m_{1,2}$, we find that
\be\label{residue_intermed}
\begin{split}
& \underset{z\rightarrow 1}{\rm Res} ~ H[T_{(1,1,1)}(SU(3))] (t; y_1 t^{1/2} z, y_2, y_1 t^{-1/2} z^{-1}; n_1, n_2) \\
&=\frac{1}{2} \frac{1}{1-t}\left[\frac{1}{1-t} \sum_{q\in\BZ} t^{\frac{1}{2}|q|} \left(\frac{y_1}{y_2}\right)^{q} \right] \left[\frac{y_1^{n_1+n_2}}{1-t} \sum_{m_{1,2}\in\BZ} \left( \frac{y_2}{y_1} \right)^{m_{1,2}} t^{\frac{1}{2}\Delta_{(2,1)}(m_{1,2};\vec n)}\right]~.
\end{split}
\ee
The factor in the first square brackets in \eref{residue_intermed} is the Coulomb branch Hilbert series of $U(1)$ with one flavor, which is mirror to a twisted hypermultiplet and evaluates to
\be \label{fac1}
\frac{1}{1-t}\sum_{q\in\BZ} t^{\frac{1}{2}|q|} (y_1 y_2^{-1})^{q} = \frac{1}{(1-t^{1/2} y_1 y_2^{-1})(1-t^{1/2} y_2 y_1^{-1})}~.
\ee
The factor in the second square brackets is the Coulomb branch Hilbert series for the quiver $[3]-(1)$ with background fluxes $\vec n$, corresponding to the partition $\vec \rho^\prime=(2,1)$. 
Therefore we conclude from the monopole formula that 
\be
\begin{split}
& \underset{z\rightarrow 1}{\rm Res} ~ H[T_{(1,1,1)}(SU(3))] (t; y_1 t^{1/2} z, y_2, y_1 t^{-1/2} z^{-1}; n_1, n_2) \\
&=\frac{1}{2} \PE \left[t^{1/2} (y_1 y_2^{-1}+y_2 y_1^{-1})+t \right]  H[T_{(2,1)}(SU(3))] (t; y_1,y_2; n_1,n_2)~,
\label{final2}
\end{split}
\ee
as deduced previously from the HL formula in \eref{final}.

\subsection{Moving the last box: from partition $\vec \rho=(\rho_1, \cdots, \rho_{d-h},\rho_{d-h+1},1^{h})$ to $\vec \rho^\prime=(\rho_1, \cdots, \rho_{d-h}, \rho_{d-h+1}+1, 1^{h-1})$}
\label{sec:boxmove}

The previous argument can be generalized to the following theories:
\be
\begin{split}
T_{\vec \rho}(SU(N)): &\quad {[N]-\cdots-(h+\rho_{d-h+1})-(h)-(h-1)-\cdots-(2)-(1)}~ \\
T_{\vec \rho^\prime}(SU(N)): &\quad {[N]-\cdots-(h+\rho_{d-h+1})-(h-1)-\cdots-(2)-(1)}~
\end{split}
\ee
where the total numbers of gauge groups are respectively $d$ and $d-1$, and
\bea N = h + \sum_{k=1}^{d-h+1} \rho_k~. \eea
This corresponds to moving the last box in a partition ending with $1$ as follows: 
\be \label{yngmove}
\begin{split}
\ytableausetup{boxsize=15pt}
&\begin{ytableau}
*(black!20) & \none[\ldots] & *(black!20) &   *(blue!20) & *(red!20)  &*(red!20)&  \none[\ldots] & *(red!20) & *(blue!60) \\
 \none[\vdots] &  \none[\vdots] & \none[\vdots] & \none[\vdots] \\
  *(black!20) & \none[\ldots] & *(black!20)&  *(blue!20) \\
    \none[\vdots]  & \none[\vdots]  & \none[\vdots]  \\
  *(black!20) & \none[\ldots] & *(black!20)  \\
   \none[\vdots]     \\
  *(black!20) 
\end{ytableau}  \\  
& \qquad\vec \rho=(\rho_1, \cdots, \rho_{d-h+1},1^{h}) \\
& \qquad \qquad \qquad \longdownarrow \\  \\
& \begin{ytableau}
*(black!20) &  \none[\ldots]  & *(black!20) & *(blue!40) & *(red!20) &*(red!20)&  \none[\ldots] & *(red!20)  \\
 \none[\vdots] &  \none[\vdots] & \none[\vdots] &  \none[\vdots] \\
  *(black!20) & \none[\ldots] & *(black!20)  & *(blue!40) \\
    \none[\vdots]  & \none[\vdots] & \none[\vdots]  & *(blue!40) \\
  *(black!20) & \none[\ldots] & *(black!20)  \\
   \none[\vdots]     \\
  *(black!20) 
\end{ytableau} \\
& \quad\vec \rho^\prime=(\rho_1, \cdots, \rho_{d-h+1}+1,1^{h-1}) 
\end{split}
\ee
The grey boxes constitute the first spectator block $(\rho_1, \cdots, \rho_{d-h+1})$, the pink boxes constitute the second spectator block $(1^{h-1})$ of the partition $(\rho_1, \cdots, \rho_{d-h+1},1^{h-1})$, and the blue boxes belong to the columns involved in the move.

Note that any partition $\sigma$ of $N$ can be obtained from $(1^N)$ by an iteration of the previous move. Therefore by repeated residue computations we may extract the Coulomb branch Hilbert series of any $T_{\vec\sigma}(SU(N))$ theory from that of $T(SU(N))$.

Let us denote the fugacities corresponding to each column of $\vec \rho$ and $\vec \rho^\prime$ by
\be \label{rho1rho2}
\begin{split}
\vec \rho: &\qquad (x_1, \ldots, x_{d-h}, {\color{blue!60} x_{d-h+1}}, {\color{red} x_{d-h+2}, \ldots, x_d},{\color{blue!100!} x_{d+1}})~, \\
\vec \rho^\prime: &\qquad (x_1,  \ldots ,x_{d-h}, {\color{blue!85} y_{d-h+1}}, {\color{red} x_{d-h+2}, \ldots, x_d} )~,
\end{split}
\ee
where the fugacities are coloured in accordance with the boxes in \eref{yngmove}. They are subject to the constraints
\bea
& \left( \sum_{k=1}^{d-h} x_k^{\rho_k} \right) (x_{d-h+1}^{\ell}) \left(  \prod_{i=1}^{h-1} x_{d-h+1+i} \right) x_{d+1} = 1~, \\
& y^{\ell+1}_{d-h+1} = x_{d-h+1}^\ell x_{d+1}~, \qquad \text{with $\ell := \rho_{d-h+1}$}~. \label{constrmove}
\eea

\begin{figure}[H]
\begin{center}
\hspace{-20pt}
\includegraphics[scale=0.65]{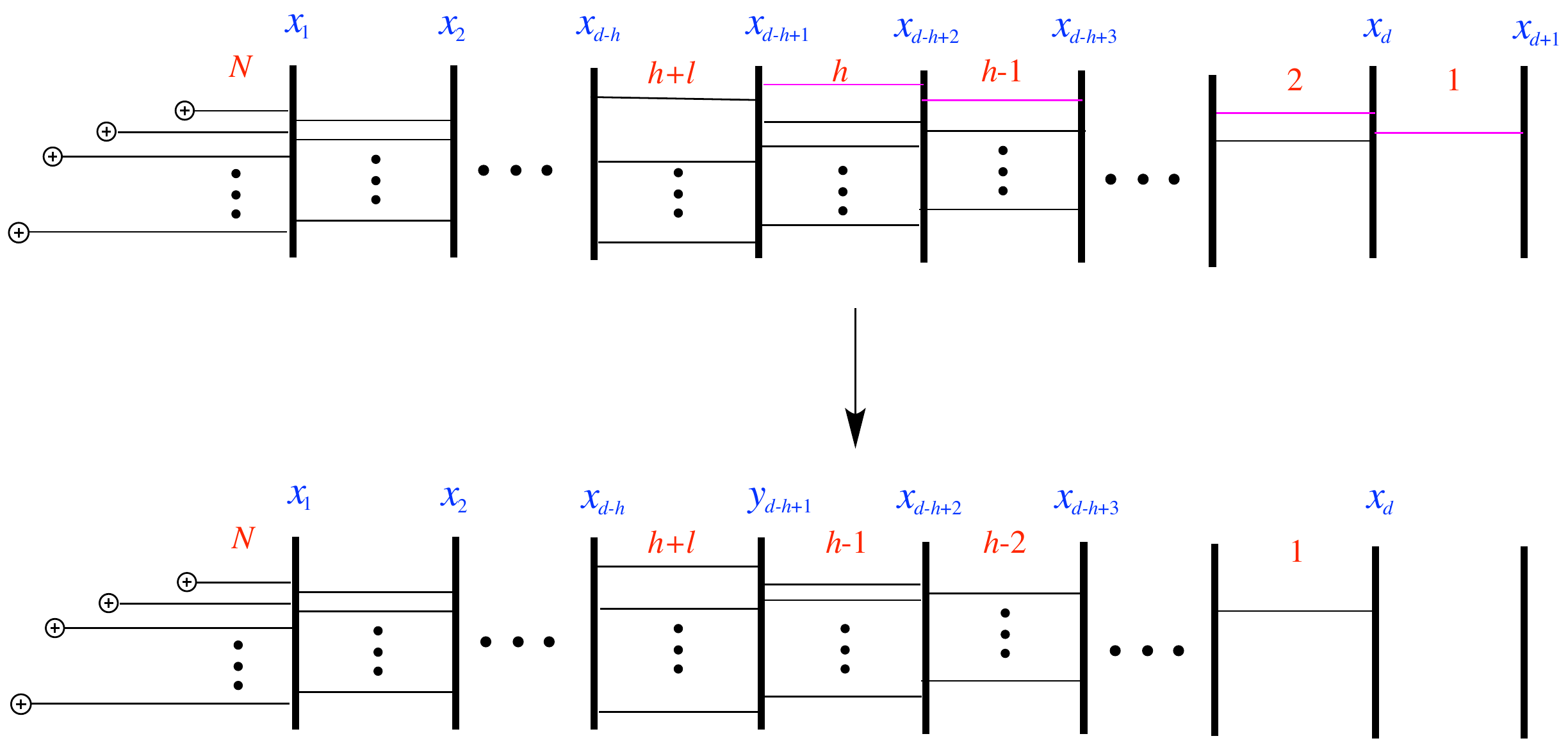}
\caption{The top and bottom brane configurations correspond to the partitions $\vec \rho$ to $\vec \rho^\prime$, respectively. The notation is as indicated in \fref{fig:TrhoG}. Here $\ell = \rho_{d-h+1}$. } 
\label{fig:partition}
\end{center}
\end{figure}

The corresponding brane configurations are depicted in \fref{fig:partition}.

Analogously to the analysis of the HL limit of superconformal indices of 4d Sicilian theories in \cite{Gaiotto:2012uq}, we find that the HL formula \eref{HLformulaT} for the Coulomb branch Hilbert series of $T_{\vec \rho}(SU(N))$ has a simple pole at $x_{d-h+1} x_{d+1}^{-1} = t^{\frac{1}{2}(1+\ell)}$. We parametrize 
\be\label{pole_gen}
x_{d-h+1}=y_{d-h+1} t^{\frac{1}{2}}z~, \qquad x_{d+1}=y_{d-h+1} (t^{\frac{1}{2}}z)^{-\ell}~,
\ee
in agreement with \eref{constrmove}, and compute the residue at the pole $z\to 1$.
The residue of the prefactor \eref{KUN} is 
\be
\begin{split}
& \underset{z \rightarrow 1}{\rm Res} ~ K_{{\vec \rho}} (x_1, \ldots, x_{d-h},y_{d-h+1} t^{\frac{1}{2}}z,x_{d-h+2}, \ldots, x_{d},y_{d-h+1} (t^{\frac{1}{2}}z)^{-\ell};t) \\
&= {\cal P}_{\vec \rho \vec \rho^\prime} (\vec x; y_{d-h+1};t) K_{{\vec \rho}^\prime} (x_1, \ldots, x_{d-h},y_{d-h+1},x_{d-h+2}, \ldots, x_{d};t) ~,
\end{split}
\ee
where
\bea \label{defPfac}
{\cal P}_{\vec \rho \vec \rho^\prime} (\vec x; y_{d-h+1};t)= \frac{1}{1+\ell} \PE \left[t+ t^{\frac{1}{2}} \sum_{i=1}^{h-1} \sum_{s=\pm 1}\left(t^{\frac{1}{2}(\ell-1)} \frac{x_{d-h+1+i}}{y_{d-h+1}}\right)^s  \right]~,
\eea
whereas the HL polynomial with arguments determined by the partition $\vec\rho$ in \eref{HLformulaT} tends to the HL polynomial with arguments determined by $\vec\rho^\prime$.
Therefore we find that
\be \label{finalgen}
\begin{split}
& \underset{z \rightarrow 1}{\rm Res} ~ H[T_{{\vec \rho}}(SU(N))] (t; x_1, \ldots, x_{d-h},y_{d-h+1} t^{\frac{1}{2}}z,x_{d-h+2}, \ldots, x_{d},y_{d-h+1} (t^{\frac{1}{2}}z)^{-\ell}; \vec n)  \\
&= {\cal P}_{\vec \rho \vec \rho^\prime} (\vec x; y_{d-h+1};t) H[T_{{\vec \rho}^\prime}(SU(N))] (t; x_1, \ldots, x_{d-h},y_{d-h+1},x_{d-h+2}, \ldots, x_{d}; \vec n)~.
\end{split}
\ee

This result can be reproduced from the monopole formula for the Coulomb branch Hilbert series along the lines of section \ref{sec:analyticSU3}. The simple pole corresponds to a noncompact flat direction in the Coulomb branch where the D3-branes depicted in purple in \fref{fig:partition} are sent to infinity. The identification of the simple pole and the computation of the residue is tedious but straightforward. We leave it to Appendix \ref{sec:analytgeneral}.

This agreement of analytic structures is a nontrivial consistency check of the conjectured equivalence of the monopole and the Hall-Littlewood formula for the Coulomb branch Hilbert series. It guarantees the equivalence of the two formulae for any $T_{\vec\rho}(SU(N))$ theories once this is established for $T(SU(N))$.


\section{Conclusions}\label{sec:conclusions}

This paper  is a first step towards the computation of the Coulomb branch Hilbert series for wide classes of $\cN=4$ gauge theories. The monopole formula introduced in \cite{Cremonesi:2013lqa} allows to determine the Hilbert series of  any good or ugly $\cN=4$ theory as an infinite sum over magnetic fluxes. Due to the presence of many sums and absolute values in the dimension formula for the monopoles, it is sometimes difficult to obtain explicit analytic expressions for the Hilbert series, especially when the gauge group becomes large. The gluing techniques that we have introduced in section \ref{sec:general} 
allows to solve this problem in many cases.  Knowing analytic formulae for the Coulomb branch Hilbert series in the presence of background fluxes for a general class of building blocks, we can determine the Hilbert series of more general theories by gluing. The mechanism requires a sum over the background monopole fluxes and involves no absolute values.

We have given a closed analytic expression for  the Coulomb branch Hilbert series with background fluxes for the class of theories $T_{\vec \rho}(G)$ which can serve as building blocks for constructing a wide classes of $\cN=4$ gauge theories. In particular, all the mirrors of Sicilian theories,  obtained from M5-branes compactified on $S^1$ times a Riemann sphere with punctures $\{ \vec \rho_i \}$,  can be obtained by gluing the corresponding  $T_{\vec \rho}(G)$ theories.  In the companion paper \cite{Cremonesi:2014vla} we will compute the Hilbert series for the Coulomb branch of the mirror of Sicilian theories of type A and D.   By mirror symmetry, this is equal to the Higgs branch Hilbert series of the Sicilian theory, which can  can be evaluated  by the Hall-Littlewood (HL) limit of the superconformal index \cite{Gadde:2011uv,Lemos:2012ph}  when the genus of the Riemann surface is zero.  We will find perfect agreement with the results in \cite{Gadde:2011uv,Lemos:2012ph}, which also involve Hall-Littlewood polynomials and were obtained in a completely different manner. The agreement with  \cite{Gadde:2011uv}  further substantiates our conjecture that the Hall-Littlewood formula computes the Coulomb branch Hilbert series of $T_{\rho}(G)$ theories.  In \cite{Cremonesi:2014vla} we will also compute the Coulomb branch Hilbert series  of mirrors of Sicilian theories with  genus greater than one, for which there is no other available method.

It would be interesting to extend our analysis to cover the more general class of theories $T_{\vec \rho}^{\vec \mu}(G)$ defined in  \cite{Gaiotto:2008ak}, where $\vec \mu$ and $\vec \rho$ are partitions related to $G$ and the dual $G^\vee$ respectively. The Coulomb and Higgs branches of these theories are not generally complete intersections. Computing their Coulomb branch Hilbert series with background fluxes would allow us to obtain the Hilbert series of an even wider class of $\cN=4$ gauge theories.


\section*{Acknowledgements}

We thank Francesco Benini, Nick Halmagyi, Yuji Tachikawa and Alessandro Tomasiello for useful discussions, and the following institutes and workshops for hospitality and partial support: the Galileo Galilei Institute for Theoretical Physics and INFN and the Geometry of Strings and Fields workshop (SC), the Simons Center for Geometry and Physics and the 2013 Summer Workshop, and Chulalongkorn University and the 3rd Bangkok Workshop on High Energy Theory (AH and NM), \'Ecole Polytechnique and the String Theory Groups of the universities of Rome ``Tor Vergata'' and of Oviedo (NM). NM is also grateful to Diego Rodr\'iguez-G\'omez, Yolanda Lozano, Raffaele Savelli, Jasmina Selmic, Hagen Triendl, Sarah Maupeu and Mario Pelliccioni for their very kind hospitality. We were partially supported by the STFC Consolidated Grant ST/J000353/1 (SC), the EPSRC programme grant EP/K034456/1 (AH), the ERC grant, Short Term Scientific Mission of COST Action MP1210, and World Premier International Research Center Initiative (WPI Initiative), MEXT, Japan (NM), and  INFN and the MIUR-FIRB grant RBFR10QS5J ``String Theory and Fundamental Interactions'' (AZ).

\appendix
\section{Notations and conventions}
Unless stated otherwise, the following notation and conventions are used throughout the paper.
\begin{itemize}
\item The plethystic exponential of a multivariate function $f(t_1, \ldots, t_n)$ that vanishes at the origin, $f(0,\ldots, 0)=0$, is defined as
\bea
\PE \left[ f(t_1, \ldots, t_n) \right] = \exp \left( \sum_{k=1}^\infty \frac{1}{k} f(t_1^k, \ldots, t_n^k) \right)~.
\eea 
\item An irreducible representation of a simple group $G$ can be denoted by its highest weight vector.
\begin{itemize}
\item With respect to a basis consisting of the fundamental weights (also known as the $\omega$-basis), we write the highest vector as $[a_1, \ldots, a_r]$ with $r = \mathrm{rank}~G$. This is the \emph{Dynkin label}. 
\item With respect to a basis of the dual Cartan subalgebra (also known as the $e$-basis or the standard basis), we denote the the highest vector by $(\lambda_1, \ldots, \lambda_r)$.  
\end{itemize}
\ei


\subsubsection*{The Weyl character formulae}

In the following, we present the character formulae that we use throughout the paper.  Our convention for characters is different from that of {\tt LiE}%
\footnote{\url{http://www-math.univ-poitiers.fr/~maavl/LiE/form.html}}. 
\bi
\item For $U(n)$, the Dynkin label $[a_1, a_2, \ldots, a_n]$ is related to $(\lambda_1,  \lambda_2, \ldots, \lambda_n)$ by the formula 
\bea
\lambda_i = a_i +\ldots+ a_n~.
\eea
The partition $\vec \lambda$ is subject to $\lambda_1 \geq \lambda_2 \geq \ldots \geq \lambda_n \geq 0$, with all $\lambda_n$ integers.
The character is given by the Schur polynomial
\bea
\chi^{U(n)}_{(\lambda_1,\ldots, \lambda_n)}(\vec y) = \frac{\det \left( y_j^{(\lambda_{i}+n-i)} \right)_{i,j=1}^n}{\det \left( y_j^{(n-i)} \right)_{i,j=1}^n}~.
\eea
For $A_{n-1}=SU(n)$, one simply restricts $\lambda_n=0$ and imposes $y_1 \cdots y_n =1$.

\item For $B_n =SO(2n+1)$, the Dynkin label $[a_1, a_2, \ldots, a_n]$ is related to $(\lambda_1,  \lambda_2, \ldots, \lambda_n)$ by the formula 
\bea
\lambda_i &= a_i + a_{i+1} + \ldots+ a_{n-1} + \frac{1}{2}a_n~, \quad 1\leq i \leq n-1~, \nn \\
\lambda_n &= \frac{1}{2} a_n~, \label{lambdaanda}
\eea
The partition $\vec \lambda$ is subject to $\lambda_1 \geq \lambda_2 \geq \ldots \geq \lambda_N \geq 0$ with all $\lambda_i$ integers or all half-integers. The character is given by 
\bea
\chi^{B_n}_{(\lambda_1,\ldots, \lambda_n)}(\vec y) = \frac{\det \left( y_j^{(\lambda_{i}+n-i+\frac{1}{2})}- y_j^{-(\lambda_{i}+n-i+\frac{1}{2})} \right)_{i,j=1}^n}{\det \left( y_j^{(n-i+\frac{1}{2})}- y_j^{-(n-i+\frac{1}{2})} \right)_{i,j=1}^n}~.
\eea
\item For $C_n =USp(2n)$, the Dynkin label $[a_1, a_2, \ldots, a_n]$ is related to $(\lambda_1,  \lambda_2, \ldots, \lambda_n)$ by the formula
\bea
\lambda_i = a_i +\ldots+ a_n~.
\eea
The partition $\vec \lambda$ is subject to $\lambda_1 \geq \lambda_2 \geq \ldots \geq \lambda_n \geq 0$, with all $\lambda_n$ integers.
 The character is given by 
\bea
\chi^{C_n}_{(\lambda_1,\ldots, \lambda_n)}(\vec y) = \frac{\det \left( y_j^{(\lambda_{i}+n-i+1)}- y_j^{-(\lambda_{i}+n-i+1)} \right)_{i,j=1}^n}{\det \left( y_j^{(n-i+1)}- y_j^{-(n-i+1)} \right)_{i,j=1}^n}~.
\eea

\item For $D_n= SO(2n)$, the Dynkin label $[a_1, a_2, \ldots, a_n]$ is related to $(\lambda_1,  \lambda_2, \ldots, \lambda_n)$ by the formula
\bea
\lambda_i &= a_i +\ldots a_{n-2} +\frac{1}{2} (a_{n-1} + a_n)~, \qquad 1\leq i \leq n-2 \nn \\
\lambda_{n-1} &=\frac{1}{2} (a_{n-1} + a_n)~, \qquad
\lambda_{n} =\frac{1}{2} (-a_{n-1} + a_n)~.
\eea
The partition $\vec \lambda$ is subject to $\lambda_1 \geq \lambda_2 \geq \ldots \geq |\lambda_n| \geq 0$. The character is given by 
\bea
\chi^{D_n}_{(\lambda_1,\ldots, \lambda_n)}(\vec y) = \frac{\det \left( y_j^{(\lambda_{i}+n-i)}- y_j^{-(\lambda_{i}+n-i)} \right)_{i,j=1}^n}{\det \left( y_j^{(n-i)}- y_j^{-(n-i)} \right)_{i,j=1}^n}~.
\eea
\end{itemize}


\section{Hall-Littlewood polynomials} \label{sec:HLpoly}

The Hall-Littlewood (HL) polynomial associated to a group $G$ and a representation $\vec \lambda$ is a polynomial labelled by the highest weight vector $\vec \lambda= \sum_{i=1}^r \lambda_i \vec{e}_i$, with $\{ \vec e_1, \ldots, \vec e_r\}$ the standard basis of the weight lattice and $r$ the rank of $G$, defined as (see \eg~ \cite{macdonald1998symmetric})%
\footnote{Note that, for $G=U(N)$, we define the Hall-Littlewood polynomial without a normalisation factor $\CN(t)$ in comparison with (5.17) of \cite{Gadde:2011uv}.  This coincides with $R_\lambda$ in Section III.1 of \cite{macdonald1998symmetric}.}
\bea
\Psi^{\vec \lambda}_{G} (x_1, \ldots, x_r; t) = \sum_{w \in W_G} {\vec x}^{w(\vec \lambda)}
\prod_{\vec \alpha \in \Delta_+(G)}  \frac{  1-t {\vec x}^{-w({\vec \alpha})} } {1- {\vec x}^{-w({\vec \alpha})} }~,
\eea
where $W_G$ denotes the Weyl group of $G$ and $\Delta_+(G)$ the set of positive roots of $G$.  Explicit details for classical groups $G$ can be listed as follows:
\bi
\item For $G=U(N)$, $W_G=S_N$ is a group of permutations of the elements in $\{\vec e_1,\ldots, \vec e_N\}$. The partition $\vec \lambda$ is subject to $\lambda_1 \geq \lambda_2 \geq \ldots \geq \lambda_N \geq 0$ with $\lambda_i \in \BZ$.  The representation $\vec \lambda$ corresponds to the Dynkin label 
\bea
\vec a = (\lambda_1 - \lambda_2, \lambda_2-\lambda_3, \ldots, \lambda_{N-1}-\lambda_{N}, \lambda_N)~.
\eea  
The corresponding HL polynomial is
\bea
\Psi^{\vec \lambda}_{U(N)} (x_1,\dots,x_N;t)=\sum_{\sigma \in S_N}
x_{\sigma(1)}^{\lambda_1} \dots x_{\sigma(N)}^{\lambda_N}
\prod_{1 \leq i<j \leq N}   \frac{  1-t x_{\sigma(i)}^{-1} x_{\sigma(j)} } {1-x_{\sigma(i)}^{-1} x_{\sigma(j)}}\,,
\eea
where the positive roots are
\bea
\Delta_+ (U(N))= \{ \vec e_i - \vec e_j \}_{1 \leq i< j \leq N}~.
\eea
For $G=A_{N-1} = SU(N)$, we set $x_1 \cdots x_N=1$ and restrict $\lambda_N=0$.
\item For $G=B_N = SO(2N+1)$, $W_G = S_N \rtimes \BZ_2^N$ under which ${\vec e}_i \rightarrow \pm {\vec e}_{\sigma(i)}$, with $\sigma \in S_N$.  The partition $\vec \lambda$ is subject to $\lambda_1 \geq \lambda_2 \geq \ldots \geq \lambda_N \geq 0$ with all $\lambda_i$ integers or all half-integers. 
The corresponding HL polynomial is
\bea
 \Psi^{\vec \lambda}_{B_N} (x_1,\dots,x_N;t) &= \sum_{s_1,\ldots, s_N= \pm 1} \; \sum_{\sigma \in S_N}
 \left( \prod_{i=1}^N x_{\sigma(i)}^{s_i \lambda_i} \frac{1-t x_{\sigma(i)}^{-s_i} } {1-x_{\sigma(i)}^{-s_i} } \right) \times \nn\\
& \qquad \left( \prod_{1 \leq i<j \leq N}   \frac{1-t x_{\sigma(i)}^{-s_i} x_{\sigma(j)}^{s_j} } {1-x_{\sigma(i)}^{-s_i} x_{\sigma(j)}^{s_j}} \cdot  \frac{1-t x_{\sigma(i)}^{-s_i} x_{\sigma(j)}^{-s_j} } {1-x_{\sigma(i)}^{-s_i} x_{\sigma(j)}^{-s_j} } \right)~,
\eea
where \bea
\Delta_+(B_N)= \{ \vec e_i + \vec e_j \}_{1 \leq i< j \leq N} \cup \{ \vec e_i - \vec e_j \}_{1 \leq i< j \leq N} \cup \{ \vec e_i\}_{1 \leq i\leq N}~.
\eea
\item For $G=C_N = USp(2N)$, $W_G = S_N \rtimes \BZ_2^N$ under which ${\vec e}_i \rightarrow \pm {\vec e}_{\sigma(i)}$, with $\sigma \in S_N$.  The partition $\vec \lambda$ is subject to $\lambda_1 \geq \lambda_2 \geq \ldots \geq \lambda_N \geq 0$, with all $\lambda_i$ integers. The corresponding HL polynomial is
\bea
 \Psi^{\vec \lambda}_{C_N} (x_1,\dots,x_N;t) &= \sum_{s_1,\ldots, s_N= \pm1} \; \sum_{\sigma \in S_N}
 \left( \prod_{i=1}^N x_{\sigma(i)}^{s_i \lambda_i} \frac{1-t x_{\sigma(i)}^{-2s_i} } {1-x_{\sigma(i)}^{-2s_i} } \right) \times \nn\\
& \qquad \left( \prod_{1 \leq i<j \leq N}   \frac{1-t x_{\sigma(i)}^{-s_i} x_{\sigma(j)}^{s_j} } {1-x_{\sigma(i)}^{-s_i} x_{\sigma(j)}^{s_j}} \cdot  \frac{1-t x_{\sigma(i)}^{-s_i} x_{\sigma(j)}^{-s_j} } {1-x_{\sigma(i)}^{-s_i} x_{\sigma(j)}^{-s_j} } \right)~,
\eea
where \bea
\Delta_+ (C_N)= \{ \vec e_i + \vec e_j \}_{1 \leq i< j \leq N} \cup \{ \vec e_i - \vec e_j \}_{1 \leq i< j \leq N} \cup \{ 2\vec e_i\}_{1 \leq i\leq N}~.
\eea
\item For $G=D_N = SO(2N)$, $W_G = S_N \rtimes \BZ_2^{N-1}$ under which ${\vec e}_i \rightarrow (-1)^{b_i} {\vec e}_{\sigma(i)}$, with $\sigma \in S_N$, $b_i = 0, 1$ and $\sum_{i=1}^N b_i$ is even.  The partition $\vec \lambda$ is subject to $\lambda_1 \geq \lambda_2 \geq \ldots \geq |\lambda_N| \geq 0$. The Dynkin label of the representation $\vec \lambda$ is $\vec a =(a_1, \ldots, a_N)$, with
\bea
\lambda_i &= a_i +\ldots a_{N-2} +\frac{1}{2} (a_{N-1} + a_N)~, \qquad 1\leq i \leq N-2 \nn \\
\lambda_{N-1} &=\frac{1}{2} (a_{N-1} + a_N)~, \qquad
\lambda_{N} =\frac{1}{2} (-a_{N-1} + a_N)~.
\eea
The corresponding HL polynomial is
\bea
 \Psi^{\vec \lambda}_{D_N} (x_1,\dots,x_N;t) &= \sum_{\substack{s_1,\ldots, s_N= \pm 1 \\ s_1 \ldots s_N =1}} \; \sum_{\sigma \in S_N}
 \left( \prod_{i=1}^N x_{\sigma(i)}^{s_i \lambda_i} \right) \times \nn\\
& \qquad \left( \prod_{1 \leq i<j \leq N}   \frac{1-t x_{\sigma(i)}^{-s_i} x_{\sigma(j)}^{s_j} } {1-x_{\sigma(i)}^{-s_i} x_{\sigma(j)}^{s_j}} \cdot  \frac{1-t x_{\sigma(i)}^{-s_i} x_{\sigma(j)}^{-s_j} } {1-x_{\sigma(i)}^{-s_i} x_{\sigma(j)}^{-s_j} } \right)~,
\eea
where \bea
\Delta_+ (D_N)= \{ \vec e_i + \vec e_j \}_{1 \leq i< j \leq N} \cup \{ \vec e_i - \vec e_j \}_{1 \leq i< j \leq N} ~.
\eea
\ei
We collect explicit expressions for HL polynomials for groups with small ranks below.
\bea
& \Psi^{n_1}_{U(1)} (x_1; t) = x_1^{n_1} \\
& \Psi^{(n_1,n_2)}_{U(2)} (x_1,x_2; t) = \frac{t x_1^{1+n_2} x_2^{n_1}-x_1^{n_2} x_2^{1+n_1}+x_1^{1+n_1} x_2^{n_2}-t x_1^{n_1} x_2^{1+n_2}}{x_1-x_2}\\
& \Psi^n_{SO(3)}(x;t) = \frac{x^{-n} \left[  -1+x^{1+2 n}+x t \left(1-x^{2 n-1}\right)\right]}{x-1} \\
& \Psi^n_{USp(2)}(x;t) = \frac{x^{-n} \left[  -1+x^{2+2 n}+x^2 t \left(1-x^{2 n-2}\right)\right]}{x^2-1} =  \Psi^{(n,0)}_{U(2)} (x,x^{-1}; t) \\
& \Psi^{(n_1,n_2)}_{SO(4)}(x_1,x_2;t) = \frac{1}{\left(x_1-x_2\right) \left(x_1 x_2-1\right)} \Big[-x_1^{1-n_2} x_2^{-n_1}-x_1^{1+n_2} x_2^{2+n_1}+x_1^{-n_1} x_2^{1-n_2}\nn \\
& \qquad +x_1^{2+n_1} x_2^{1+n_2} +t \Big\{ x_1^{2-n_2} x_2^{1-n_1}+x_1^{-n_2} x_2^{1-n_1}+x_1^{n_2} x_2^{1+n_1}+x_1^{2+n_2} x_2^{1+n_1} \nn \\
& \qquad -x_1^{1-n_1} x_2^{2-n_2}-x_1^{1-n_1} x_2^{-n_2}-x_1^{1+n_1} x_2^{n_2} \left(1+x_2^2\right) \Big \}+t^2 \Big\{ -x_1^{1+n_2} x_2^{n_1}+x_1^{n_1} x_2^{1+n_2} \nn \\
& \qquad +x_1 x_2 \Big(-x_1^{-n_2} x_2^{1-n_1}+x_1^{1-n_1} x_2^{-n_2}\Big)\Big \} \Big]~.
& 
\eea


\section{Examples of non-maximal $B$ and $C$ punctures}\label{sec:nonmaximal} 

In this appendix, we focus on non-trivial partitions $\vec \rho$ providing many examples for $SO$ and $USp$ groups.  

For groups with low ranks, there are certain isomorphism between their Lie algebras, \eg~ $SO(5)$ and $USp(4)$, $SO(6)$ and $SU(4)$.  We use such isomorphisms as a tool to check the Hall-Littlewood formula \eref{mainHL} in many examples; these checks are similar in spirit to what we performed in section \ref{sec:checkTSO6andTSU4}.  

Furthermore, in many examples below, we use the Hilbert series as a tool to understand the relationships between the nilpotent orbit and the Higgs/Coulomb branches of the theories.  Let us summarize some highlights below:
\bi
\item In appendix \ref{sec:veryeven44} we consider the ``very even'' partition $\vec \rho=(4,4)$ of $SO(8)$, which corresponds to two distinct nilpotent orbits of $SO(8)$ \cite{collingwood1993nilpotent, Chacaltana:2011ze, Chacaltana:2012zy}.  We study how the Coulomb branches of the theories corresponding to these two orbits are related. 
\item In appendix \ref{sec:3311VS3221}, we study two distinct partitions of $SO(8)$, namely $\vec \rho_1=(3,3,1,1)$ and $\vec \rho_2=(3,2,2,1)$.  Although these two partitions are different, their images under the Spaltenstein map are identical \cite{Chacaltana:2012zy}.  Physically, the latter describe the Higgs branches of $T_{\vec \rho_1}(SO(8))$ and $T_{\vec \rho_2}(SO(8))$ as the moduli spaces of the same Hitchin system \cite{Chacaltana:2011ze, Chacaltana:2012zy}.  In particular, we show that the Higgs branch Hilbert series of two theories are indeed equal.
\ei

\subsection{$T_{(3,1,1,1)}(SO(6))$}

The quiver for $T_{(3,1,1,1)}(SO(6))$ is
\bea \label{T3111}
T_{(3,1,1,1)}(SO(6)): \qquad [SO(6)]-(USp(2))-(O(2))~.
\eea
The mirror of this theory is given below:
\be  \label{mirT3111}
\begin{tikzpicture}[font=\scriptsize]
\begin{scope}[auto,%
  every node/.style={draw, minimum size=1cm}, node distance=1cm];
\node[circle] (SO2) at (0, 0) {$SO(2)$};
\node[circle, right=of SO2] (USp2) {$USp(2)$};
\node[circle, right=of USp2] (SO3)  {$SO(3)$};
\node[circle, right=of SO3] (USp2p)  {$USp(2)$};
\node[rectangle, below=of USp2] (SO1)  {$SO(1)$};
\node[rectangle, below=of USp2p] (SO3f)  {$SO(3)$};
\end{scope}
\draw (SO2) -- (USp2)
(USp2) -- (SO3)
(USp2) -- (SO1)
(SO3) -- (USp2p)
(USp2p) -- (SO3f);
\end{tikzpicture}
\ee
The mirror is obtained using the brane configuration as discussed in \cite{Feng:2000eq}.

\subsubsection*{Hall-Littlewood formula}

The Coulomb branch Hilbert series of $T_{3,1,1,1} (SO(6))$ is given by
\be \label{HT3111SO6}
\begin{split}
&H[T_{(3,1,1,1)} (SO(6))] (t; x; n_1,n_2,n_3)  \\
& = t^{2n_1+n_2} (1-t)^3 K^{SO(6)}_{(3,1,1,1)}(x, t) \Psi^{(n_1,n_2,n_3)}_D (t, 1, x^2;t)~,
\end{split}
\ee
where $x$ is the fugacity associated with the topological charge of $SO(2)$ gauge group in \eref{T3111} and
\bea
K^{SO(6)}_{(3,1,1,1)}(x_1, t) = \PE \left[t( \chi^{SO(3)}_{[1]}(x)) +t^2 ( \chi^{SO(3)}_{[1]}(x)+1)   \right]~,
\eea
with
\bea
 \chi^{SO(3)}_{[1]}(x) = x^2+1+x^{-2}~.
\eea
The argument of the HL polynomial in (\ref{HT3111SO6}) and the factor $K^{SO(6)}_{(3,1,1,1)}$ follow from  the decompositions
of the fundamental and adjoint representations of $SO(6)$
\be
\begin{split}
\chi^{SO(6)}_{[1,0,0]}(\vec a) & = \sum_{i=1}^3 \left(a_i + a_i^{-1}\right) = (t+\frac{1}{t} + x^2 +\frac{1}{x^2} +2)   =   \chi^{SU(2)}_{[2]}(t^{\frac{1}{2}}) + \chi^{SO(3)}_{[1]}(x) ~,  \\
\chi^{SO(6)}_{[0,1,1]} (\vec a) &= \chi^{SO(3)}_{[1]}(x) \chi^{SU(2)}_{[2]}(t^{\frac{1}{2}}) + \chi^{SU(2)}_{[2]}(t^{\frac{1}{2}}) + \chi^{SO(3)}_{[1]}(x)  ~. 
\end{split}
\ee
As predicted in  (\ref{sympunc}), the Coulomb branch symmetry is enhanced to $SO(3)$. 

For $n_1=n_2=0$, we obtain the Hilbert series
\be \label{T3111HS000}
\begin{split}
&H[T_{(3,1,1,1)} (SO(6))] (t; x; 0,0,0) \\
&= (1-t^3)(1-t^4)\PE \left[ t \chi^{SO(3)}_{[1]}(x) + t^2 \chi^{SO(3)}_{[1]}(x) \right]~,
\end{split}
\ee
which agrees with the Higgs branch Hilbert series of the mirror quiver \eref{mirT3111}.

\subsubsection*{Comparison with $T_{(2,2)} (SU(4))$}

We observe that the Hilbert series \eref{T3111HS000} is equal to the Coulomb branch Hilbert series of $T_{(2,2)} (SU(4))$ theory, given by (4.7) of \cite{Hanany:2011db}.  
As a check of formula \eref{HT3111SO6}, we compare this to the HL formula for $T_{(2,2)}(SU(4))$, given by \eref{HLformulaT}, with the background charges $n_1,n_2,n_3$ turned on:
\be
\begin{split}
&H[T_{(2,2)}(SU(4))] (t, x_1,x_2; n_1,n_2,n_3,0) \\
&=  t^{\frac{1}{2}(3n_1+n_2-n_3)}  (1-t)^4  K^{U(4)}_{(2,2)}(t; \vec x) \Psi^{(n_1,n_2,n_3,0)}_{U(4)} (\vec xt^{\frac{1}{2} w_{(2,2)}} ; t)~,
\end{split}
\ee
where
\bea
\vec x t^{\frac{1}{2} w_{(2,2)}} &= (x_1 t^{1/2}, x_1 t^{-1/2}, x_2 t^{1/2}, x_2 t^{-1/2}) \\
K^{U(4)}_{(2,2)}(t; \vec x) &= \PE \left[ (t+t^2) ( 2+x_1 x_2^{-1}+ x_1^{-1} x_2)  \right]~.
\eea
Note that the prefactor $K^{U(4)}_{(2,2)}(t; \vec x)$ corresponds to the following decomposition of \eref{decompadj}:
\be
\begin{split}
\chi^{U(4)}_{\Adj} (\vec a) &= \sum_{i,j =1}^4 {a_i}a_j^{-1}  \\
&= ( 2+x_1 x_2^{-1}+ x_1^{-1} x_2) \left[ \chi^{SU(2)}_{[0]}(t^{1/2})+ \chi^{SU(2)}_{[2]}(t^{1/2}) \right]~, \quad \vec a = \vec x  t^{\frac{1}{2} w_{(2,2)}}~.
\end{split}
\ee

Indeed, we find that for any $a_1, a_2, a_3 \geq 0$,
\bea
H[T_{(3,1,1,1)} (SO(6))] \left(t;x; \vec m (\vec a) \right) = H[T_{2,2}(SU(4))](t; x, x^{-1}; \vec n(\vec a) ) ~,
\eea
with 
\be
\begin{split}
\vec m (\vec a) &= \left(\frac{1}{2}a_1+a_2+ \frac{1}{2}a_3, \; \frac{1}{2}a_1+\frac{1}{2}a_3, \; -\frac{1}{2}a_1+\frac{1}{2}a_3 \right)~, \qquad  \\
\vec n (\vec a) &= (a_1+a_2+a_3, \; a_2+a_3, \; a_3, \;0)~.
\end{split}
\ee

\subsection{$T_{(4,4)}(SO(8))$ and the very even partition $(4,4)$} \label{sec:veryeven44}

In this appendix, we consider the partition $(4,4)$ of $SO(8)$.  This partition is ``very even'', therefore it corresponds to two different nilpotent orbits of $SO(8)$ (see, \eg~ Recipe 5.2.6 of \cite{collingwood1993nilpotent} and \cite{Chacaltana:2011ze, Chacaltana:2012zy}).  These two types of puncture $(4,4)$ are related by an outer automorphism of $SO(8)$ that interchanges the two spinor representations $[0,0,0,1]$ and $[0,0,1,0]$; we distinguish these punctures by subscripts $I$ and $II$.  Even though the distinction of these two types is not apparent in the quiver diagram, the Hall-Littlewood formulae for these two partitions are not equal, even though they can be related.

\subsubsection*{Quiver and mirror theory}

For both types of the $(4,4)$ puncture, the quiver diagram of $T_{(4,4)}(SO(8))$ is given by $USp(4)$ gauge theory with 4 flavors, namely
\bea
T_{(4,4)}(SO(8)): \qquad (USp(4))-[SO(8)]~. \label{quiv:T44D4}
\eea
This theory is a `bad theory', since the number of flavors is 4, less than $2(2)+1 =5$.  The Coulomb branch Hilbert series cannot be computed from the monopole formula, but we expect that the Hall-Littlewood formula gives the correct result. 

The mirror theory of this quiver can be determined using brane configurations as in Figure 13 of \cite{Feng:2000eq}; the quiver diagram is given by
\bea  \label{mirT44p}
\begin{tikzpicture}[font=\scriptsize]
\begin{scope}[auto,%
  every node/.style={draw, minimum size=1cm}, node distance=1cm];
\node[circle] (SO2) at (0, 0) {$SO(2)$};
\node[circle, right=of SO2] (USp2) {$USp(2)$};
\node[circle, right=of USp2] (SO4)  {$SO(4)$};
\node[circle, right=of SO4] (USp2p)  {$USp(2)$};
\node[circle, right=of USp2p] (SO2p)  {$SO(2)$};
\node[rectangle, below=of SO4] (USp2f)  {$USp(2)$};
\end{scope}
\draw (SO2) -- (USp2)
(USp2) -- (SO4)
(SO4) -- (USp2p)
(USp2p) -- (SO2p)
(SO4) -- (USp2f);
\end{tikzpicture}
\eea

\subsubsection*{The Coulomb branch Hilbert series}

The global symmetry group of the $(4,4)$ puncture is $USp(2)$.  Two types of punctures corresponds to different embeddings of $USp(2)$ in $SO(8)$.  From \eref{decompfund}, we consider the following decomposition:
\bea
\chi^{SO(8)}_{[1,0,0,0]}(\vec a) = \sum_{i=1}^4 \left(a_i + a_i^{-1}\right) = (t^{3/2}+t^{1/2}+t^{-1/2}+t^{-3/2}) (x+x^{-1})~.
\eea
There are two inequivalent choices of the fugacity maps corresponding to the two types of (4,4) puncture:%
\footnote{In this paper we take $a_3 = x^{-1} t^{1/2}$, differently from Fig. 8 of \cite{Lemos:2012ph}.  Our choice is to be consistent with $K_{(4,4)}^{SO(8)}(t; x)$ given in \eref{K44SO8}.}
\be \label{embedding44}
\begin{split}
(I): &\qquad a_1= x t^{3/2}, \quad a_2 = x t^{1/2}, \quad a_3 = x^{-1} t^{1/2}, \quad a_4 = x t^{-3/2}~,  \\
(II): & \qquad a_1= x t^{3/2}, \quad a_2 = x t^{1/2}, \quad a_3 = x^{-1} t^{1/2}, \quad a_4 = x^{-1} t^{3/2}~.
\end{split}
\ee
For the two types of (4,4) puncture, the HL formula for the Coulomb branch Hilbert series of \eref{quiv:T44D4} is given by
\begin{align}
\begin{split}
H[T_{(4,4)_I}(SO(8))] (t; x; \vec n) &= t^{3 n_1+2 n_2 +n_3} (1-t)^4 K_{(4,4)}^{SO(8)} (t; x) \times \\
& \qquad \Psi^{(n_1,n_2,n_3,n_4)} _{D} (t^{3/2} x, t^{1/2} x, t^{1/2} x, t^{-3/2} x; t)~, 
\end{split} \\
\begin{split}
H[T_{(4,4)_{II}}(SO(8))] (t; x; \vec n) &= t^{3 n_1+2 n_2 +n_3} (1-t)^4 K_{(4,4)}^{SO(8)} (t; x) \times \\
& \qquad \Psi^{(n_1,n_2,n_3,n_4)} _{D} (t^{3/2} x, t^{1/2} x, t^{1/2} x, t^{3/2} x^{-1}; t)~,
\end{split}
\end{align}
where
\bea \label{K44SO8}
K_{(4,4)}^{SO(8)}(t,x) 
= \PE \left[ t \chi_{[2]}(x) + t^2 +t^3 \chi_{[2]}(x) +t^4 \right]~.
\eea
Note that \eref{K44SO8} is consistent with \eref{embedding44}, namely
\be
\begin{split}
\chi^{SO(8)}_{[0,1,0,0]} (\vec a) &= \chi^{USp(2)}_{[2]}(x) \chi^{SU(2)}_{[0]}(t^{\frac{1}{2}}) + \chi^{SU(2)}_{[2]}(t^{\frac{1}{2}}) + \chi^{USp(2)}_{[2]}(x) \chi^{SU(2)}_{[4]}(t^{\frac{1}{2}})   \\
& \qquad + \chi^{SU(2)}_{[6]}(t^{\frac{1}{2}})~. 
\end{split}
\ee

For $\vec n =(0,0,0,0)$, we obtain the Coulomb branch Hilbert series of quiver \eref{quiv:T44D4} and the Higgs branch Hilbert series of quiver \eref{mirT44p}:
\be
\begin{split}
& H[T_{(4,4)_I}(SO(8))] (t; x; 0,0,0,0) = H[T_{(4,4)_{II}}(SO(8))] (t; x; 0,0,0,0) \\
&= \PE \left[ t^2 \chi_{[2]} (x) + t^3 \chi_{[2]} (x) - t^4 - t^6\right]~.
\end{split}
\ee
Note that the space is 4 complex dimensional as required.

For $SO(8)$, the vector representation $[1,0,0,0]$ corresponds to $\vec n=(1,0,0,0)$, and the two spinor representations $[0,0,1,0]$ and $[0,0,0,1]$ correspond to $\vec n = \frac{1}{2} (1,1,1,-1)$ and $\frac{1}{2}(1,1,1,1)$. 
The outer automorphism that relates the two spinor representations exchanges $(4,4)$ punctures of types I and II.  In general we find that
\bea
H[T_{(4,4)_I}(SO(8))] \left(t; x; n_1,n_2,n_3, n_4\right) = H[T_{(4,4)_{II}}(SO(8))] \left(t; x; n_1,n_2,n_3, -n_4 \right)~.
\eea

\bigskip

\subsection{$T_{(3,3,1,1)}(SO(8))$ and $T_{(3,2,2,1)}(SO(8))$} \label{sec:3311VS3221}

In this appendix we consider $T_{(3,3,1,1)}(SO(8))$ and $T_{(3,2,2,1)}(SO(8))$ theories, which are mirrors to the second and third example on pp. 24-25 of \cite{Chacaltana:2012zy}. Their quiver diagrams are given below:
\bea
T_{(3,3,1,1)}(SO(8)): & \qquad (O(2))-(USp(4))-[SO(8)]~, \\
T_{(3,2,2,1)}(SO(8)): & \qquad (SO(4))-(USp(4))-[SO(8)] ~.
\eea
Note that the quiver for $T_{(3,3,1,1)}(SO(8))$ is a `good theory', whereas the quiver for $T_{(3,2,2,1)}(SO(8))$ is a `bad theory' since the number of flavors under the $SO(4)$ gauge group is $2$, smaller than $4-1=3$.

The partitions $\vec \rho= (3,3,1,1)$ and $\vec \rho=(3,2,2,1)$ define the Nahm poles of the above theories.  The Hitchin pole of each theory can be obtained by a procedure consisting of a transposition and a series of D-collapses \cite{Chacaltana:2011ze, Chacaltana:2012zy}, whereby the box in the bottom row of the left most column is moved to the next right column.  The D-collapsing is to be repeated until the number of columns that contain even boxes is even.%
\footnote{This condition corresponds to the decomposition of the $2N$ dimensional representation into irreducible representations of $SU(2)$ with dimensions $n_i$: $2N \rightarrow n_1+ n_2 + \ldots +n_k$.  In this decomposition, each even $n_i$ appeas even times.}  The resulting D-partition $\tilde{\vec \rho}$ defines the Hitchin pole of the theory. 

Below we show step-by-step the procedure to obtain the Hitchin poles for both theories.  Let us start from $\vec \rho= (3,3,1,1)$.
\bea
\ytableausetup
{boxsize=0.7em}
\ytableausetup
{aligntableaux=bottom}
\begin{array}{cccc}
\ydiagram{4,2,2}  &  \qquad {\begin{ytableau} ~& ~& ~\\ ~& ~& ~\\ ~\\  *(blue!20)  \end{ytableau}} & \qquad {\begin{ytableau} ~& ~& ~\\ ~& ~& *(blue!20)~\\ ~ & ~ \end{ytableau}}  & \qquad  \ydiagram{4,2,2}\\
 {\vec \rho}=(3,3,1,1) \quad &\xrightarrow{\text{transp.}} \quad (4,2,2) \quad &\xrightarrow{\text{D-coll.}} \quad (3,3,2) \quad &\xrightarrow{\text{D-coll.}}  \quad \tilde{\vec {\rho}} = (3,3,1,1) \\
\end{array}
\eea
where in each D-collapse the blue box is moved to the next column, according to the rule given in \cite{Chacaltana:2011ze}.  Similarly for $\vec \rho= (3,2,2,1)$ we have
\bea
\ytableausetup
{boxsize=0.7em}
\ytableausetup
{aligntableaux=bottom}
\begin{array}{cccc}
\ydiagram{4,3,1}  &  \qquad {\begin{ytableau} ~& ~& ~\\ ~& ~\\~& ~\\  *(blue!20)  \end{ytableau}} & \qquad {\begin{ytableau} ~& ~& ~\\ ~& ~& *(blue!20)~\\ ~ & ~ \end{ytableau}}  & \qquad  \ydiagram{4,2,2}\\
 {\vec \rho}=(3,2,2,1) \quad &\xrightarrow{\text{transp.}} \quad (4,3,1) \quad &\xrightarrow{\text{D-coll.}} \quad (3,3,2) \quad &\xrightarrow{\text{D-coll.}}  \quad \tilde{\vec {\rho}} = (3,3,1,1) \\
\end{array}
\eea

As a result, the Hitchin poles of the two theories are identical, namely $\tilde{\vec \rho} =(3,3,1,1)$.  In other words, the Higgs branches of $T_{(3,3,1,1)}(SO(8))$ and $T_{(3,2,2,1)}(SO(8))$ correspond to the moduli space of the same Hitchin system.%
\footnote{We thank Yuji Tachikawa for pointing this out to us.}  
Indeed it can be shown, using Hilbert series, that the hypermultiplet moduli spaces of the quivers 
\bea
(O(2))-[USp(4)]~,  \qquad \qquad (SO(4))-[USp(4)]~ \label{O2USp4VSSO4USp4}
\eea
are identical.
We give details of the computation in Appendix \ref{app:3311_3221}.
Thus, upon gluing with $[USp(4)]-[SO(8)]$ via the $USp(4)$ group, we obtain the same Higgs branch Hilbert series for both $T_{(3,3,1,1)}(SO(8))$ and $T_{(3,2,2,1)}(SO(8))$.  

Let us now turn to the Coulomb branch Hilbert series of $T_{(3,3,1,1)}(SO(8))$ and $T_{(3,2,2,1)}(SO(8))$.  The Coulomb branches of these theories are different; the former is 3 quaternionic dimensional, whereas the latter is 4 quaternionic dimensional. The $G_{\vec\rho}$ global symmetries associated with the punctures ${(3,3,1,1)}$ and ${(3,2,2,1)}$ are $SO(2)\times SO(2)$ and $USp(2)$ respectively.  The HL formulae for the Hilbert series are given by
\begin{align} \label{H3311}
\begin{split}
H[T_{(3,3,1,1)}(SO(8))] (t; x_1, x_2; \vec n) &= t^{3 n_1+2 n_2 +n_3} (1-t)^4 K_{(3,3,1,1)}^{SO(8)} (t; x) \times \\
& \qquad \Psi^{(n_1,n_2,n_3,n_4)} _{D} ( x_1 t^{-1}, x_1 t, x_1, x_2; t)~, 
\end{split}\\
\begin{split}
H[T_{(3,2,2,1)}(SO(8))] (t; x; \vec n) &=  t^{3 n_1+2 n_2 +n_3} (1-t)^4 K_{(3,2,2,1)}^{SO(8)} (t; x) \times \\
& \qquad \Psi^{(n_1,n_2,n_3,n_4)} _{D} ( t^{1/2} x, t^{-1/2} x,t ,1; t)~,
\end{split}
\end{align}
where the notations are explained below:
\bi
\item $(x_1, x_2)$ are fugacities for $SO(2) \times SO(2)$ and $x$ is a fugacity for $USp(2)$. They are related to the following embeddings.  For punctures ${(3,3,1,1)}$ and ${(3,2,2,1)}$, the decompositions \eref{decompfund} are respectively
\bea
\chi^{SO(8)}_{[1,0,0,0]} (\vec a) &=\sum_{i=1}^4 \left(a_i+a_i^{-1}\right) = \chi^{SU(2)}_{[2]}(t^{1/2}) (x_1+x_1^{-1}) + (x_2+x_2^{-1})~, \\
\chi^{SO(8)}_{[1,0,0,0]} (\vec b) &= \sum_{i=1}^4 \left(b_i+b_i^{-1}\right) = \chi^{SU(2)}_{[2]}(t^{1/2}) + \chi^{SU(2)}_{[1]}(t^{1/2}) (x+x^{-1}) +1~.
\eea
We pick 
\bea
\vec a= ( x_1 t^{-1}, x_1 t, x_1, x_2)~, \qquad \vec b= ( t^{1/2} x, t^{-1/2} x,t ,1);
\eea
these are the argument of the above Hall-Littlewood polynomials.
\item The prefactor $K_{(3,3,1,1)}^{SO(8)}$ is given by 
\bea
K_{(3,3,1,1)}^{SO(8)} (t; x_1,x_2) &= \PE \left[ 2 t + \{x_1^2+1+x_1^{-2}+(x_1+x_1^{-1})(x_2+x_2^{-1}) \} t^2 +t^3\right]~;
\eea
this corresponds to the following decomposition in \eref{decompadj}:
\be \label{decomp3311}
\begin{split}
\chi^{SO(8)}_{[0,1,0,0]} (\vec a) &= 2 +\{x_1^2+1+x_1^{-2}+(x_1+x_1^{-1})(x_2+x_2^{-1}) \} \chi^{SU(2)}_{[2]}(t^{1/2})   \\
& \qquad + \chi^{SU(2)}_{[4]}(t^{1/2})~.
\end{split}
\ee
\item The prefactor $ K_{(3,2,2,1)}^{SO(8)}$ is given by
\bea 
 K_{(3,2,2,1)}^{SO(8)} (t; x) &= \PE \left[ \chi^{USp(2)}_{[2]}(x)t + 2\chi^{USp(2)}_{[1]}(x) t^{3/2} + 3t^2 +\chi^{USp(2)}_{[1]}(x)t^{5/2} \right]~;
\eea
this corresponds to the following decomposition in \eref{decompadj}:
\be \label{decomp3221}
\begin{split}
\chi^{SO(8)}_{[0,1,0,0]} (\vec b) &= \chi^{USp(2)}_{[2]}(x) +2\chi^{USp(2)}_{[1]}(x) \chi^{SU(2)}_{[1]}(t^{1/2}) +3\chi^{SU(2)}_{[2]}(t^{1/2})  \\
& \qquad +\chi^{USp(2)}_{[1]}(x) \chi^{SU(2)}_{[3]}(t^{1/2})~.
\end{split}
\ee
\ei

For $\vec n=(0,0,0,0)$, we have
\bea
&H[T_{(3,3,1,1)}(SO(8))] (t; x_1, x_2; \vec 0) \nn \\
& \qquad = \PE \left[2 t +\{x_1^2+x_1^{-2}+(x_1+x_1^{-1})(x_2+x_2^{-1}) \} t^2 +t^3 - 2t^4 -t^6\right]~, \\
&H[T_{(3,2,2,1)}(SO(8))] (t; x; \vec 0)  \nn \\
& \qquad = \PE \left[\chi^{USp(2)}_{[2]}(x) t +2 \chi^{USp(2)}_{[1]}(x) t^{3/2} +2 t^2+ \chi^{USp(2)}_{[1]}(x) t^{5/2} - 2t^4 -t^6 \right]~.
\eea
These formulae show that the Coulomb branches are complete intersections of complex dimension $6$ and $8$ respectively.

Let us compare the results with the prediction of the monopole formula \eref{Hilbert_series}.  Since $T_{(3,2,2,1)}(SO(8))$ is a bad theory, the monopole formula diverges in this case.  For $T_{(3,3,1,1)}(SO(8))$, only one fugacity corresponding to topological charge of the gauge group $SO(2)$ can be made manifest; this corresponds to the fugacity $x_2$ in \eref{H3311}. We precisely reproduce   \eref{H3311} with $x_1$ set to unity.

\subsection{$T_{(3,1,1)}(USp(4))$ and $T_{(2,2)}(SO(5))$ } \label{sec:T311Usp4}

The quiver diagrams for $T_{(3,1,1)}(USp(4))$ and $T_{(2,2)}(SO(5))$ are
\bea
T_{(3,1,1)}(USp(4)): &\qquad [USp(4)]-(O(2)) \\
T_{(2,2)}(SO(5)): &\qquad [SO(5)]-(USp(2))-(O(1))~.
\eea
In this appendix we show that the Hilbert series of the Coulomb branch of these quivers, which are both 1 quaternionic dimensional in agreement with (2.3) of \cite{Chacaltana:2012zy}, are equal.

Note that $(3,1,1)$ is a $B$-partition for $SO(5)$, and so the global symmetry $G_{\vec\rho}$ associated with this puncture is $SO(2)$, according to \eref{sympunc}.  Similarly, $(2,2)$ is a $C$-partition for $USp(4)$, and the corresponding symmetry is therefore also $SO(2)$.

Using formula \eref{mainHL}, we obtain for $T_{(2,2)}(USp(4))$
\bea \label{T311USp4}
H[T_{(3,1,1)}(USp(4))]  (t; x; n_1, n_2)= t^{2n_1+n_2} (1-t)^2 K^{SO(5)}_{(3,1,1)} (x; t)  \Psi^{n_1,n_2}_{SO(5)} (t, x; t)~,
\eea
where the argument $(t, x)$ of the HL polynomial comes from the decomposition \eref{decompfund}
\be
\begin{split}
\chi^{SO(5)}_{[1,0]}(\vec a) &=1+a_1+a_2+a_1^{-1}+a_2^{-1} \\
&= (t+1+t^{-1})+(x+x^{-1})~, \qquad \vec a = (t, x),
\end{split}
\ee
with $x$ a fugacity of $SO(2)$, and the prefactor corresponding to the decomposition \eref{decompadj} is
\bea
 K^{SO(5)}_{(3,1,1)} (x; t) =  \PE \left[ t + t^2(1+x+x^{-1})\right]~.
\eea

Similarly, using formula \eref{mainHL} we obtain for $T_{(2,2,1)}(SO(5))$,
\be
\begin{split} \label{T22SO5}
&H[T_{(2,2)}(SO(5))]  (t; x; n_1, n_2) \\
& = t^{\frac{1}{2}(3n_1+n_2)} (1-t)^2 K^{C_2}_{(2,2)} (x; t)  \Psi^{n_1,n_2}_{C_2} (t^{1/2} x, t^{1/2} x^{-1}; t)~,
\end{split}
\ee
where the argument $(t^{1/2} x, t^{1/2} x^{-1})$ of the HL polynomial comes from the decomposition \eref{decompfund}:
\be
\begin{split}
\chi^{USp(4)}_{[1,0]}(\vec a) &=a_1+a_2+a_1^{-1}+a_2^{-1}  \\
&= (t^{1/2}+t^{-1/2})(x+x^{-1})~, \qquad \vec a = (t^{1/2} x, t^{1/2} x^{-1}),
\end{split}
\ee
and the prefactor corresponding to the decomposition \eref{decompadj} is
\bea
 K^{C_2}_{(2,2)} (x; t) =  \PE \left[ t + t^2(1+x^2+x^{-2})\right]~.
\eea

The two Hilbert series can be equated as follows:
\bea
H[T_{(3,1,1)}(USp(4))]  (t; x^2; n_1, n_2) = H[T_{(2,2)}(SO(5))]  (t; x; n_1+n_2, n_1-n_2)~.
\eea

For reference we present the result for $n_1=n_2=0$,
\bea
H[T_{(3,1,1)}(USp(4))]  (t; x; \vec 0) = 
\PE \left[ t + t^2 (x+x^{-1}) -t^4\right]~.
\eea
This is the Hilbert series for $\BC^2/\widehat{D}_3 = \BC^2/\BZ_4$, as expected for the Coulomb branch of $U(1)$ gauge theory with $4$ flavors.

Let us compare the results with the prediction of the monopole formula \eref{Hilbert_series}.  For $T_{(3,1,1)}(USp(4))$, the monopole formula gives the same answer as \eref{T311USp4}. However, for $T_{(2,2)}(SO(5))$, it is not possible to refine the Hilbert series with respect to a topological charge, since there is no factor of $U(1)$ in the quiver diagram;  the unrefined monopole formula gives the same Hilbert series as \eref{T22SO5}, with $x=1$.

\subsection{$T_{(2,2,1)}(USp(4))$ and $T_{(2,1,1)}(SO(5))$}

The quiver diagrams for $T_{(2,2,1)}(USp(4))$ and $T_{(2,1,1)}(SO(5))$ are
\bea
T_{(2,2,1)}(USp(4)): &\qquad [USp(4)]-(O(4)) \\
T_{(2,1,1)}(SO(5)): &\qquad [SO(5)]-(USp(2))-(O(3))~.
\eea
Note that both are `bad' theories in the sense of \cite{Gaiotto:2008ak}, since in $T_{(2,2,1)}(USp(4))$ the global symmetry $USp(4)$ that connects to gauge group $O(4)$ has rank 2, which is smaller than $4-1=3$, and in $T_{(2,1,1)}(SO(5))$ the global symmetry $USp(2)$ that connects to gauge group $O(3)$ has rank 1, which is smaller than $3-1=2$.  Thus the monopole formula \eref{Hilbert_series} diverges for both theories.

In this appendix, we show that the Hilbert series of the Coulomb branch of these quivers, which are both 2 quaternionic dimensional,  are equal.  As discussed in \eref{sympunc}, the Coulomb branch global symmetry $G_{\vec\rho}$ corresponding to both partitions is $USp(2)$.

From formula \eref{mainHL}, we obtain
\begin{align}
\begin{split}
&H[T_{(2,2,1)}(USp(4))]  (t; x; n_1, n_2) \\
&\qquad = t^{2n_1+n_2} (1-t)^2 K^{B_2}_{(2,2,1)} (x; t)  \Psi^{n_1,n_2}_{B_2} (t^{\frac{1}{2}} x, t^{\frac{1}{2}} x^{-1}; t)~,
\end{split}\\
\begin{split}
&H[T_{(2,1,1)}(SO(5))]  (t; x; n_1, n_2) \\
& \qquad = t^{\frac{1}{2}(3n_1+n_2)} (1-t)^2 K^{C_2}_{(2,1,1)} (x; t)  \Psi^{n_1,n_2}_{C_2} (t^{1/2}, x; t)~,
\end{split}
\end{align}
where the prefactors are
\be
\begin{split}
K^{B_2}_{(2,2,1)} (x; t) &= K^{C_2}_{(2,1,1)} (x; t) \\
&= \PE \left[ t \chi^{SU(2)}_{[2]}(x) +t^{3/2} \chi^{SU(2)}_{[1]}(x) +t^2 \right]~.
\end{split}
\ee
The two Hilbert series can be equated as follows:
\bea
H[T_{(2,2,1)}(USp(4))]  (t; x; n_1, n_2) = H[T_{(3,1,1)}(SO(5))]  (t; x; n_1+n_2, n_1-n_2)~.
\eea

The result for $n_1=n_2=0$ is
\bea
H[T_{(2,2,1)}(USp(4))]  (t; x; \vec 0) = 
\PE \left[ t \chi^{SU(2)}_{[2]} (x)+ t^{3/2} \chi^{SU(2)}_{[1]} (x) - t^4 \right]~.
\eea


\section{Hilbert series of hypermultiplet moduli spaces}

In this appendix we compute the Higgs branch Hilbert series of the mirror of various examples considered in the paper.  

\subsection{Hypermultiplet spaces of $(O(2))-[USp(4)]$ and $(SO(4))-[USp(4)]$} \label{app:3311_3221}
In appendix \ref{sec:3311VS3221}, we provide certain evidence that the hypermultiplet moduli spaces of $T_{(3,3,1,1)}(SO(8))$ and $T_{(3,2,2,1)}(SO(8))$ are identical.  In this appendix, we compute Hilbert series for the hypermultiplet moduli spaces of the uncommon part of the two theories, namely $(O(2))-[USp(4)]$ and $(SO(4))-[USp(4)]$ and show that they equal to each other.  After `gluing'%
\footnote{See \cite{Benvenuti:2010pq, Hanany:2011db} for more details on the computations of Higgs branch Hilbert series when two or more theories are `glued' together.} the resulting Hilbert series with that of $[USp(4)]-[SO(8)]$ via the $USp(4)$ group, we then expect the same Hilbert series for $T_{(3,3,1,1)}(SO(8))$ and $T_{(3,2,2,1)}(SO(8))$ as required. In the following we work with the variable
\bea
\tau = t^{1/2}~.
\eea

\subsubsection{The Higgs branch of $(O(2))-[USp(4)]$}

The quiver $(O(2))-[USp(4)]$ is almost identical to the ADHM quiver for 2 $USp(4)$ instantons on $\BC^2$ (see Fig. 7 of \cite{Hanany:2012dm}), except that the former has no symmetric hypermultiplet under $O(2)$ gauge group. Thus, the Hilbert series can be computed in a similar way as Section 4 of \cite{Hanany:2012dm}.

\paragraph{The contribution from the positive parity of $O(2)$.}  This is the Higgs branch of $(SO(2))-[USp(4)]$, whose Hilbert series is
\bea
g_+(\tau; \vec x) = \oint_{|z|=1} \frac{{\rm d} z}{2 \pi i z}  (1-\tau^2) \PE\left[ \chi^{USp(4)}_{[1,0]} (\vec x) (z+z^{-1}) \tau\right]~.
\eea
The first few terms in the $USp(4)$ character expansion of $g_+(\tau; \vec x)$ are
\be
\begin{split}
g_+(\tau; \vec x) &=  1+ (\chi^{C_2}_{[2,0]}(\vec x)+\chi^{C_2}_{[0,1]}(\vec x)) \tau^2  \\
& \qquad +(\chi^{C_2}_{[4,0]}(\vec x)+\chi^{C_2}_{[0,2]}(\vec x)+\chi^{C_2}_{[2,1]}(\vec x)) \tau^4 + \ldots~, \label{g+USp4}
\end{split}
\ee
and the unrefined Hilbert series is
\bea
g_+(\tau; \{ x_i =1 \}) = \frac{1 + 9 \tau^2 + 9 \tau^4 + \tau^6}{(1 - \tau^2)^6}~.
\eea

It should be observed that the $USp(4)$ representations appearing in \eref{g+USp4} embed into $SU(4)$ representations, namely 
\bea
g_+(\tau; \vec y) =\sum_{m=0}^\infty \chi^{SU(4)}_{[m,0,m]} (\vec y) \tau^{2m}~, \label{g+SU4}
\eea
where a fugacity map between $SU(4)$ and $USp(4)$ is%
\footnote{Here we take $\chi^{SU(4)}_{[1,0,0]} (\vec y)= \sum_{i=1}^4 y_i$ and $\chi^{USp(4)}_{[1,0]} (\vec x)= \sum_{i=1}^2 (x_i+ x_i^{-1})$.}
\bea
y_1 = x_1^{-1}~, \quad y_2= x_1~, \quad y_3 = x_2^{-1}~, \quad y_4 = x_2~.
\eea
Note that \eref{g+SU4} is in fact the Hilbert series of the reduced moduli space of 1 $SU(4)$ instanton on $\BC^2$, or of the Higgs branch of $U(1)$ gauge theory with $4$ flavors.

\paragraph{The contribution from the negative parity of $O(2)$.}   This is given by
\be
\begin{split}
g_-(\tau; \vec x) &= \frac{1+\tau^2}{ \prod_{i=1}^2 \det( {\bf 1}_{2} - \tau x_i \sigma_-(z))\det( {\bf 1}_{2} - \tau x_i^{-1} \sigma_-(z)) }  \\
&= (1+\tau^2) \PE[ \tau^2 \chi^{USp(4)}_{[1,0]} (x_1^2, x_2^2) ]~,
\end{split}
\ee
where $ {\bf 1}_{2}$ denotes a two-by-two identity matrix and 
\bea
\sigma_-(z) = \begin{pmatrix} 0 & z \\ z^{-1} & 0 \end{pmatrix}~. \label{parityO2}
\eea

\paragraph{Combining $g_+$ and $g_-$.}  The Higgs branch Hilbert series of $(O(2))-[USp(4)]$ is therefore
\be
\begin{split}
& g[(O(2))-[USp(4)]](\tau; \vec x)
= \frac{1}{2} \left(g_+(\tau; \vec x) + g_-(\tau; \vec x)  \right)  \\
&\qquad = 1+ [2,0]_{\vec x} \tau^2 +( [4,0]_{\vec x} +[0,2]_{\vec x})t^2 +( [6,0]_{\vec x} +[2,2]_{\vec x})\tau^3  \\
& \hspace{1.5cm} +( [8,0]_{\vec x} +[4,2]_{\vec x}+[0,4]_{\vec x})\tau^4+\ldots~,
\end{split}
\ee
where we abbreviate $[a_1, a_2]_{\vec x} = \chi^{USp(4)}_{[a_1,a_2]} (\vec x)$.  The unrefined Hilbert series is
\bea
g[(O(2))-[USp(4)]](\tau; \{ x_i =1\}) = \frac{1+4 \tau^2+4 \tau^4+\tau^6}{\left(1-\tau^2\right)^6}~. \label{O2C2}
\eea

\subsubsection{The Kibble branch of $(SO(4))-[USp(4)]$}

Let us denote the half-hypermultiplets in the theory by $Q_a^i$, where $a=1,\ldots, 4$ is an $SO(4)$ gauge index and $i=1, \ldots, 4$ is a $USp(4)$ flavor index.  The superpotential is $W= Q_a^i Q_b^j \varphi^{ab} J_{ij}$, where $J$ is a symplectic matrix and $\varphi$ is an adjoint field under $SO(4)$. Using {\tt Macaulay2} \cite{M2}, we find the space of $F$-term solutions (\ie~ the $F$-flat space) defined by 
\bea
0 = \partial_{\phi^{ab}} W = Q_a^i Q_b^j J_{ij}
\eea
is a 11 complex dimensional space; {\it not} $16-\frac{1}{2}(4 \times 3) =10$ complex dimensional if the gauge group $SO(4)$ is completely broken.    Hence we conclude that at a generic point of the hypermultiplet moduli space (also known as the {\it Kibble branch} \cite{Hanany:2010qu}), the gauge symmetry $SO(4)$ is broken to $SO(2)$. The remaining unbroken symmetry on the Kibble branch is indeed as expected, since $(SO(4))-[USp(4)]$ is a `bad' theory.

The Hilbert series of the $F$-flat space can be computed using {\tt Macaulay2}:
\be
\begin{split}
{\cal F}^\flat (\tau; \vec x; \vec z) &= \PE \left[ \tau [1,0]_{\vec x} [1,0]_{\vec z} \right] \times  \\
& \Big[ 1 -([2,0]_{\vec z} + [0,2]_{\vec z}) \tau^2 +([2,2]_{\vec z} +[2,0]_{\vec z}+[0,2]_{\vec z}+[0,1]_{\vec x})\tau^4   \\
& -([1,1]_{\vec z}[1,0]_{\vec x}) \tau^5 -([2, 2]_{\vec z} + 1)\tau^6+([1,1]_{\vec z}[1,0]_{\vec x}) \tau^7- [0,1]_{\vec z} \tau^8 \Big]~,
\end{split}
\ee
where we abbreviate $[a_1, a_2]_{\vec x} = \chi^{USp(4)}_{[a_1,a_2]} (\vec x)$ and $[b_1, b_2]_{\vec z} = \chi^{SO(4)}_{[b_1,b_2]} (\vec z)$.  The unrefined Hilbert series of the $F$-flat space is
\bea
{\cal F}^\flat (\tau; \{x_i=1 \}; \{z_i=1 \}) &= \frac{1 + 5 \tau+ 9 \tau^2 + 5 \tau^3}{(1 - \tau)^{11}}~.
\eea
Note that the $F$-flat space is indeed $11$ complex dimensional.

Integrating over the Haar measure of $SO(4)$, we obtain the Kibble branch Hilbert series of $(SO(4))-[USp(4)]$
\bea
 g[(SO(4))-[USp(4)]](\tau; \vec x) &= \int {\rm d} \mu_{SO(4)} (\vec z) \; {\cal F}^\flat (\tau; \vec x; \vec z) \nn \\
&= 1+ [2,0]_{\vec x} \tau^2 +( [4,0]_{\vec x} +[0,2]_{\vec x})\tau^2 +( [6,0]_{\vec x} +[2,2]_{\vec x})\tau^3 \nn \\
& \quad +( [8,0]_{\vec x} +[4,2]_{\vec x}+[0,4]_{\vec x})\tau^4+\ldots~.
\eea
The corresponding unrefined Hilbert series is
\bea
g[(SO(4))-[USp(4)]](\tau; \{ x=1 \})  = \frac{1+4 \tau^2+4 \tau^4+ \tau^6}{\left(1-\tau^2\right)^6}~,
\eea
which is is equal to \eref{O2C2}.

\subsection{The Higgs branch of the mirror of $T_{(3,1,1)}(USp(4))$ and the baryonic generating function} \label{app:HiggsT311Usp4}

In this section of the appendix, we discuss the computation of the Higgs branch Hilbert series and the baryonic generating function of the mirror theory of $T_{(3,1,1)}(USp(4))$ theory, whose quiver is given by \eref{mirT311Usp4}.

Let us now compute the baryonic generating on the Higgs branch of this mirror theory.  Let us denote the chiral fields in the quiver as follows:
{\small
\bea
\begin{array}{lll}
[SO(2)]-(USp(2)): &\quad Q^{i_1}_{~a_1} & \quad \text{$i_1=1,2$ of $[SO(2)]$}, ~\text{$a_1=1,2$ of $(USp(2))$,} \\
(USp(2))-(SO(2)): &\quad X^{a_2}_{~a_1} & \quad \text{$a_2=1,2$ of $(SO(2))$,} \\
(SO(2))-(USp(2)): &\quad Y^{a_2}_{~a_3} & \quad \text{$a_3=1,2$ of $(USp(2))$,}\\
(USp(2))-[SO(2)]: &\quad q^{j_2}_{~b_3} & \quad \text{$j_2=1,2$ of $[SO(2)]$}, ~\text{$b_3=1,2$ of $(USp(2))$}
\end{array}
\eea}
The superpotential is given by
\be
\begin{split}
W&= M^{SO(2)}_{i_1 i'_1} \phi_1^{a_1 b_1} Q^{i_1}_{~a_1}  Q^{i'_1}_{b_1} + \epsilon^{a_1 a'_1}(\phi_2)_{a_2 a'_2} X^{a_2}_{~a_1} X^{a'_2}_{~a'_1} -\phi_1^{a_1 a'_1} M^{SO(2)}_{a_2 a'_2} X^{a_2}_{~a_1} X^{a'_2}_{~a'_1} \\
& \quad  + \epsilon^{a_3 a'_3}(\phi_2)_{a_2 a'_2} Y^{a_2}_{~a_3} Y^{a'_2}_{~a'_3} -\phi_3^{a_3 a'_3} M^{SO(2)}_{a_2 a'_2} Y^{a_2}_{~a_3} Y^{a'_2}_{~a'_3}+ M^{SO(2)}_{j_2 j'_2} \phi_3^{a_3 b_3} Q^{j_2}_{~a_3}  Q^{j'_2}_{b_3}~,
\end{split}
\ee
where $M^{SO(2)}$ is a matrix associated with the bilinear form of $SO(2)$.  In order for the Lie algebra to contain a nonzero diagonal matrix, which is important for fugacity assignments, we take%
\footnote{If $M^{SO(2)}$ is taken to be the identity matrix, the Lie algebra would consist of anti-symmetric matrices, and there would be no nonzero diagonal matrix in the Lie algebra.}
\bea
M^{SO(2)} = \begin{pmatrix} 0& 1 \\ 1& 0 \end{pmatrix}~.
\eea

Since the mirror theory is a `bad' theory,  we expect that the space of $F$-terms solutions (\ie~ the $F$-flat space)
\bea
\{ \partial_{\phi_1} W =0, ~ \partial_{\phi_2} W =0,~ \partial_{\phi_3} W =0 \}~,
\eea
has many branches.  
In order to determine the relevant branch (9 complex dimensional space), we perform the {\bf primary decomposition}, which can be done using mathematical packages such as {\tt STRINGVACUA} \cite{Gray:2008zs} or {\tt SINGULAR} \cite{DGPS}.  After such a decomposition, we find using {\tt Macaulay2} \cite{M2} the Hilbert series of the relevant branch of the $F$-flat space.  Since the result is too long to be reported here, we present the first few terms in the series expansion in $\tau$:
\be
\begin{split}
& \fflat(t; x, y; z_1, b, z_2)  \\
&=  1+ \tau \Big\{ ( x^{\pm1}+ b^{\pm1})[1]_{z_1} +( y^{\pm1}+ b^{\pm1})[1]_{z_2}  \Big\}  \\
& \quad + \tau^2 \Big \{ (b^{\pm2}+x^{\pm2}+b^{\pm1}x^{\pm1}+1 )[2]_{z_1} + (x \rightarrow y, \; z_1 \rightarrow z_2)  \\ 
& \qquad \quad + (x^{\pm1} y^{\pm 1}+b^{\pm1} y^{\pm 1}+x^{\pm1} b^{\pm 1}+b^{\pm2}+2 )[1]_{z_1}[1]_{z_2} \\
& \qquad \quad + (b x^{-1})^{\pm 1} + (b y^{-1})^{\pm1} +1\Big\} + \ldots~,
\end{split}
\ee
where $x,  y$ are the fugacities of the two $SO(2)$ flavor symmetries and $z_1, b, z_2$ are respectively the fugacitites of the gauge groups $USp(2),  SO(2),  USp(2)$.  We have used the shorthand notation $a^{\pm n} = a^n+a^{-n}$. The unrefined $F$-flat Hilbert series is
\bea
\fflat(\tau; 1, 1; 1, 1, 1)  = \frac{(1+\tau) (1+3 \tau)^2}{(1-\tau)^9}~.
\eea

Let us next integrate over the two $USp(2)$ gauge groups, but not the $SO(2)$ gauge group:
\bea
&\frac{1}{(2 \pi i)^2}\oint_{|z_1|=1} \frac{1-z_1^2}{z_1} \mathrm{d} z_1 \oint_{|z_2|=1} \frac{1-z_2^2}{z_2} \mathrm{d} z_2\; \fflat(\tau; x, y; z_1, b, z_2)  \nn \\
&= \PE \left[ \{ 1+ b (x^{-1}+y^{-1}) + b^{-1} (x+y) \} \tau^2  -2\tau^4  \right]~.
\eea
The baryonic generating function for this theory is obtained by ungauging the $SO(2)$ group; this amounts to multiplying the above function by $(1-\tau^2)^{-1}$ to remove the contribution of the $F$-term from this $SO(2)$ group:
\bea  \label{GmirT311USp4}
&\CG[\text{Mirror of $T_{(3,1,1)}(USp(4))$}/SO(2)](t; x, y; b) \nn \\
&= \PE \left[ \{ 2+(b_1+b_1^{-1}) + (b_2+b_2^{-1}) \} \tau^2 -  2\tau^4 \right]~.
\eea
The Higgs branch Hilbert series is then given by
\bea \label{HiggsmirT311USp4}
&\oint_{|b|=1} \frac{{\rm d} b}{2 \pi i b} (1-\tau^2)\CG[\text{mirror of $T_{(3,1,1)}(USp(4))/SO(2)$}](\tau; x, y; b) \nn \\
&= \PE \left[ \tau^2 + \tau^4( x y^{-1} + y x^{-1}) -\tau^8 \right]~.
\eea


\section{Derivation of \eref{finalgen} from the monopole formula} \label{sec:analytgeneral}

The monopole formula for the Coulomb branch Hilbert series for $T_{\vec \rho} (SU(N))$ reads
\be \label{stdrho1SUN}
\begin{split}
& H[T_{\vec \rho}(SU(N))] (t; \vec x; \vec n) \\
& =x_1^{\sum_{j=1}^N n_j} \sum_{m_{1,1} \in \BZ} ~\sum_{m_{2,2}\ge m_{1,2} > -\infty} \cdots ~\sum_{m_{N-\rho_1,N-\rho_1}\ge \cdots \ge m_{1,N-\rho_1} > -\infty} \\
& \quad \left(\frac{x_2}{x_1}\right)^{\sum\limits_{i=1}^{N-\rho_1} m_{i,N-\rho_1}} \cdots \left(\frac{x_{d-h+1}}{ x_{d-h}}\right)^{\sum\limits_{i=1}^{h+\ell} m_{i,h+\ell}} \left(\frac{x_{d-h+2}}{ x_{d-h+1}}\right)^{\sum\limits_{i=1}^{h} m_{i,h}} \cdots \left(\frac{x_{d+1}}{ x_{d}}\right)^{m_{1,1}}  \times \\
& \quad t^{\frac{1}{2} \Delta{(\rho_1, \cdots, \rho_{d-h})} + \frac{1}{2} \Delta{(\ell,1^h)}} P_{U(N-\rho_1)}(t;\{ m_{i,N-\rho_1}  \}) \cdots P_{U(2)}(t;\{ m_{i,2} \}) P_{U(1)}(t)~,
\end{split}
\ee
where the dimension formula $ \Delta{(\rho_1, \cdots, \rho_{d-h})} $  corresponds to the quiver tail $[N]-\cdots-(h+\ell+\rho_{d-h})-(h+\ell)$ including the contribution of the $U(h+\ell)$ vector multiplet, and $\Delta{(\ell,1^h)}$ corresponds to the tail $[h+\ell]-(h)-\cdots-(2)-(1)$ without the contribution of the $U(h+\ell)$ vector multiplet. Explicitly, the latter is given by
\be
\begin{split}
  \Delta_{(\ell,1^h)} &= \sum_{j=1}^{h-1} \sum_{i=1}^j \sum_{i'=1}^{j+1} | m_{i,j}-m_{i',j+1} | +\sum_{i=1}^h \sum_{i'=1}^{h+\ell} | m_{i,h}-m_{i',h+\ell} | \\
& \quad -2 \sum_{j=2}^h \sum_{1\leq i<i'\leq j} (m_{i,j}-m_{i',j})~.
\end{split}
\ee

We will show that the monopole formula \eref{stdrho1SUN} for the Coulomb branch Hilbert series has a simple pole at $x_{d-h+1}x_{d+1}^{-1} = t^{\frac{1}{2}(1+\ell)}$, and compute the residue at $z\to 1$ when the fugacities associated to the partition $\rho$ satisfy \eref{pole_gen}, to reproduce \eref{finalgen}.

The pole is due to the region in which $m_{1,1}, m_{2,2}, \ldots, m_{h,h}$ are much larger than other $m_{i,j}$ and all $n_i$. 
In the brane picture at the top of \fref{fig:partition}, this limit corresponds to considering the $h$ rightmost pairs of adjacent NS5-branes, and moving one D3-brane attached to each such pair far out on the Coulomb branch. This limit leaves the brane picture at the bottom of \fref{fig:partition} at finite distance.
Setting
\bea
m_{h-k,h-k}=m_{h,h} + \sum_{i=1}^k q_i~, \qquad k = 1, \ldots, h-1~,
\label{mandq}
\eea
we obtain
\bea \label{Deltaconv}
 \Delta{(\ell,1^h)} = \Delta{(\ell+1,1^{h-1})} + \widehat{\Delta}~,
\eea
where 
\be
\begin{split}
\widehat{\Delta}&= \sum_{j=1}^{h-1} \sum_{i'=1}^j (m_{j,j}-m_{i',j+1}) + \sum_{j=2}^{h-1} \sum_{i=1}^{j-1} (m_{j+1,j+1} - m_{i,j}) + \sum_{i=1}^{h-1} |m_{i,i}-m_{i+1,i+1}|  \\
 & \quad + \sum_{i'=1}^{h+\ell} (m_{h,h} - m_{i',h+\ell}) - 2 \sum_{j=2}^h \sum_{i=1}^{j-1} (m_{j,j}-m_{i,j})  \\
 &= -\sum_{i=1}^{h+\ell} m_{i,h+\ell} + \sum_{i=1}^{h-1} |q_i| + m_{1,1}+ \ell m_{h,h} + \sum_{i=1}^{h-1} m_{i,h}  \\
 &= -\sum_{i=1}^{h+\ell} m_{i,h+\ell} + \sum_{i=1}^{h-1} |q_i| +(\ell+1) m_{h,h}+ \sum_{i=1}^{h-1}q_i+ \sum_{i=1}^{h-1} m_{i,h} 
\end{split}
\ee
and $\Delta{(\ell+1,1^{h-1})}$ corresponds to the tail $[h+\ell]-(h-1)-\cdots-(1)$ without the contribution of the $U(h+\ell)$ vector multiplet. Note that the magnetic charges of the $U(j-1)$ gauge group in the $[h+\ell]-(h-1)-\cdots-(1)$ tail of $T_{\vec \rho^\prime}(SU(N))$, with $j=2,\dots,h$, arise from the magnetic charges of the $U(j)$ gauge group in the original $[h+\ell]-(h-1)-\cdots-(1)$ tail of $T_{\vec \rho}(SU(N))$ which are kept finite:
\bea \label{mapmono}
\text{GNO charges for the $U(j-1)$ group of $T_{\vec \rho^\prime}(SU(N))$}: \qquad (m_{i,j})_{i=1}^{j-1}~.
\eea

Next, we focus on the changes in the topological factors in the second line of \eref{stdrho1SUN}.  The factor that involves the fugacities $x_{d-h+1}$ and $x_{d+1}$ associated to the blue boxes becomes, using \eref{pole_gen} and \eref{mandq}, 
\be \label{reltopfac}
\begin{split}
&(x_{d-h+1})^{\sum_{i=1}^{h+\ell} m_{i,h+\ell}-\sum_{i=1}^{h} m_{i,h}} \cdot x_{d+1}^{m_{1,1}} = \\ 
&=  (t^{1/2} z)^{\sum_{i=1}^{h+\ell} m_{i,h+\ell}-\sum_{i=1}^{h-1} m_{i,h} -(\ell+1) m_{h,h}- \ell \sum_{i=1}^{h-1}q_i} \times  \\
& \quad y_{g-h+1}^{\sum_{i=1}^{h+\ell} m_{i,h+\ell}-\sum_{i=1}^{h-1} m_{i,h} + \sum_{i=1}^{h-1}q_i}~.
\end{split}
\ee
The factor that involves the fugacities associated to the $h-1$ pink boxes becomes
\be  \label{reltopfac2}
\begin{split}
& \prod_{j=1}^{h-1} (x_{d-h+1+j})^{\sum_{i=1}^{h+1-j} m_{i,h+1-j} - \sum_{i=1}^{h-j} m_{i,h-j}} = \\
&= \left \{ \prod_{j=1}^{h-2} (x_{d-h+1+j})^{\sum_{i=1}^{h-j} m_{i,h+1-j} - \sum_{i=1}^{h-j-1} m_{i,h-j}} \times x_d^{m_{1,2}} \right \}  \prod_{j=1}^{h-1} x_{d-h+1+j}^{-q_j}~,
\end{split}
\ee
where the quantity in the curly bracket $\{ \cdots \}$ becomes parts of the new topological factors after the box is moved.
The topological factors associated to the grey boxes are not affected.

Finally we consider the fate of the classical factors of the $U(j)$ gauge groups, for $j=1,\dots,h$, in the limit where $m_{1,1}, m_{2,2}, \ldots, m_{h,h}$ are much larger than other $m_{i,j}$ and all $n_i$. The limit breaks $U(j)\to U(j-1)\times U(1)$ along the Coulomb branch, and 
\be \label{limit_class}
P_{U(j)} (t; \{m_{i,j}\}_{i=1}^{j-1}, m_{j,j} ) = P_{U(j-1)} (t; \{m_{i,j}\}_{i=1}^{j-1} ) \times P_{U(1)}(t)~. 
\ee

Combining the relevant terms together, we summarize each factor below: 
\bi
\item The power of $t^{1/2}$. We combine \eref{Deltaconv} and \eref{reltopfac} and yield
\bea
\Delta{(\rho_{g-h+1}+1,1^{h-1})} +\sum_{i=1}^{h-1} |q_i|-(\ell-1)\sum_{i=1}^{h-1} q_i~.
\eea
\item The power of $z$. This comes from \eref{reltopfac},
\bea
\sum_{i=1}^{h+\ell} m_{i,h+\ell}-\sum_{i=1}^{h-1} m_{i,h} -(\ell+1) m_{h,h}- \ell \sum_{i=1}^{h-1}q_i~.
\eea
\item The power of $y_{d-h+1}$. This comes from \eref{reltopfac},
\bea
\sum_{i=1}^{h+\ell} m_{i,h+\ell}-\sum_{i=1}^{h-1} m_{i,h} + \sum_{i=1}^{h-1}q_i~.
\eea
\item The factors containing $x$'s in the $(1^{h-1})$ block. These come from \eref{reltopfac2},
\bea
\left \{ \prod_{j=1}^{h-2} (x_{d-h+1+j})^{\sum_{i=1}^{h-j} m_{i,h+1-j} - \sum_{i=1}^{h-j-1} m_{i,h-j}} \times x_d^{m_{1,2}} \right \}  \prod_{j=1}^{h-1} x_{d-h+1+j}^{-q_j}~.
\eea
\item The classical factors. These come from \eref{limit_class},
\be
\prod_{j=1}^h P_{U(j)} (t, \{m_{i,j}\}_{i=1}^j ) = \PE[ h t ] ~ \prod_{j=2}^{h} P_{U(j-1)} (t, \{m_{i,j}\}_{i=1}^{j-1} )~. \label{summ3}
\ee
\ei

Finally, we perform the summations over the large magnetic charges $\{m_{j,j}\}_{j=1}^h$, or equivalently over $m_{h,h}$ and $q_1,\dots,q_{h-1}$:
\ben
\item The summation over $m_{h,h}$: the summand depends on $m_{h,h}$ only via $z^{-(\ell+1)m_{h,h}}$. This is responsible for a simple pole when $z\to 1$, with residue
\be
\res_{z \rightarrow 1} \sum_{m_{h,h} \geq L} z^{-(1+\ell) m_{h,h}} =  \frac{1}{1+\ell} ~. \label{summ1}
\ee
The remaining factors are finite as $z\to 1$, that we set in the following.  
\item The summations over $q_i$ yield
\be\label{summ2}
\begin{split}
&\prod_{i=1}^{h-1} \sum_{q_i \in \BZ} t^{\frac{1}{2} |q_i|}  \left(t^{-\frac{1}{2}(\ell-1)} \frac{y_{d-h+1}}{x_{d-h+1+i}} \right)^{q_i} \\
&= \prod_{i=1}^{h-1} \PE \left[-t +t^{\frac{1}{2}} \sum_{s = \pm1} \left(t^{-\frac{1}{2}(\ell-1)} \frac{y_{d-h+1}}{x_{d-h+1+i}} \right)^s  \right] \\
&= \PE \left[ -(h-1)t + t^\frac{1}{2} \sum_{i=1}^{h-1} \sum_{s =\pm 1}  \left(t^{-\frac{1}{2}(\ell-1)} \frac{y_{d-h+1}}{x_{d-h+1+i}} \right)^s  \right]~. 
\end{split}
\ee
\een
Combining \eref{summ1} and \eref{summ2} with the classical factor $\PE[h t]$ of $U(1)^h$ in \eref{summ3}, we recover ${\cal P}_{\vec \rho \vec \rho^\prime}$ in \eref{defPfac}. The remaining factors and summations combine with the factors and summations of the spectator part of the quiver to reconstruct the monopole formula for the Coulomb branch Hilbert series of $T_{\vec \rho^\prime}(SU(N))$. Hence we have reproduced formula \eref{finalgen} from the monopole formulae.


\section{From $U(N_c)$ gauge theory with $N_f$ flavors to $N_f-1$ flavors}
\label{sec:reduceflav}

In this appendix we show how one can obtain the Coulomb branch Hilbert series of $U(N_c)$ gauge theory with $N_f-1$ flavors from that with $N_f$ flavors by means of a residue computation. The idea can be explained physically as follows. We gauge a $U(1)$ subgroup of the $SU(N_f)$ flavor symmetry, introducing a flat direction from the Coulomb branch of the $U(1)$ gauge group. Associated to the flat direction there is a pole in the Coulomb branch Hilbert series, whose residue is related to the Hilbert series of the leftover $U(N_c)$ gauge theory with $N_f-1$ flavors.

We start from the Hilbert series for $U(N_c)$ gauge theory with $N_f$ flavors,
\bea
& H[U(N_c),N_f] (t; z, z_0; n_1, \ldots, n_{N_f}) \nn \\
& = z_0^{\sum_{i=1}^{N_f} n_i} \sum_{m_1 \geq \ldots \geq m_{N_c} > -\infty} t^{\Delta_{N_c,N_f}(\vec m,  \vec n)} P_{U(N_c)} (t; m_1, \ldots, m_{N_c}) z^{m_1+\ldots+m_{N_c}}~,
\eea
where 
\bea
\Delta_{N_c,N_f}(\vec m,  \vec n) &= \frac{1}{2} \sum_{i=1}^{N_c} \sum_{f=1}^{N_f} |m_i-n_f| - \sum_{1 \leq i <j \leq N_c} |m_i - m_j|~.
\eea
If we gauge the $U(1)$ subgroup of $SU(N_f)$ associated to the magnetic flux $n_{N_{f}}$, and let $w$ be the fugacity of the corresponding topological symmetry, the Hilbert series becomes
\be
\begin{split}
& H(t; z_0, z, w; n_1, \ldots, n_{N_f-1}) \nn \\
& = \frac{1}{1-t} ~ z_0^{\sum_{i=1}^{N_f-1} n_i} \sum_{n_{N_f} \in\bZ}   \sum_{m_1 \geq \ldots \geq m_{N_c} > -\infty} t^{\Delta(\vec m,  \vec n)} P_{U(N_c)} (t; \vec m) z^{m_1+\ldots+m_{N_c}} w^{n_{N_f}}~.
\end{split}
\ee

Now we set
\be
 z = t^{1/2} y x~, \qquad w = t^{-\frac{1}{2}N_c} x^{-1}
\ee
and consider the $x\to 1$ limit. Along the lines of the previous subsection, we find that there is a simple pole due to the region where $n_{N_f}\gg m_i, n_j$,  with $i=1,\dots,N_c$ and $j=1,\dots,N_f-1$, and the residue is
\be
\begin{split}
\res_{x \rightarrow 1} & ~H(t;\;z_0,\;  t^{\frac{1}{2}} y x, \; t^{-\frac{1}{2} N_c} x^{-1}; n_1, \ldots, n_{N_f-1}) \nn \\
 &= \frac{1}{1-t} ~ H[U(N_c),N_f-1] (t; y, n_1, \ldots, n_{N_f-1})~,
\end{split}
\ee
where $H[U(N_c),N_f-1] $ is the Coulomb branch Hilbert series of the $U(N_c)$ gauge theory with $N_f-1$ hypermultiplets.

\bibliographystyle{ytphys}
\bibliography{ref_v2}

\end{document}